\documentclass[letterpaper,11pt]{article}%
\usepackage{amsfonts}
\usepackage{chicago}
\usepackage{amsmath}
\usepackage{graphicx}
\usepackage{amssymb}
\usepackage[letterpaper]{geometry}
\usepackage[onehalfspacing]{setspace}%
\providecommand{\U}[1]{\protect\rule{.1in}{.1in}}
\providecommand{\U}[1]{\protect\rule{.1in}{.1in}}
\newenvironment{proof}[1][Proof]{\textbf{#1.} }{\ \rule{0.5em}{0.5em}}

\begin{document}

\title{Identification robust inference for moments based analysis of linear dynamic
panel data models }
\author{Maurice J.G. Bun\thanks{Economics and Research Division (De Nederlandsche
Bank) and Amsterdam School of Economics (University of Amsterdam). Email:
m.j.g.bun@uva.nl. The research of the author has been funded by the NWO
Vernieuwingsimpuls research grant `Causal Inference with Panel Data'.}
\and Frank Kleibergen\thanks{Amsterdam School of Economics, University of
Amsterdam, Roetersstraat 11, 1018 WB Amsterdam, The Netherlands. Email:
f.r.kleibergen@uva.nl. \newline\textbf{Acknowledgements: }We thank
Manuel\ Arellano, Richard Blundell, Steve Bond, Peter Boswijk, Geert Dhaene,
Frank Windmeijer and participants of seminars at Bristol, CEMFI, CORE, the
Cowles Summer Conference at Yale, the EC$^{2}$ meeting in Maastricht,
Groningen, Leuven, Monash, the 19-th International conference on panel data in
London, Oxford, the Tinbergen Institute in\ Amsterdam, Toulouse and UCL for
helpful comments and discussion.}}
\maketitle

\begin{abstract}
We use identification robust tests to show that difference, level and
non-linear moment conditions, as proposed by Arellano and Bond (1991),
Arellano and Bover (1995), Blundell and Bond (1998) and Ahn and Schmidt (1995)
for the linear dynamic panel data model, do not separately identify the
autoregressive parameter when its true value is close to one and the variance
of the initial observations is large. We prove that combinations of these
moment conditions, however, do so when there are more than three time series
observations. This identification then solely results from a set of,
so-called, robust moment conditions. These robust moments are spanned by the
combined difference, level and non-linear moment conditions and only depend on
differenced data. We show that, when only the robust moments contain
identifying information on the autoregressive parameter, the discriminatory
power of the Kleibergen (2005) LM test using the combined moments is identical
to the largest rejection frequencies that can be obtained from solely using
the robust moments. This shows that the KLM test implicitly uses the robust
moments when only they contain information on the autoregressive
parameter.\medskip

JEL\ codes: C12, C23, C26

\end{abstract}

\newpage

\section{Introduction}

It is common to estimate the parameters of linear dynamic panel data models
using the Generalized Method of Moments (GMM, Hansen (1982)). The moment
conditions for the linear dynamic panel data model either analyze it in first
differences using lagged levels of the series as instruments, in levels using
lagged first differences as instruments or using a product of levels and first
differences. We refer to the first set of moment conditions as Dif(ference)
moment conditions, see Arellano and Bond (1991), the second set as Lev(el)
moment conditions, see Arellano and Bover (1995), Blundell and Bond (1998) and
the third set as N(on-)L(inear) moment conditions, see Ahn and Schmidt (1995).

The Dif, Lev and NL moment conditions can be used separately to identify the
parameters of dynamic panel data models. To exhaust all information, however,
two particular combinations of Dif, Lev and NL moment conditions have been
proposed. We refer to the combined Dif and Lev moment conditions as the
Sys(tem) moment conditions and the combination of the Dif and NL moment
conditions as the A(hn-)S(chmidt) moment conditions.\footnote{Note that in a
combination of all three sets of moments conditions, the NL\ moment conditions
are redundant.} The Sys moment conditions exhaust all information on the
autoregressive parameter that is present under mean stationarity, see Arellano
and Bover (1995) and Blundell and Bond (1998). The AS moment conditions
exhaust all information whilst not assuming mean stationarity, see Ahn and
Schmidt (1995).

We analyze the identification of the autoregressive parameter by the various
sets of moment conditions for a range of true values including the case of
highly persistent panel data. All moment conditions involve first differences
of the series to remove individual specific effects. The first difference
operator removes information in the time series at the unit root value of the
autoregressive parameter. It is well known that the Dif moment conditions
therefore do not identify the autoregressive parameter when its true value is
(close to) one, since lagged levels are then weak predictors of first
differences. This has led to the development of the NL and Lev, and hence AS
and Sys, moment conditions which were originally considered to identify the
autoregressive parameter when the panel data are highly persistent.

To show the identification issues at specific values of the autoregressive
parameter, we use identification robust tests, i.e. the GMM-A(nderson-)R(ubin)
statistic of Anderson and Rubin (1949)\nocite{AR49} and Stock and Wright
(2000),\nocite{sw00} and the K(leibergen) L(agrange) M(ultiplier) statistic of
Kleibergen (2005)\nocite{kf00a}. At values of the parameters where
identification issues occur, the rejection frequency of these tests provenly
coincides with the significance level so the identification issues are
relatively easy to detect by inspecting the power curves. Using power curves
of the KLM test, we show that Dif, Lev\ and NL moment conditions separately do
not identify the autoregressive parameter for persistent values of it when
paired with a large variance of the initial observations. The same holds for
the Sys moment conditions with three times series observations. The power
curves further show that Sys and AS moment conditions generally identify the
autoregressive parameter when the number of time series observations exceeds three.

We formally prove these identification results using an asymptotic sampling
scheme in which we jointly let the variance of the initial observations and
the number of cross section observations go to infinity. For a range of
relative convergences rates of the variance of the initial observations
compared to the cross section sample size, the Dif, Lev and NL sample moments
and their derivatives diverge. Both the population moment and the Jacobian
identification condition are then ill defined which implies that the
autoregressive parameter is not separately identified by the Dif, NL or Lev
moment conditions. These results confirm and extend earlier findings in Madsen
(2003)\nocite{mad03}, \nocite{bonwind03} Bond $et$ $al.$
(2005),\nocite{bondnaugwind05} Hahn $et$ $al.$ (2007),\nocite{hhk07} Kruiniger
(2009)\nocite{krui09} and Phillips (2018).\nocite{phillip2018}

Using our asymptotic sampling scheme, we also prove that AS and Sys moment
conditions identify the autoregressive parameter irrespective of the variance
of the initial observation when the number of time series observations exceeds
three. When the variance of the initial observations is large, the
identification results from a set of, so-called, robust sample moments that
are a combination of the Dif, Lev and NL\ sample moments (other than AS and
Sys) and\ only depend on differenced data. These robust sample moments are
spanned by the Sys sample moments and also by the AS sample moments. They
identify the autoregressive parameter irrespective of the variance of the
initial observation and including the case of highly persistent data. They are
a subset of the moment conditions in Kruiniger (2002)\nocite{krui02}, which
are derived under the additional assumption of time series homoskedasticity.

Despite these positive identification results for the Sys and AS moments, the
large sample distributions of corresponding one step and two step GMM
estimators are known to be non-standard when the variance of the initial
observation is large and the autoregressive parameter is close to one. This
makes it hard to infer if and how standard GMM inference using the original AS
or Sys sample moments exploits the information contained in the robust sample
moments that they encompass. The non-standard limiting behavior results since
the identification of the autoregressive parameter is then of, so-called,
second order since the Jacobian of the robust sample moments is rank deficient
but the Hessian is not, see $e.g.$ Dovonon and Renault (2013),\nocite{dr13}
Dovonon and Hall (2018)\nocite{dh16} and Dovonon $et$ $al.$
(2020).\nocite{dhk17} It explains the large biases of the one step and two
step GMM estimators and the size distortions of their corresponding
t-statistics when the series are persistent, see $e.g.$ Madsen
(2003)\nocite{mad03}, Bond and Windmeijer (2005),\nocite{bonwind03} Bond $et$
$al.$ (2005),\nocite{bondnaugwind05} Dhaene and Jochmans
(2016),\nocite{dhjoch12} Hahn $et$ $al.$ (2007),\nocite{hhk07} Kruiniger
(2009)\nocite{krui09} and Bun and Windmeijer (2010).\nocite{bw10} Because of
the second order identification, GMM estimators based on the robust sample
moments also have non-standard asymptotic distributions when the data are
persistent, see Dovonon $et$ $al.$ (2020).\nocite{dhk17}

We therefore analyze how identification robust test statistics exploit the
identifying information in the robust sample moments. We prove that the
identification robust KLM test procedure based on either AS or Sys sample
moments exploits all the identifying information contained in the robust
sample moments. We do so by first determining the (infeasible) optimal
weighted average of the robust sample moments that maximizes the
discriminatory power of a GMM-AR test of the autoregressive parameter in
settings where only the robust sample moments contain identifying information.
Next we determine the discriminatory power of KLM\ tests, based on AS or Sys
moment conditions, under such settings and prove that it equals that of the
GMM-AR test using the optimal weighted average of the robust sample moments.
KLM tests using AS or Sys moment conditions thus resort to just using the
robust sample moments when only the latter contain information on the
autoregressive parameter. It is therefore not necessary to explicitly use the
robust sample moments, which provide identification under mild conditions,
since they are implicitly used in the KLM test based on either AS or Sys
sample moments.

The paper is organized as follows. Section 2 introduces the linear dynamic
panel data model and the different moment conditions we use to identify its
parameters. It also discusses identification robust statistics, specifically
the KLM\ test, that we use to illustrate the identification issues that occur
at persistent values of the autoregressive parameter. In Section 3, we use a
representation theorem, akin to the cointegration representation theorem, see
Engle and Granger (1987)\nocite{eg87} and Johansen (1991),\nocite{JS91} to pin
down the identification properties of the different moment conditions. This
theorem also allows us to obtain the robust sample moments. In Section 4, we
define the GMM-AR test that uses the (infeasible) optimal weighted average of
the robust sample moments and derive the large sample distribution of the
KLM\ test using AS or Sys moment conditions under settings where only the
robust sample moments contain information on the autoregressive parameter. The
fifth (final) section concludes. Proofs of theorems and definitions of sample
moments are provided in the Appendix. We use the following notation throughout
the paper: vec($A)$ stands for the (column) vectorization of the $k\times n$
matrix $A,$ vec($A)=(a_{1}^{\prime}\ldots a_{n}^{\prime})^{\prime}$ for
$A=(a_{1}\ldots a_{n}),$ $P_{A}=A(A^{\prime}A)^{-1}A^{\prime}$ is a projection
on the columns of the full rank matrix $A$ and $M_{A}=I_{N}-P_{A}$ is a
projection on the space orthogonal to $A.$ Convergence in probability is
denoted by \textquotedblleft$\underset{p}{\rightarrow}$\textquotedblright%
,\ convergence in distribution by \textquotedblleft$\underset{d}{\rightarrow}%
$\textquotedblright\ and \textquotedblleft$\underset{a}{=}$\textquotedblright%
\ means asymptotically equivalent.

\section{Identification robust GMM inference for dynamic panel data models}

In this section, we briefly describe the dynamic panel data model and the
different sets of moment conditions. Thereafter we discuss identification
robust GMM inference including the construction of confidence intervals.
Finally, we illustrate the identification issues that occur when using the
different moment conditions for dynamic panel data models, by computing power
curves based on the identification robust KLM statistic.

\subsection{Model and moment conditions}

We analyze the first-order autoregressive linear dynamic panel data model%
\begin{equation}%
\begin{array}
[c]{cll}%
y_{it}= & c_{i}+\theta y_{it-1}+u_{it},\qquad & i=1,\ldots,N,\text{
}t=2,\ldots,T,
\end{array}
\label{panelar1}%
\end{equation}
with $T$ the number of time periods and $N$ the number of cross section
observations. We assume that the initial observation $y_{i1}$ is observed and
that the vector of observations $(y_{i1},...,y_{iT})$\ for individual $i$\ is
independently distributed across the $N$\ individuals. We will later on make
further assumptions on the initial observations to properly define the process
in (\ref{panelar1}). For expository purposes, we analyze the simple dynamic
panel data model in (\ref{panelar1}) which can be extended with additional
lags of $y_{it}$ and explanatory variables.\footnote{The extension to other
explanatory variables would depend on the nature of these. For some settings
such an extension would be trivial but for others not so.}\ Estimation of the
parameter $\theta$ by means of least squares leads to an inconsistent
estimator in samples with a finite value of $T$ and large $N,$ see $e.g.$
Nickell (1981).\nocite{nick81} We therefore estimate it using GMM. We obtain
the GMM moment conditions from the unconditional moment assumptions:%
\begin{equation}%
\begin{array}
[c]{rll}%
E[u_{it}]= & 0,\qquad & t=2,\ldots,T,\text{ }\\
E[u_{it}u_{is}]= & 0, & s\neq t;\text{ }s,\text{ }t=2,\ldots,T,\text{ }\\
E[u_{it}c_{i}]= & 0, & t=2,\ldots,T,\\
E[u_{it}y_{i1}]= & 0, & t=2,\ldots,T.
\end{array}
\label{marexpec}%
\end{equation}

Under these assumptions, the moments of the $T(T-1)$ interactions of $\Delta
y_{it}$ and $y_{it}:$%
\begin{equation}%
\begin{array}
[c]{c}%
E[\Delta y_{it}y_{ij}],\qquad j=1,\ldots,T,\text{ }t=2,\ldots,T
\end{array}
\label{samplemom}%
\end{equation}
can be used to construct functions which identify the parameter of interest
$\theta.$ We do not use products of $\Delta y_{it}$ to identify $\theta$ since
we would need further assumptions, $i.e.$ homoskedasticity or initial
condition assumptions, see $e.g.$ Han and Phillips (2010). \nocite{hp10}

Two different sets of moment conditions, which are functions of the moments in
(\ref{samplemom}), are commonly used to identify $\theta:$

\begin{enumerate}
\item Difference (Dif) moment conditions:
\begin{equation}%
\begin{array}
[c]{c}%
E[y_{ij}(\Delta y_{it}-\theta\Delta y_{it-1})]=0,\qquad j=1,\ldots,t-2;\text{
}t=3,\ldots,T,
\end{array}
\label{difmom}%
\end{equation}
as proposed by $e.g.$ Anderson and Hsiao (1981)\nocite{ah81} and Arellano and
Bond (1991).\nocite{ab91} The Dif moment conditions solely result from the
conditions in (\ref{marexpec}).

\item Level (Lev) moment conditions:%
\begin{equation}%
\begin{array}
[c]{c}%
E[\Delta y_{it-1}(y_{it}-\theta y_{it-1})]=0,\qquad t=3,\ldots,T,\text{ }%
\end{array}
\label{levmom}%
\end{equation}
as proposed by Arellano and Bover (1995), see also Blundell and Bond
(1998).\nocite{abov95} Besides the conditions in (\ref{marexpec}), the Lev
moment conditions use
\begin{equation}
E\left[  \Delta y_{it}c_{i}\right]  =0, \label{ASS2}%
\end{equation}
which implies that the original data in levels have constant correlation over
time with the individual-specific effects. The Lev moment conditions
(\ref{levmom}) hold under the following conditions regarding the initial
observations $y_{i1}$ $(i=1,...,N):$%
\begin{equation}
y_{i1}=\mu_{i}+u_{i1}, \label{init1}%
\end{equation}%
\begin{equation}
\mu_{i}=c_{i}/(1-\theta), \label{init3}%
\end{equation}%
\begin{equation}%
\begin{array}
[c]{rl}%
E[u_{i1}]= & 0,\\
E[u_{i1}c_{i}]= & 0,\qquad\\
E[u_{i1}u_{it}]= & 0,\qquad t>1.
\end{array}
\label{init2}%
\end{equation}
The specification of the initial observations in (\ref{init1})-(\ref{init2})
is often referred to as mean stationarity. In our analysis we maintain the
assumption of mean stationarity.
\end{enumerate}

The Dif and Lev moments can be used separately or jointly to identify
$\theta.$ When we use the moment conditions in (\ref{difmom}) and
(\ref{levmom}) jointly, we refer to them as system (Sys) moment
conditions,\footnote{We could extend the Lev moment conditions to $\frac{1}%
{2}(T-1)(T-2)$ sample moments by including additional interactions of $\Delta
y_{it-j}$ and $y_{it}-\theta y_{it-1},$ for $j=2,\ldots,t-2.$ It can be shown,
however, that all conditions on top of those in (\ref{levmom}) can be
constructed as linear combinations of the Dif conditions in (\ref{difmom}) and
the Lev conditions in (\ref{levmom}).} see Arellano and Bover (1995) and
Blundell and Bond (1998).\nocite{bb98} Another set of nonlinear (NL) moment
conditions, which just like the Dif moments only use the conditions in
(\ref{marexpec}), results from Ahn and Schmidt (1995):\nocite{as95}%
\begin{equation}%
\begin{array}
[c]{c}%
E[(y_{it}-\theta y_{it-1})(\Delta y_{it-1}-\theta\Delta y_{it-2})]=0\qquad
t=4,\ldots,T.
\end{array}
\label{ahnschmidt}%
\end{equation}
The NL moments can be used separately or jointly with the Dif moments to
identify $\theta.$ When we use the moment conditions in (\ref{difmom}) and
(\ref{ahnschmidt}) jointly, we refer to them as Ahn-Schmidt (AS) moment conditions.

Ahn and Schmidt (1995) show that their AS moment conditions exhaust the
information on $\theta$ in the moment conditions (\ref{marexpec}) and are
therefore complete. Mean stationarity adds one moment condition (\ref{ASS2})
to the moment conditions in (\ref{marexpec}). Hence, the complete set of
moment conditions under (\ref{marexpec}) and (\ref{ASS2}) equals the AS moment
conditions and (\ref{ASS2}). Upon rewriting we can show that these combined
moment conditions are identical to the Sys moment conditions so they are
complete under (\ref{marexpec}) and (\ref{ASS2}).

\subsection{Identification robust GMM tests}

In GMM, we consider a $k$-dimensional vector of moment conditions, see Hansen
(1982):\nocite{han82}%
\begin{equation}
E[f_{i}(\theta_{0})]=0,\qquad i=1,\ldots,N, \label{gmmmom}%
\end{equation}
where $f_{i}(\theta)$ is a $k$-dimensional (continuous and continuously
differentiable) function of the observed data for individual $i$ and the
unknown parameter vector $\theta$ whose functional expression is identical for
all individuals. There is a unique true value of the $p$-dimensional vector
$\theta$ where the moment conditions are satisfied, which we denote by
$\theta_{0},$ and $k$ is at least as large as $p.$ We only analyze the
first-order autoregressive panel data model so $p=1$ for our setting. The
population moments in (\ref{gmmmom}) are estimated using the sample moments,%
\begin{equation}%
\begin{array}
[c]{c}%
f_{N}(\theta)=\frac{1}{N}\sum_{i=1}^{N}f_{i}(\theta).
\end{array}
\label{sammom}%
\end{equation}
The $k\times p$ dimensional matrix $q_{N}(\theta)$ contains the derivative of
$f_{N}(\theta)$ with respect to $\theta:$%
\begin{equation}%
\begin{array}
[c]{c}%
q_{N}(\theta)=\frac{\partial}{\partial\theta^{\prime}}f_{N}(\theta)=\frac
{1}{N}\sum_{i=1}^{N}q_{i}(\theta),
\end{array}
\label{dermom}%
\end{equation}
with $q_{i}(\theta)=\frac{\partial}{\partial\theta^{\prime}}f_{i}(\theta).$
Specifications of the sample moment functions $f_{N}(\theta)$ and
$q_{N}(\theta)$ for the Dif, Lev, Sys, NL and AS moment conditions are
provided in the Appendix.

Statistical inference based on the two step GMM estimator is known to be of
poor quality in the case of weak identification, which leads to an
inconsistent estimator with non-standard behavior of its corresponding
t-statistic, see e.g. Phillips (1989),\nocite{p89} Staiger and Stock
(1997)\nocite{stst97} and Stock and Wright (2000). The non-standard limiting
behavior of one and two step GMM estimators for dynamic panel data models in
the case of weak identification has been documented in \textit{e.g.} Madsen
(2003),\nocite{mad03} Kruiniger (2009)\nocite{krui09} and Phillips
(2018).\nocite{phillip2018}

In this study we therefore use identification robust GMM statistics to
overcome the aforementioned problems. The main advantage of identification
robust statistics is that, unlike conventional two step GMM statistics, their
limiting distributions are unaffected by the identification strength. Define
$\theta^{\ast}$ as the hypothesized value under the null hypothesis. A
particularly simple to compute identification robust GMM statistic to test
H$_{0}:\theta=\theta^{\ast}$ is the GMM extension of the Anderson-Rubin
statistic, see Anderson and Rubin (1949) and Stock and Wright
(2000):\nocite{sw00}%
\begin{equation}%
\begin{array}
[c]{c}%
GMM\text{-}AR(\theta^{\ast})=Nf_{N}(\theta^{\ast})^{\prime}\hat{V}_{ff}%
(\theta^{\ast})^{-1}f_{N}(\theta^{\ast}),
\end{array}
\label{gmmar}%
\end{equation}
with $\hat{V}_{ff}(\theta)$ the Eicker-White covariance matrix estimator:%
\begin{equation}%
\begin{array}
[c]{cc}%
\hat{V}_{ff}(\theta)= & \frac{1}{N}\sum_{i=1}^{N}(f_{i}(\theta)-f_{N}%
(\theta))(f_{i}(\theta)-f_{N}(\theta))^{\prime}.
\end{array}
\label{eickwhit}%
\end{equation}
The GMM-AR statistic equals the continuous updating objective function (Hansen
\textit{et al.}, 1996\nocite{hhy96}) evaluated in $\theta^{\ast}.$ A possible
drawback of the GMM-AR statistic is its lower power in the case of
overidentified models. The KLM statistic of Kleibergen (2005) partly overcomes
this. The KLM statistic is a quadratic form of the score of the GMM-AR
statistic with respect to $\theta$:\nocite{kf00a}%
\begin{equation}%
\begin{array}
[c]{cl}%
KLM(\theta^{\ast})= & Nf_{N}(\theta^{\ast})^{\prime}\hat{V}_{ff}(\theta^{\ast
})^{-1}\hat{D}_{N}(\theta^{\ast})\left[  \hat{D}_{N}(\theta^{\ast})^{\prime
}\hat{V}_{ff}(\theta^{\ast})^{-1}\hat{D}_{N}(\theta^{\ast})\right]  ^{-1}\\
& \hat{D}_{N}(\theta^{\ast})^{\prime}\hat{V}_{ff}(\theta^{\ast})^{-1}%
f_{N}(\theta^{\ast}),
\end{array}
\label{klm}%
\end{equation}
with $\hat{D}_{N}(\theta)$ a $k\times p$ dimensional matrix,%
\begin{equation}%
\begin{array}
[c]{c}%
\text{vec}(\hat{D}_{N}(\theta))=\text{vec}(q_{N}(\theta))-\hat{V}_{qf}%
(\theta)\hat{V}_{ff}(\theta)^{-1}f_{N}(\theta),
\end{array}
\label{dnspec}%
\end{equation}
and
\begin{equation}%
\begin{array}
[c]{cc}%
\hat{V}_{qf}(\theta)= & \frac{1}{N}\sum_{i=1}^{N}\left(  \text{vec}%
[q_{i}(\theta)-q_{N}(\theta)]\right)  (f_{i}(\theta)-f_{N}(\theta))^{\prime}.
\end{array}
\label{vthetaf}%
\end{equation}

The limiting distributions of the identification robust GMM-AR and
KLM\ statistics apply under less restrictive assumptions than those of the
traditional test statistics based on two step GMM. The GMM-KLM and GMM-AR
statistics converge under H$_{0}$ to $\chi^{2}(p)$ and $\chi^{2}(k)$
distributed random variables even when the Jacobian, $J(\theta_{0}%
)=E(q_{i}(\theta_{0})),$ does not have a full rank value, see Stock and Wright
(2000), Kleibergen (2005) and Newey and Windmeijer (2009).\nocite{NewWind09}
Other identification robust statistics for GMM are proposed in Kleibergen
(2005), Andrews (2016)\nocite{and15} and Andrews and Mikusheva
(2016)\nocite{am16} which all provide extensions of the conditional likelihood
ratio statistic of Moreira (2003)\nocite{mor01} to GMM. The conditional
likelihood ratio statistic is optimal for the homoskedastic linear
instrumental variables regression model with one included endogenous variable,
see Andrews $et$ $al.$ (2006).\nocite{andms05} None of its extensions to GMM
has, however, shown to be optimal for our setting of the dynamic linear panel
autoregression so we just use the easier to implement GMM-AR and KLM
statistics.\footnote{Andrews et al. (2006) establish the optimality of the
likelihood ratio test for the iid linear instrumental variables regression
model using the Neymann-Pearson lemma. We cannot do so here since the
identification of $\theta$ depends on other nuisance parameters besides the
Jacobian, like the initial observations, so it is not obvious how optimality
can be established.}

The identification robust tests can be inverted to obtain corresponding
identification robust confidence sets. The $100\times(1-\alpha)\%$ confidence
set for $\theta$ (denoted by $\text{CS}_{\theta}\text{(}\alpha) $ below)
consists of all values of $\theta^{\ast}$ for which the respective
identification robust test does not reject using its $100\times\alpha\%$
asymptotic critical value:%
\begin{equation}%
\begin{array}
[c]{c}%
\text{CS}_{\theta}\text{(}\alpha)=\left\{  \theta^{\ast}:IRT\text{(}%
\theta^{\ast})\leq CDF_{IRT}(\alpha)\right\}  ,
\end{array}
\label{conf set}%
\end{equation}
with $IRT(\theta^{\ast})$ the identification robust statistic evaluated at
$\theta^{\ast}$ and $CDF_{IRT}(\alpha)$ the $(1-\alpha)\times100$-th
percentile of the limiting distribution of $IRT$($\theta_{0}).$

The identification robust tests are not quadratic functions of $\theta^{\ast}$
so they cannot directly be inverted to obtain the confidence set.\footnote{An
exception is the GMM-AR statistic in the homoskedastic linear instrumental
variables regression model, see Dufour and Taamouti (2003).\nocite{duftaa03}}
The confidence sets resulting from them do therefore not have the usual
expression of an estimator plus or minus a multiple of the standard error.
Instead, we have to specify a $p$-dimensional grid of values of $\theta^{\ast
}$ and compute the identification robust statistic for every value of
$\theta^{\ast}$ on the grid to determine if it is less than the appropriate
critical value so $\theta^{\ast}$ is part of the confidence set.

Specifically, the confidence set in (\ref{conf set}) can have three distinct shapes:

\begin{enumerate}
\item Bounded and convex: there is a closed compact set of values of
$\theta^{\ast}$ for which the identification robust test statistic does not
exceed the critical value.

\item Unbounded: this occurs either when there are no values of $\theta^{\ast
}$ for which the identification robust test statistic exceeds the critical
value (unbounded and convex), or when there are bounded sets of values of
$\theta^{\ast}$ for which the identification robust test statistic exceeds the
critical value (unbounded and disjoint).

\item Empty: the identification robust test statistic exceeds the critical
value for all values of $\theta^{\ast}.$
\end{enumerate}

Bounded and convex confidence sets occur when the parameters of interest are
well identified. Unbounded confidence sets are indicative of weak
identification so if we then test H$_{0}:\theta=\theta^{\ast}$ at a very
large, possibly infinite, value of $\theta^{\ast}$ using an identification
robust test at, say, the 5\% significance level, it does not necessarily
reject. For such instances, we thus often do not reject the hypothesis of an
infinite value of $\theta$ so we obtain an unbounded 95\% confidence set. In
Dufour (1997, Theorems 3.3 and 3.6), \nocite{duf97} it is shown that any size
correct procedure used to test parameters which can be non-identified must
have a positive probability of producing an unbounded 95\% confidence set.
Conversely, also any test procedure, like, for example, the Wald $t$-test,
which can not generate an unbounded 95\% confidence set, can not be a size
correct test procedure when the tested parameter can be non-identified. Empty
confidence sets occur when the model is misspecified so there is no value of
$\theta$ for which the moment condition holds. Since the GMM-AR statistic
tests whether all moment conditions hold, it also tests misspecification. It
can therefore result in empty confidence sets but the KLM test cannot since it
is equal to zero at the continuous updating estimator of Hansen $et$ $al.$
(1996),\nocite{hhy96} which is the minimizer of the GMM-AR\ statistic.

The identification robust statistics conduct tests on the full parameter
vector $\theta.$ Valid $(1-\alpha)\times100\%$ confidence sets for the
individual elements of $\theta$ then result by projecting the joint
$p$-dimensional $(1-\alpha)\times100\%$ confidence set for $\theta$ on the $p
$ different axes. These projection based confidence sets are size correct so
they contain the true value of $\theta$ with a probability which is at least
$(1-\alpha)\times100\%$ irrespective of the strength of identification.
Projection based confidence sets can face computational issues when $p$ is
rather large given the large number of points on the $p$-dimensional grid for
which the statistic then has to be computed.

Confidence sets for the individual elements of $\theta$ can also be obtained
by plugging in an estimator for the remaining elements of $\theta$ after which
the (conditional) limiting distribution can be sharpened using the usual
degrees of freedom correction of the $\chi^{2}$ limiting distributions. The
resulting confidence sets only have correct coverage when these remaining
parameters are well identified, see Kleibergen (2005). Just in some isolated
cases, such, as for example, when using the GMM-AR statistic in the
homoskedastic linear instrumental variables regression model or in the linear
factor model for determining risk premia in finance, can we prove that these
confidence sets are valid without requiring the partialled out parameters to
be well identified, see Guggenberger $et$ $al.$ (2012, 2019), Kleibergen and
Zhan (2020), Kleibergen (2020)\nocite{kf2020} and Kleibergen $et$ $al.$
(2020).\nocite{gkm17}\nocite{gkmc12}\nocite{kz19}\nocite{kkz19}

\subsection{Using identification robust tests to highlight identification
issues}

Identification robust GMM tests are size correct irrespective of the
identification strength. Therefore, their rejection frequencies can be used in
a straightforward manner to illustrate the identification issues at particular
values of the autoregressive parameter in the dynamic panel data model. The
conventional $t$-test based on the two step GMM estimator is not suitable for
this purpose as it is size distorted in the case of weak identification and,
hence, rejection frequencies would not equal the significance level.

To illustrate the identification issues for the different moment conditions,
we compute the rejection frequencies of 5\% significance KLM\ tests of
H$_{0}:\theta=0.5$ for a range of (true data generating) values $\theta_{0}.$
We do so by simulating data from the panel autoregressive model in
(\ref{panelar1}) with three or four time series observations, so $T=3$ or $4,$
and two hundred and fifty individuals, so $N=250.$ The individual specific
effects $c_{i}$ and idiosyncratic errors $u_{it}$ are independently generated
from $N(0,\sigma_{c}^{2})$ and $N(0,1)$ distributions respectively. We vary
the value of $\sigma_{c}^{2}$ to show the sensitivity of the identification of
$\theta$ using the panel moment conditions to the variance of the initial
observations. We assume mean stationarity so (\ref{init1})-(\ref{init2}) hold.

We consider four KLM tests based on Dif, Lev, Sys and AS moment conditions,
which have been calculated according to equation (\ref{klm}) using
$\theta^{\ast}=0.5$. The figures in Panels 1 and 2 show the rejection
frequencies of KLM\ tests of H$_{0}:\theta=0.5$ with 5\% significance for four
values of $\sigma_{c}^{2}$ and a range of true values $\theta_{0}$. Panel 1
does so for three times series observations while Panel 2 covers four time
series observations. The simulation experiment is designed such that the
variance of the initial observations becomes very large when $\theta_{0}$ gets
close to one and $\sigma_{c}^{2}$ exceeds zero.%

\[%
\begin{array}
[c]{c}%
\text{Panel 1. Rejection frequencies of KLM test of H}_{0}:\theta=0.5\text{
with 5\% significance }\\
\text{using different moment conditions for }T=3,\text{ }N=250\text{ and
}\sigma_{c}^{2}=0\text{ (dashed), }\\
\text{0.5 (solid), one (dashed-dot) and two (dotted)}\\%
\begin{array}
[c]{ccc}%
\raisebox{-0pt}{\includegraphics[
height=2.2139in,
width=2.9386in
]%
{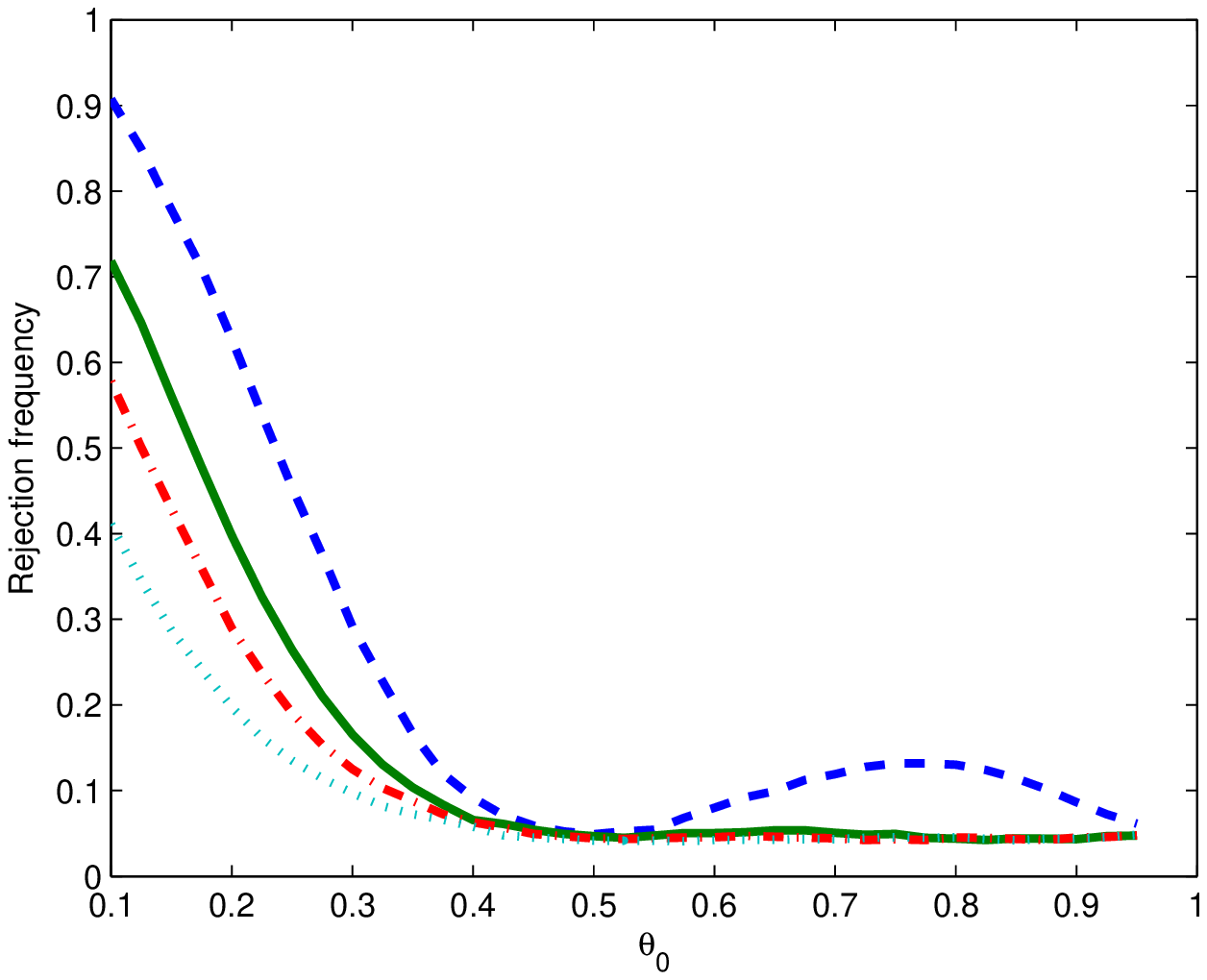}%
}
&  &
\raisebox{-0pt}{\includegraphics[
height=2.2139in,
width=2.9386in
]%
{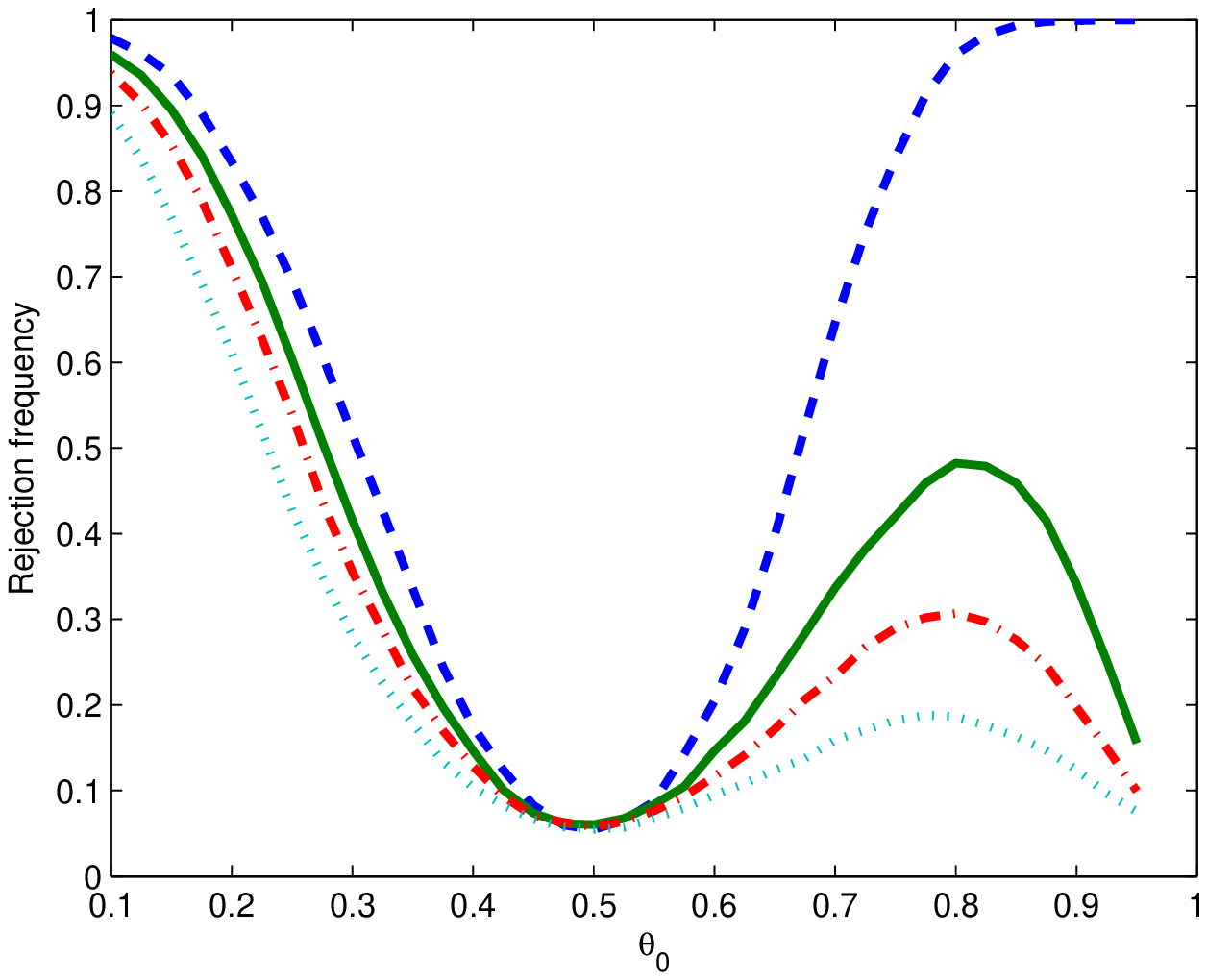}%
}
\\
\text{Figure 1.1: Dif moment conditions} &  & \text{Figure 1.2: Lev moment
conditions}%
\end{array}
\\%
\begin{array}
[c]{c}%
\raisebox{-0pt}{\includegraphics[
height=2.2139in,
width=2.9386in
]%
{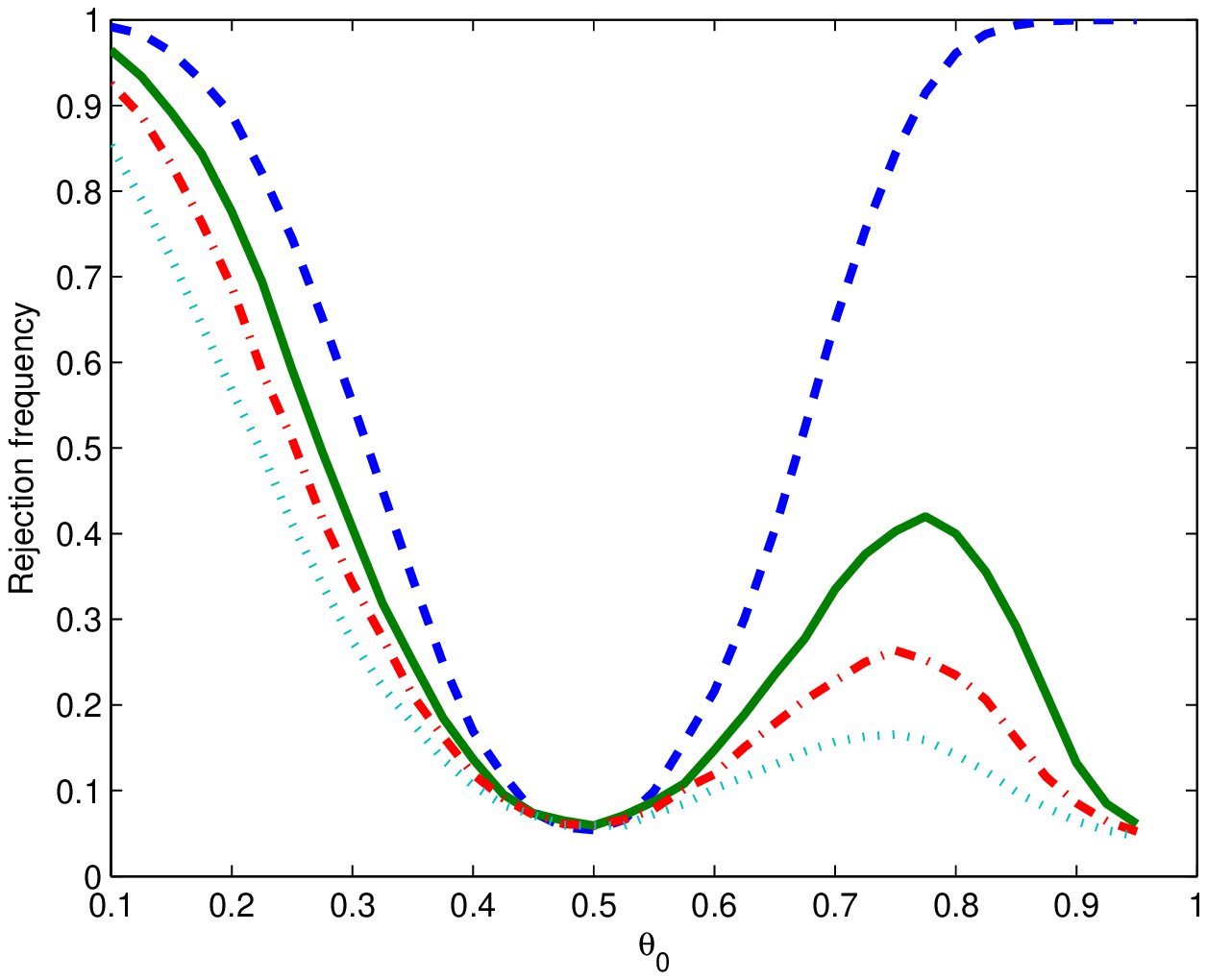}%
}
\\
\text{Figure 1.3: Sys moment conditions}%
\end{array}
\end{array}
\]
\bigskip

Figures 1.1 and 2.1 show that the rejection frequencies of the KLM\ test with
Dif moment conditions for $\theta_{0}$ close to one converges to the
significance level of 5\%. It is well known that the Jacobian of the Dif
moment conditions is zero when $\theta_{0}$ equals one so they then do not
identify $\theta.$ The KLM\ test is identification robust which explains why
the rejection frequency equals the significance level both at the hypothesized
value of $\theta^{\ast}=0.5$ and when $\theta_{0}$ is close to one for all
values of $\sigma_{c}^{2}$. The latter results since the Dif moment conditions
do then not identify $\theta$, hence the KLM test has no discriminating power
so the power of the KLM test equals the significance level.%

\[%
\begin{array}
[c]{c}%
\text{Panel 2. Rejection frequencies of KLM test of H}_{0}:\theta=0.5\text{
with 5\% significance }\\
\text{using different moment conditions for }T=4,\text{ }N=250\text{ and
}\sigma_{c}^{2}=0\text{ (dashed), }\\
\text{0.5 (solid), one (dashed-dot) and two (dotted)}\\%
\begin{array}
[c]{ccc}%
\raisebox{-0pt}{\includegraphics[
height=2.2139in,
width=2.9386in
]%
{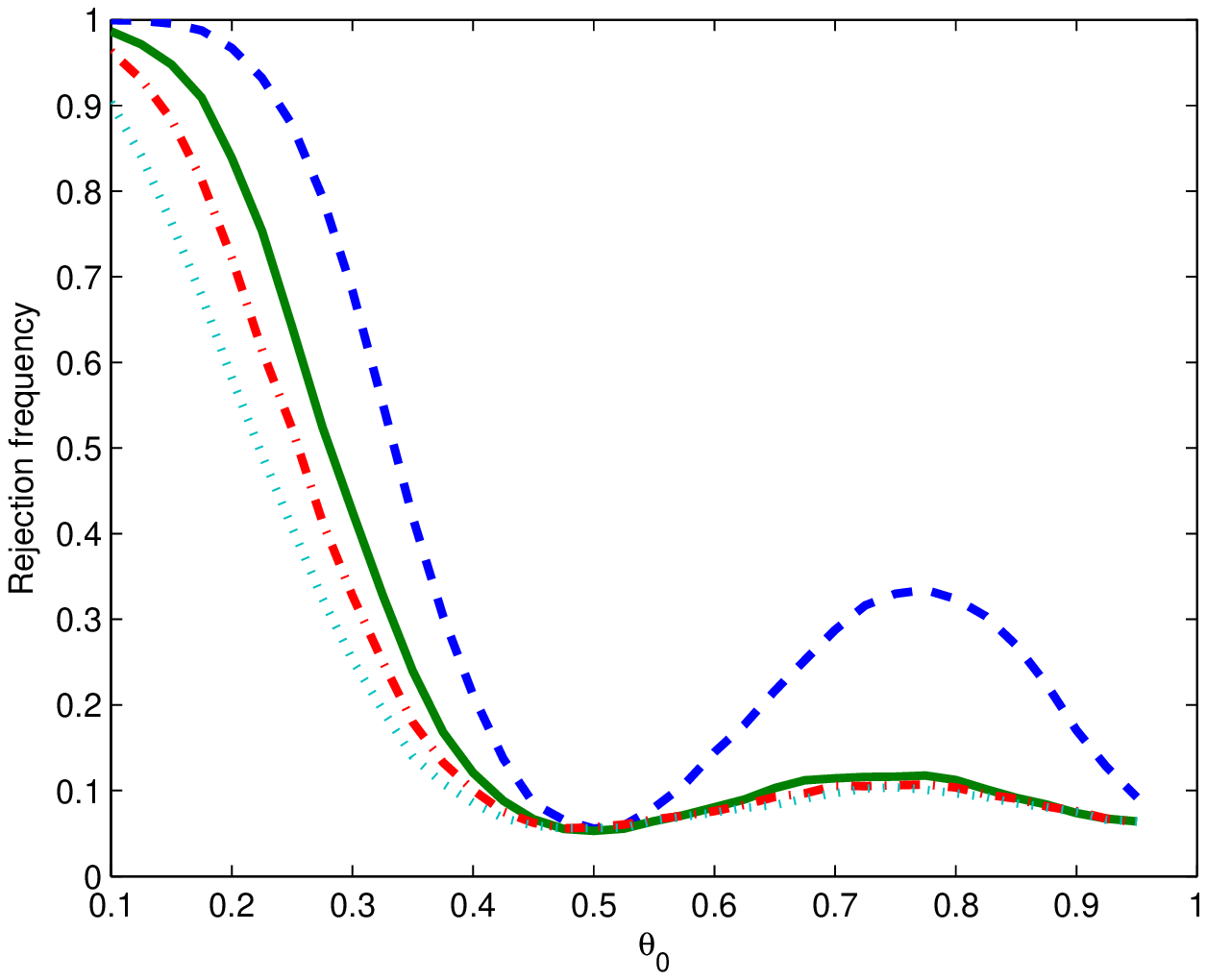}%
}
&  &
\raisebox{-0pt}{\includegraphics[
height=2.2139in,
width=2.9386in
]%
{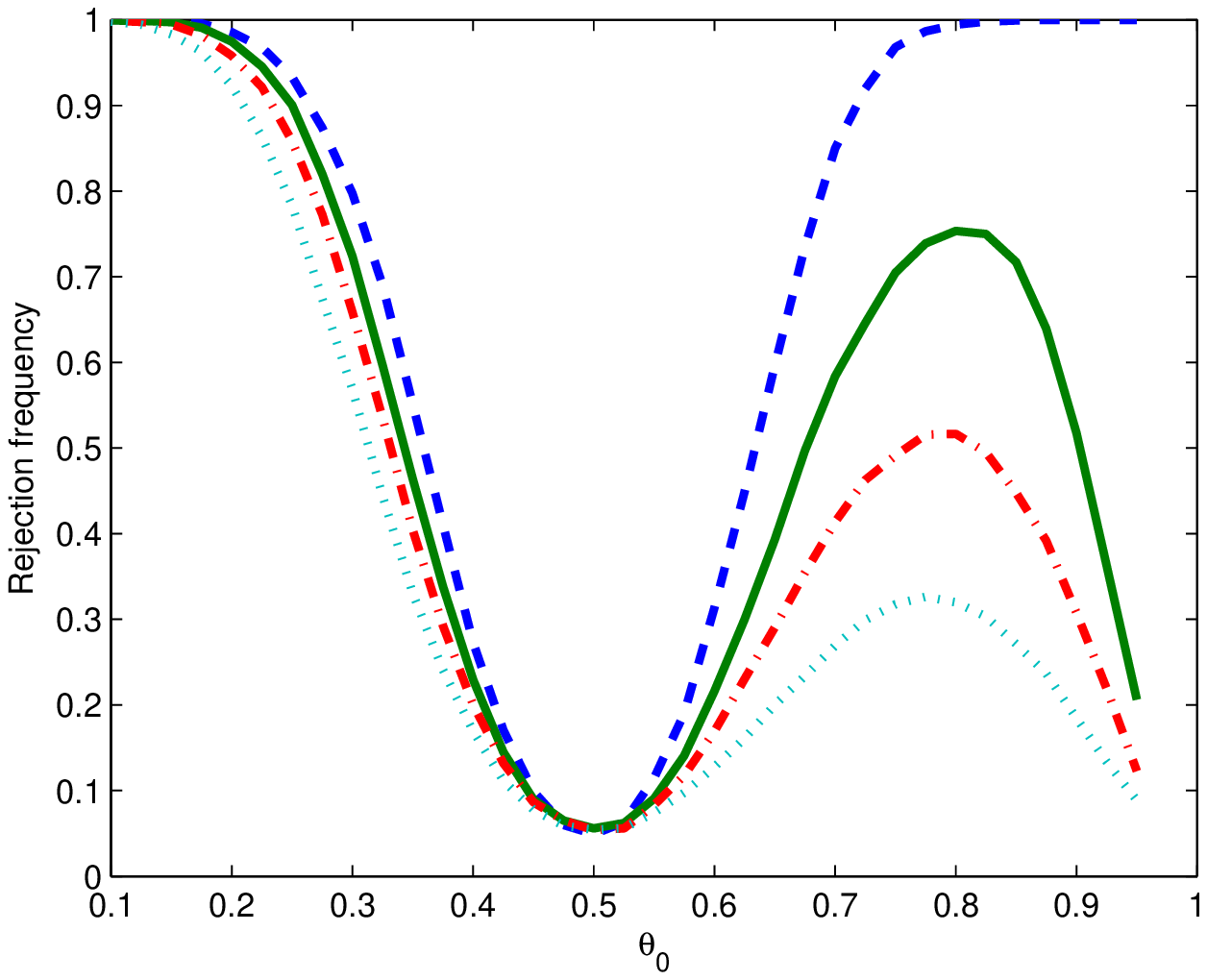}%
}
\\
\text{Figure 2.1: Dif moment conditions} &  & \text{Figure 2.2: Lev moment
conditions}%
\end{array}
\\%
\begin{array}
[c]{ccc}%
\raisebox{-0pt}{\includegraphics[
height=2.2139in,
width=2.9386in
]%
{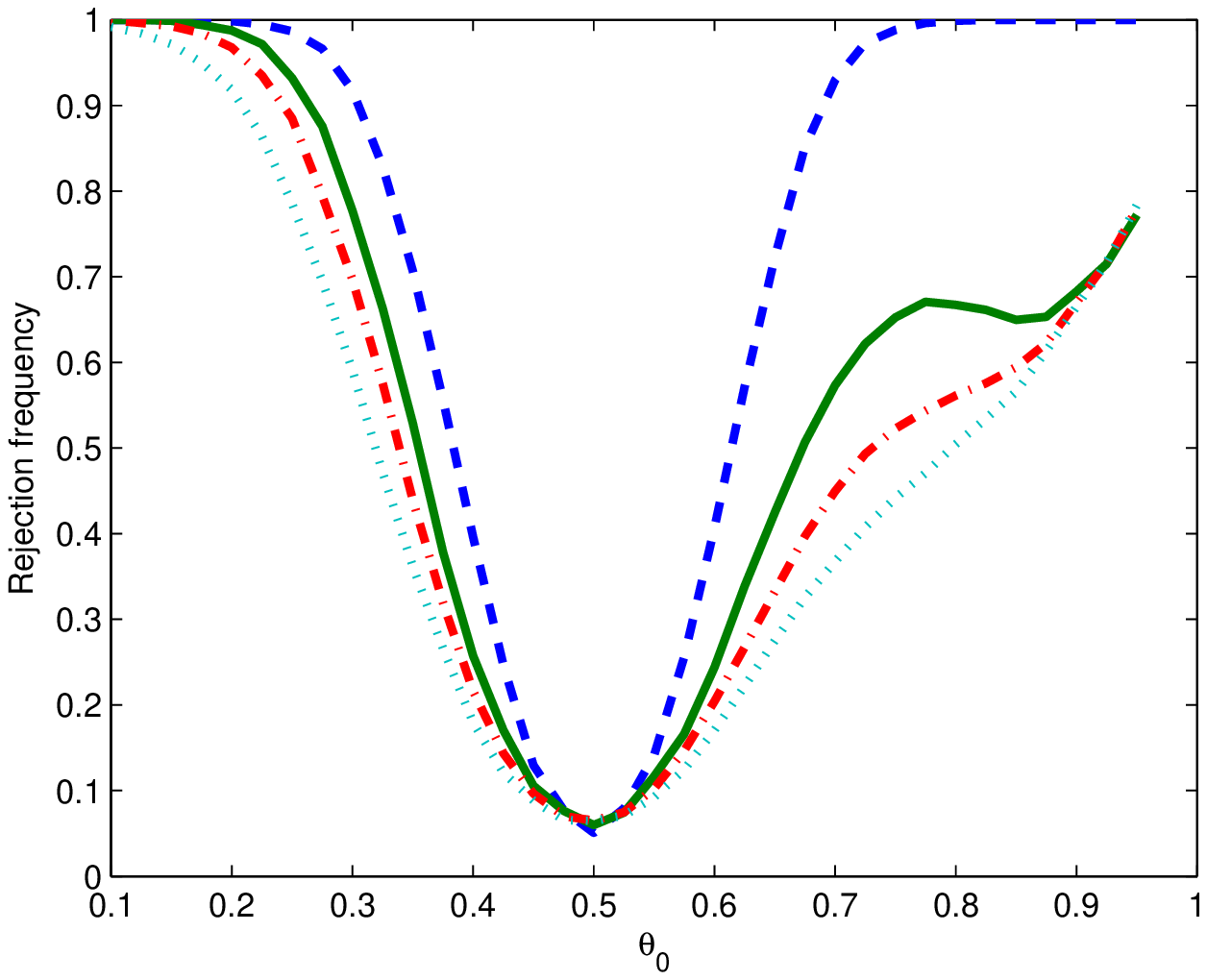}%
}
&  &
\raisebox{-0pt}{\includegraphics[
height=2.2139in,
width=2.9386in
]%
{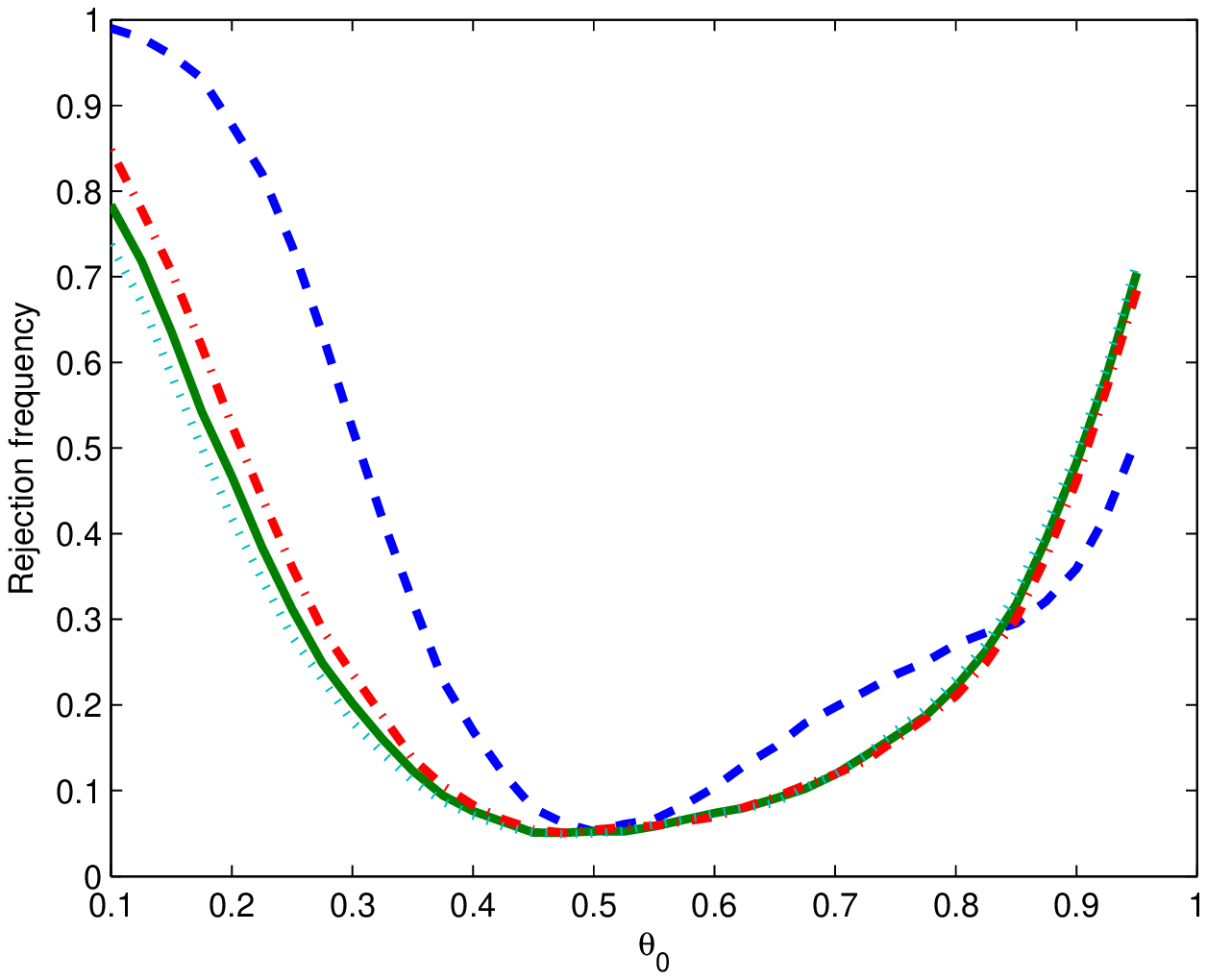}%
}
\\
\text{Figure 2.3: Sys moment conditions} &  & \text{Figure 2.4: AS moment
conditions}%
\end{array}
\end{array}
\]
\bigskip

Figures 1.2 and 2.2 show the rejection frequencies of 5\% significance tests
of H$_{0}:\theta=0.5$ using the KLM\ test with Lev moment conditions.
Interestingly, these figures show that the Lev moment conditions only identify
$\theta$ when the true value $\theta_{0}$ is close to one when $\sigma_{c}%
^{2}=0.$ Non-zero values of $\sigma_{c}^{2}$ correspond with a large variance
of the initial observations when $\theta_{0}$ is close to one and Figures 1.2
and 2.2 show that the Lev moment conditions do not identify $\theta$ in this
case. This contradicts the common perception that the Lev moment conditions
generally identify $\theta$ irrespective of the setting of nuisance
parameters, like, the variance of the initial observations.

Figures 1.3 and 2.3 show the rejection frequencies of 5\% significance tests
of H$_{0}:\theta=0.5$ using the KLM\ test with Sys moment conditions.
Surprisingly, these figures show that the Sys moment conditions do not
identify $\theta$ when $\theta_{0}$ is close to one and $\sigma_{c}^{2}>0$
when $T=3$ but do so when $T=4.$

Figure 2.4 shows the rejection frequencies of 5\% significance tests of
H$_{0}:\theta=0.5$ using the KLM\ test with AS moment conditions. These
rejection frequencies show that the AS moment conditions, which are not
defined for $T=3,$ identify $\theta$ when its true value is close to one and
the variance of the initial observations is very large. Interestingly, the
rejection frequencies of KLM\ tests of H$_{0}$ using the Sys and AS moment
conditions are very close when $\theta_{0}$ is near one when paired with large
variances of the initial observations.

Summarizing, Panels 1 and 2 illustrate a few stylized facts that concern the
identification of $\theta$ for the DGP used in the simulation experiment:

\begin{enumerate}
\item Dif moment conditions do not identify $\theta$ when $\theta_{0}$ is
close to one for general $T.$

\item Lev moment conditions do not identify $\theta$ when $\theta_{0}$ is
close to one for large variances of the initial observations for general $T.$

\item Sys moment conditions do not identify $\theta$ when $\theta_{0}$ is
close to one for large variances of the initial observations when $T=3.$

\item Sys and AS moment conditions identify $\theta$ when $\theta_{0}$ is
close to one for large variances of the initial observations when $T$ exceeds
$3.$

\item The rejection frequencies of KLM\ tests of H$_{0}$ using AS\ and Sys
moment conditions when $\theta_{0}$ is close to one and the variance of the
initial observations is large are almost identical.
\end{enumerate}

Except for the first stylized fact, a theory backing them up is lacking so we
aim to provide one in the sections ahead. In doing so, we show that all
information regarding $\theta,$ when its true value is close to one and the
variance of the initial observations is large, is contained in a set of,
so-called, robust moment conditions which are a combination of either the AS
or Sys moment conditions. We furthermore show that the KLM test based on the
original AS or Sys moment conditions, as reported in Panels 1 and 2, makes
optimal use of these robust sample moments when only they contain information
on $\theta$.

Alongside the identification issues we can infer from the rejection
frequencies in Panels 1 and 2, they are also indicative of the different kind
of confidence sets that can result from the identification robust tests as
discussed previously. For example, the low rejection frequencies occurring for
$\theta_{0}$ around one, that result from the identification issues, show that
the 95\% confidence sets for $\theta$ are then typically very wide, possibly
unbounded, when $\theta_{0}$ has such a value paired with a large variance of
the initial observations. To visualize this further, Panel 3 contains the (one
minus the) $p$-value plots of KLM tests using AS, Dif, Lev and Sys moment
conditions for four data sets using the same DGPs as in Panels 1-2 with
$N=250$ and $\theta_{0}=0.95.\footnote{We note that the figures in Panel 3
show (one minus) the p-value for one realized data set and do not show the
simulated empirical distribution function of the test under the null
hypothesis which is sometimes also referred to as a $p$-value plot, see
Davidson and MacKinnon (2002). \nocite{DMK02}}$ The DGPs used for the four
figures differ over the values of $T$ and $\sigma_c^2.$ The intersections of
the depicted $p$-value plots with the line at 0.95 indicate the 95\%
confidence sets of KLM\ tests with the respective moment condition.

In Figures 3.1 and 3.3, $\sigma_{c}^{2}=0$ so identification issues only occur
at $\theta_{0}$ close to one when using the Dif moment conditions. Since
$\theta_{0}$ is 0.95, this explains why the $p$-value plots of the KLM\ test
with the Dif moments conditions do not cross the line at 0.95 in Figures 3.1
and 3.3 so the resulting 95\% confidence sets are very wide. The $p$-value
plots in Figures 3.1 and 3.3 of KLM tests with Sys and Lev moment conditions
show that they lead to bounded 95\% confidence sets since these moment
conditions have no identification issues when $T=3$ and $\sigma_{c}^{2}=0.$

In Figure 3.2, where $T=3$ and $\sigma_{c}^{2}=0.5,$ none of the $p$-value
plots crosses the line at 0.95 so 95\% confidence sets that result from
KLM\ tests with Dif, Lev and Sys moment conditions are all very wide and
possibly unbounded. This is indicative of the identification issues when $T=3$
and $\sigma_{c}^{2}=0.5$ for true values of $\theta$ close to one.

In Figure 3.4, where $T=4$ and $\sigma_{c}^{2}=0.5,$ KLM tests with Sys and AS
moment conditions both result in finite 95\% confidence sets while the KLM
test with Dif and Lev moment conditions leads to very wide possibly unbounded
confidence sets. Hence, Sys and AS moment conditions have no identification
issues while Dif and Lev moment conditions do. The AS moment conditions are
quadratic functions of $\theta$ which explains the somewhat unusual shape of
their $p$-value plots in Figures 3.3 and 3.4.\bigskip%

\[%
\begin{array}
[c]{c}%
\text{Panel 3. One minus }p\text{-value plots of KLM tests using different
moments conditions: Sys }\\
\text{(solid), AS (dotted), Lev (dashed), Dif (dash-dot) for }\theta
_{0}=0.95\text{ and }N=250.\text{ }\\
\\%
\begin{array}
[c]{ccc}%
\raisebox{-0pt}{\includegraphics[
height=2.2139in,
width=2.9386in
]%
{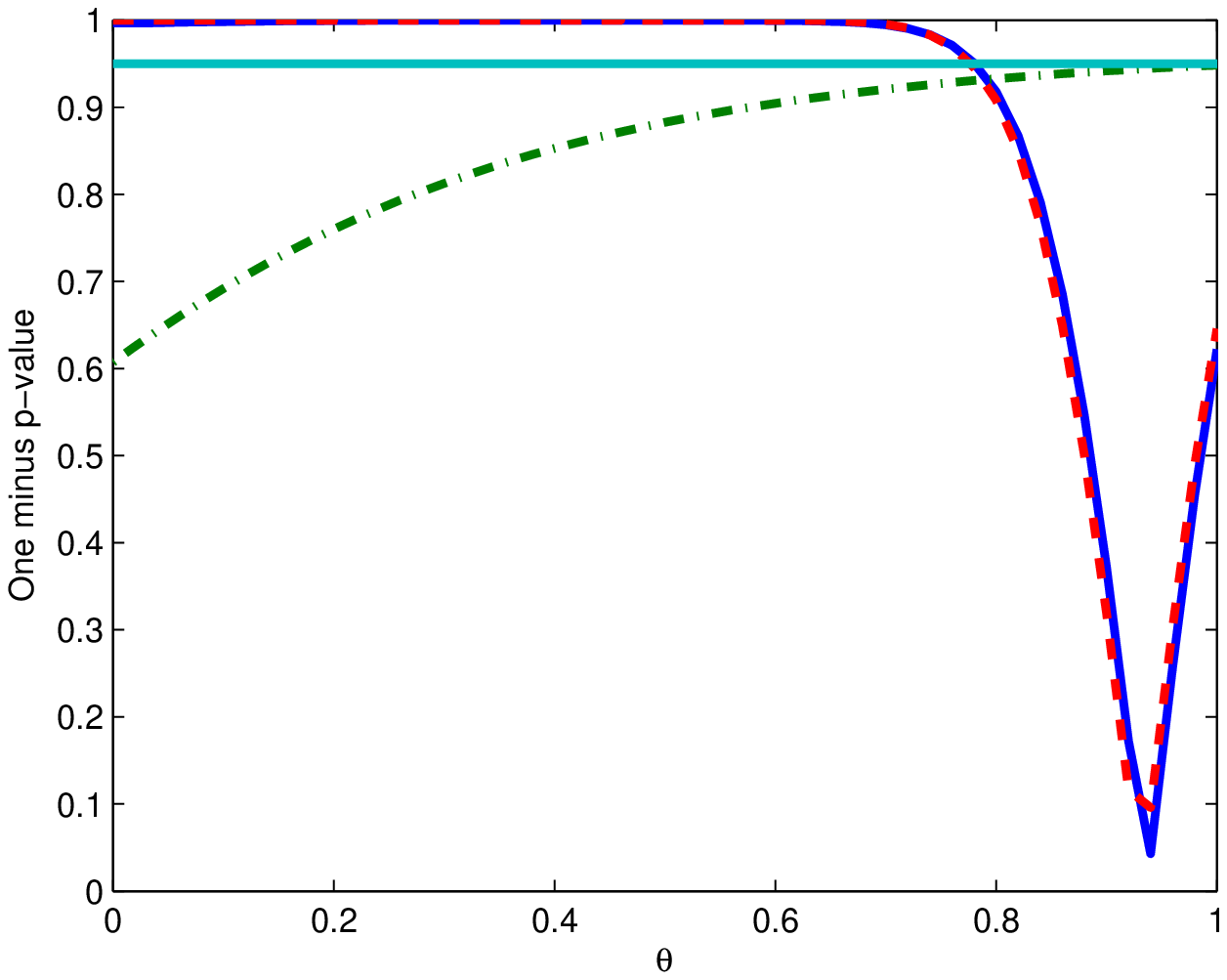}%
}
&  &
\raisebox{-0pt}{\includegraphics[
height=2.2139in,
width=2.9386in
]%
{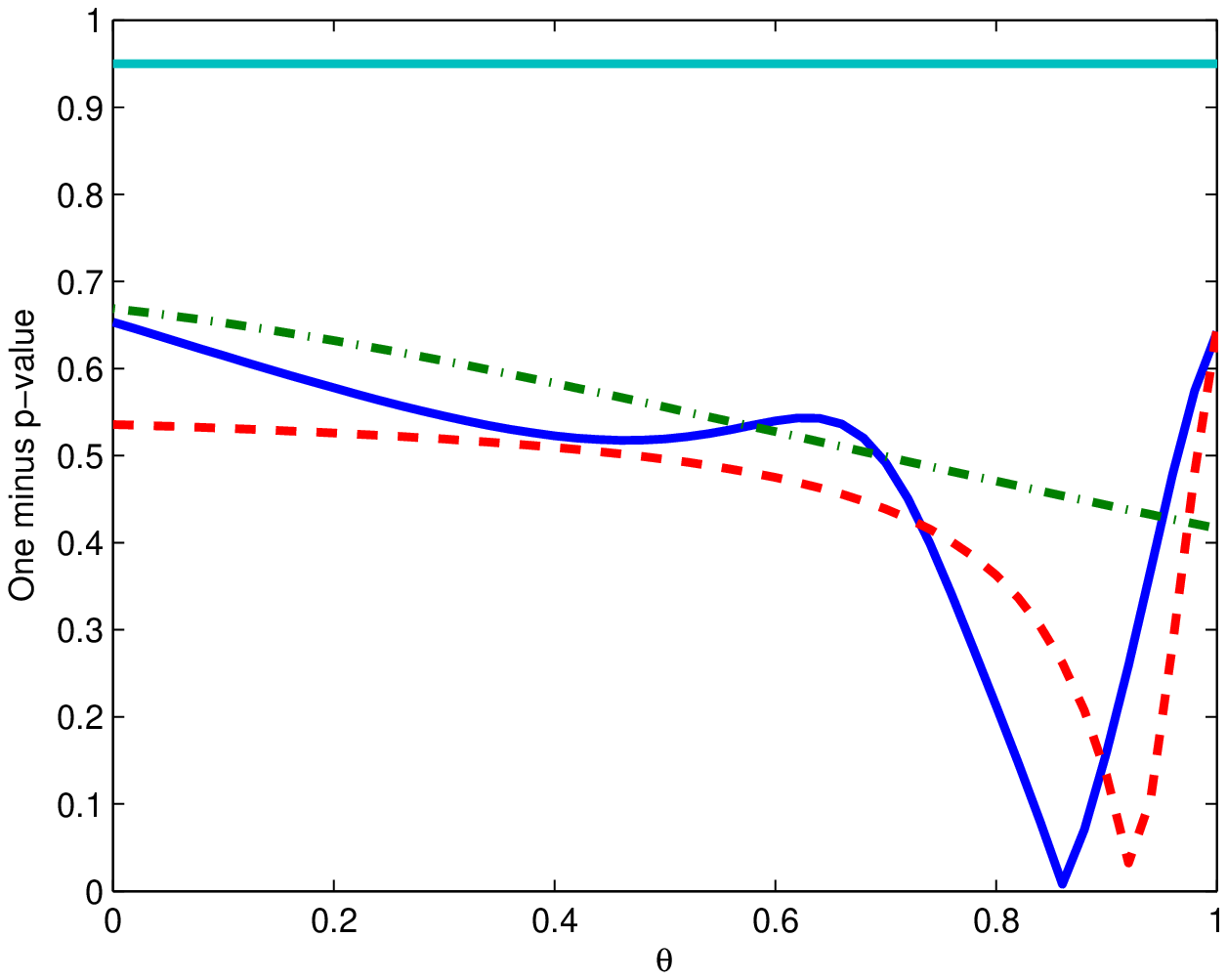}%
}
\\
\text{Figure 3.1: }T=3,\text{ }\sigma_{c}^{2}=0 &  & \text{Figure 3.2:
}T=3,\text{ }\sigma_{c}^{2}=0.5
\end{array}
\\%
\begin{array}
[c]{ccc}%
\raisebox{-0pt}{\includegraphics[
height=2.2139in,
width=2.9386in
]%
{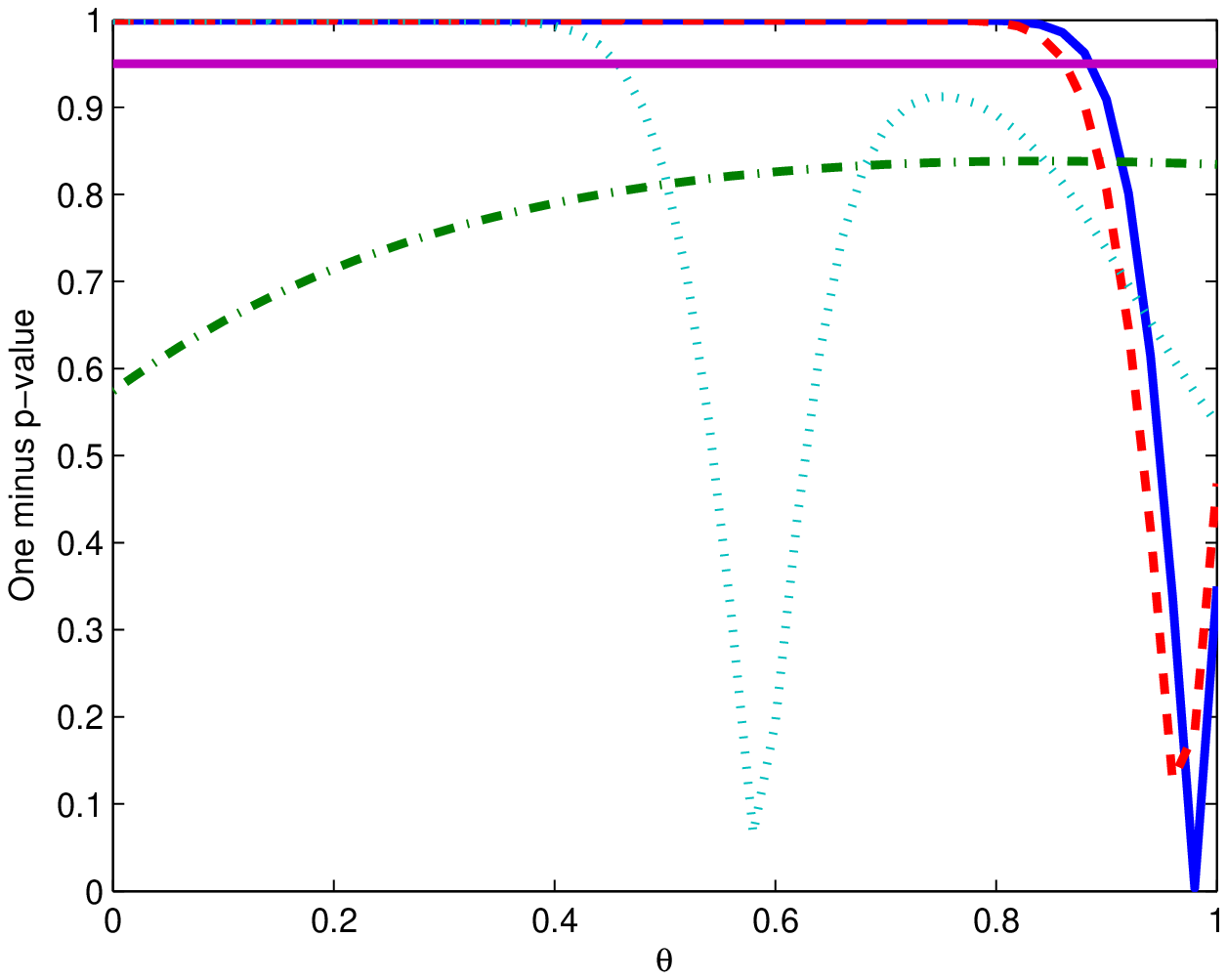}%
}
&  &
\raisebox{-0pt}{\includegraphics[
height=2.2139in,
width=2.9386in
]%
{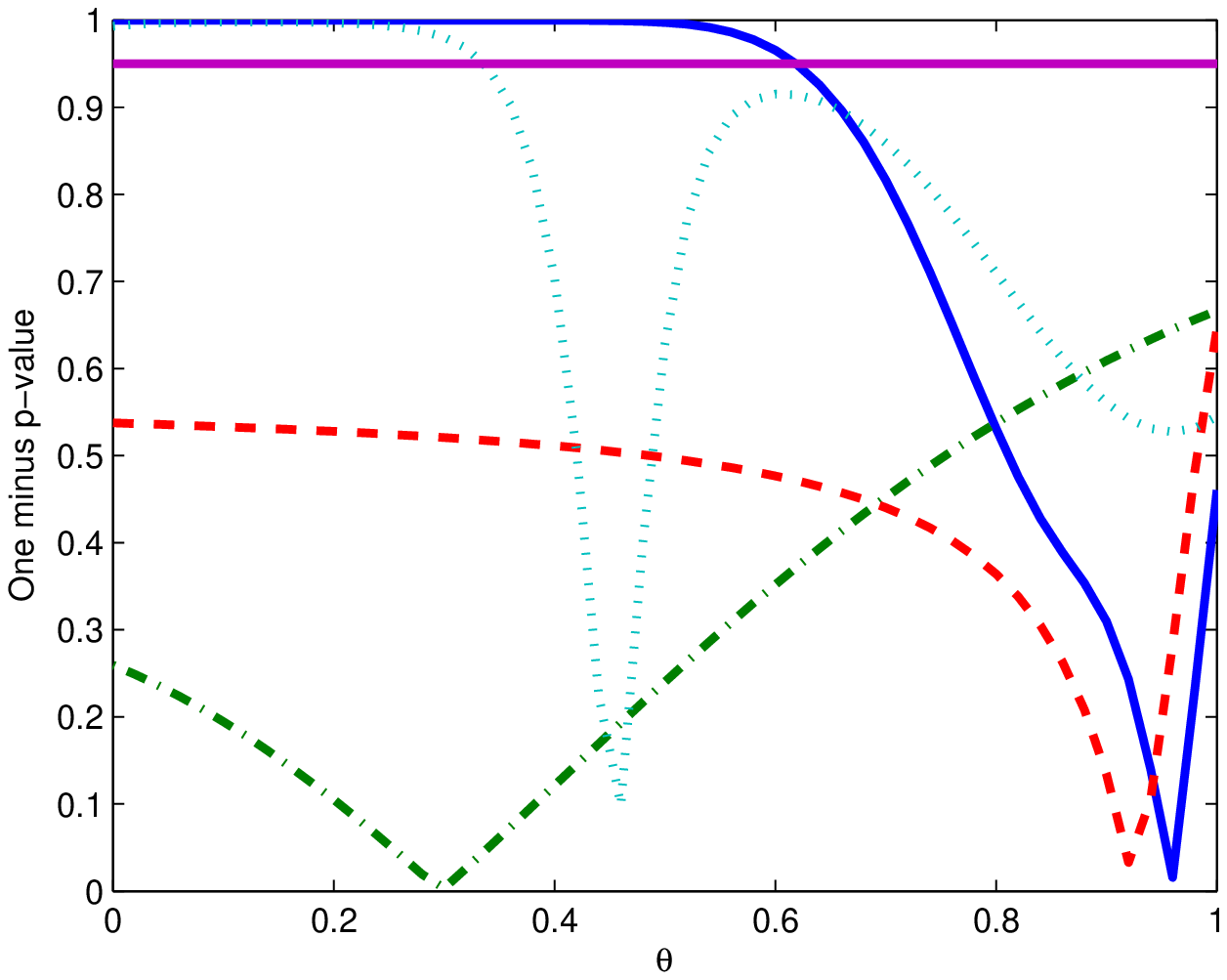}%
}
\\
\text{Figure 3.3: }T=4,\text{ }\sigma_{c}^{2}=0 &  & \text{Figure 3.4:
}T=4,\text{ }\sigma_{c}^{2}=0.5
\end{array}
\end{array}
\]

\section{Identification from different moment conditions}

Stylized facts 1-4 illustrated by the figures in Panels 1-3 show the
identification issues that occur for the autoregressive parameter $\theta$
when the variance of the initial observations is large and $\theta_{0}$, i.e.
the true value in the DGP,\ is close to one. To pin these identification
issues down precisely, we use an asymptotic sampling scheme which consists of
joint drifting sequences for the autoregressive parameter and the variance of
the initial observation. We indicate this dependence on the sample size
$N$\ by $\theta_{0,N}$\ and $h_{N}(\theta_{0,N})=\frac{1}{\sqrt{var(y_{i1})}}%
$\ respectively. The true value of $\theta,$ previously denoted by $\theta
_{0},$ is from now on therefore denoted by $\theta_{0,N}$. Assumptions 1 and 2
group the different requirements needed to obtain our results.

\paragraph{Assumption 1.}

\noindent\textbf{a. }\textit{The drifting sequences of the autoregressive
parameter and variance of the initial observations are such that: }%
\begin{equation}%
\begin{array}
[c]{rl}%
\lim_{N\rightarrow\infty}\theta_{0,N}= & 1\\
\multicolumn{1}{c}{\lim_{N\rightarrow\infty}h_{N}(\theta_{0,N})=} & d_{1},
\end{array}
\label{limseq}%
\end{equation}
\textit{with }$d_{1}$\textit{\ a finite, possibly zero constant. }

\noindent\textbf{b. }\textit{The initial observations satisfy the mean
stationarity conditions in (\ref{init1})-(\ref{init2}).}

\noindent\textbf{c.}\textit{\ The joint limit behavior of the variance of
}$u_{i1}$ \textit{and} $(1-\theta_{0,N})$\textit{\ is such that}%
\begin{equation}%
\begin{array}
[c]{c}%
\lim_{N\rightarrow\infty}(1-\theta_{0,N})\sigma_{1,N}^{2}=d_{2},
\end{array}
\label{ass1}%
\end{equation}
\textit{with }$\sigma_{1,N}^{2}=$\textit{var(}$u_{i1})$\textit{, }$d_{2}%
$\textit{\ a finite, possibly zero constant and }$(1-\theta_{0,N})^{\frac
{1}{2}}u_{i1}$ \textit{is a random variable with finite fourth order moments.
}

\noindent\textbf{d. }\textit{The variance of the product of the initial
observation }$y_{i1}$\textit{\ and the disturbances }$u_{it}$\textit{\ is such
that}%
\begin{equation}%
\begin{array}
[c]{c}%
var(u_{it}y_{i1})=\sigma_{t}^{2}var(y_{i1}),\text{ }t=2,\ldots,T,
\end{array}
\label{condvary1}%
\end{equation}
\textit{with }$\sigma_{t}^{2}=var(u_{it}),$ $t=2,\ldots,T.$

\noindent\textbf{e. }\textit{The errors }$u_{i1}/\sigma_{1,N},$ $u_{i2}%
\ldots,u_{iT}$\textit{\ and }$c_{i},$ $i=1,\ldots,N,$\textit{\ are
independently distributed within individuals and over the different
individuals and have mean zero, finite variance and finite fourth order
moments and satisfy the conditions in (\ref{marexpec}).\smallskip}

Assumption 1a concerns the joint limit behavior of the variance of the initial
observations and $\theta_{0,N}$. By the definition of $\mu_{i}$ in
(\ref{init3}) and Assumption 1a, $\mu_{i}$ is also drifting with the sample
size since it is a function of $\theta_{0,N}$ and so are $y_{i1}$ and
$\sigma_{1,N}^{2}.$ Assumption 1b specifies that the initial observations
follow the mean stationarity assumption, which is necessary for the Lev and
Sys moment conditions to hold. Assumptions 1c-e are mainly technical
assumptions, which are needed to obtain our theoretical results. Assumption 1c
sets an upper bound on the rate at which the variance of $u_{i1}$ can diverge.
It implies that the variance of $u_{i1}$ is at most proportional to
$(1-\theta_{0,N})^{-1}$ (so covariance stationarity is allowed for).
Assumption 1d holds under independence of $u_{it}$ and $y_{i1}$ but it can
also hold under less stringent conditions. In the sequel, we analyze the
identification of $\theta$ when the variance of the initial observations gets
large compared to that of the subsequent disturbances. Assumption 1d enables
such settings. Assumption 1e is a technical assumption which is needed to use
a central limit theorem.

Assumption 1a allows the variance of the initial observations to be large
jointly with a large value for the autoregressive parameter. When $d_{1}$\ in
(\ref{limseq}) equals zero, the rate at which $h_{N}(\theta_{0,N})$\ goes to
zero, or the variance of the initial observation goes to infinity, is key to
the identification of $\theta$ from the sample moment conditions. We therefore
put down two alternative assumptions regarding the joint convergence of the
sample size and the variance of the initial observations under which there is
identification or identification is problematic for specific moment conditions.

\paragraph{Assumption 2.}

\noindent\textbf{a. }$d_{1}=0$ \textit{and the drifting sequence of the
variance of the initial observation is such that:}%
\begin{equation}%
\begin{array}
[c]{c}%
h_{N}(\theta_{0,N})\sqrt{N}\underset{N\rightarrow\infty}{\rightarrow}0.
\end{array}
\label{convinfzero}%
\end{equation}

\noindent\textbf{b. }$d_{1}\neq0$ \textit{or the drifting sequence of the
variance of the initial observation is such that:}
\begin{equation}%
\begin{array}
[c]{c}%
h_{N}(\theta_{0,N})\sqrt{N}\underset{N\rightarrow\infty}{\rightarrow}\infty.
\end{array}
\label{convinfinity}%
\end{equation}

Identification generically holds under Assumption 2b but can become
problematic under Assumption 2a and then depends on the particular moment
condition and number of time series observations as we show later on. In the
intermediate case where $h_{N}(\theta_{0,N})\sqrt{N}$ converges to a finite,
but non-zero constant, we are in a case similar to that discussed in the weak
instrument literature where the sample Jacobian converges to a random variable
which leads to inconsistent estimators with non-standard behavior of their
corresponding t-statistics. Because of the practical similarities with
Assumption 2a, however, we do not separately discuss it.

Since any assumption about the convergence rates of the sample size and the
variance of the initial observations is to a large extent arbitrary, also the
identification of $\theta$ by these conditions is arbitrary for DGPs for which
the true value of $\theta$ is close to one and the variance of the initial
observations is infinite when the true value of $\theta$ equals one. Some
plausible DGPs, all of which accord with mean stationarity (\ref{init1}%
)-(\ref{init2}), for the initial observations belong to this category:

\begin{description}
\item[DGP 1.] $\sigma_{c}^{2}=$var($c_{i}),$ $\sigma_{1,N}^{2}=\sigma_{1}%
^{2},$ $h(\theta_{0,N})^{-2}=\sigma_{c}^{2}/(1-\theta_{0,N})^{2}+\sigma
_{1}^{2},$ so when $\theta_{0,N}\underset{N\rightarrow\infty}{\rightarrow}1,$
$(1-\theta_{0,N})^{-1}h(\theta_{0,N})\underset{N\rightarrow\infty}%
{\rightarrow}\sigma_{c}^{-1}.$

\item[DGP 2.] $\sigma_{c}^{2}=$var($c_{i}),$ $\sigma_{1,N}^{2}=\frac
{\sigma^{2}}{1-\theta_{0,N}^{2}},$ $\sigma^{2}=var(u_{it}),$ $t=2,\ldots,T,$
$h(\theta_{0,N})^{-2}=\sigma_{c}^{2}/(1-\theta_{0,N})^{2}+\sigma^{2}%
/(1-\theta_{0,N}^{2}),$ so when $\theta_{0,N}\underset{N\rightarrow\infty
}{\rightarrow}1,$ $(1-\theta_{0,N})^{-1}h(\theta_{0,N})\underset
{N\rightarrow\infty}{\rightarrow}\sigma_{c}^{-1}.$

\item[DGP 3.] $\sigma_{\mu}^{2}=$var($\mu_{i}),$ $\sigma_{1,N}^{2}%
=\frac{\sigma^{2}}{1-\theta_{0,N}^{2}},$ $\sigma^{2}=var(u_{it}),$
$t=2,\ldots,T,$ $h(\theta_{0,N})^{-2}=\sigma_{\mu}^{2}+\sigma^{2}%
/(1-\theta_{0,N}^{2}),$ so when $\theta_{0,N}\underset{N\rightarrow\infty
}{\rightarrow}1,$ $(1-\theta_{0,N}^{2})^{-\frac{1}{2}}h(\theta_{0,N}%
)\underset{N\rightarrow\infty}{\rightarrow}\sigma^{-1}.$

\item[DGP 4.] $\sigma_{\mu}^{2}=$var($\mu_{i}),$ $\sigma_{1,N}^{2}=\sigma
^{2}\frac{1-\theta_{0,N}^{2(g+1)}}{1-\theta_{0,N}^{2}},$ $\sigma
^{2}=var(u_{it}),$ $t=2,\ldots,T,$ $h(\theta_{0,N})^{-2}=\sigma_{\mu}%
^{2}+\sigma^{2}\frac{1-\theta_{0,N}^{2(g+1)}}{1-\theta_{0,N}^{2}},$ so when
$\theta_{0,N}\underset{N\rightarrow\infty}{\rightarrow}1,$ $\left(
\frac{1-\theta_{0,N}^{2}}{1-\theta_{0,N}^{2(g+1)}}\right)  ^{-\frac{1}{2}%
}h(\theta_{0,N})\underset{N\rightarrow\infty}{\rightarrow}\sigma^{-1}.$

\item[DGP 5.] $\sigma_{c}^{2}=$var($c_{i}),$ $\sigma_{1,N}^{2}=\sigma^{2}%
\frac{1-\theta_{0,N}^{2(g+1)}}{1-\theta_{0,N}^{2}},$ $\sigma^{2}=var(u_{it}),$
$t=2,\ldots,T,$ $h(\theta_{0,N})^{-2}=\sigma_{c}^{2}/(1-\theta_{0,N}%
)^{2}+\sigma^{2}\frac{1-\theta_{0,N}^{2(g+1)}}{1-\theta_{0,N}^{2}},$ so when
$\theta_{0,N}\underset{N\rightarrow\infty}{\rightarrow}1,$ $\left(
1-\theta_{0,N}\right)  ^{-1}h(\theta_{0,N})\underset{N\rightarrow\infty
}{\rightarrow}\sigma_{c}^{-1}.$
\end{description}

\noindent DGPs 4 and 5 characterize an autoregressive process of order one
that has started $g$ periods in the past while the initial observations that
result from DGP 2 and 3 result from an autoregressive process that has started
an infinite number of periods in the past. DGPs 2 and 3 are also used by
Blundell and Bond (1998) and Arellano and Bover (1995) use DGP 2, but these
studies keep the variance of the initial observations fixed.

For DGPs 1-5 to imply Assumption 2a, the limiting sequence $\theta_{0,N}$ has
to be such that:
\begin{equation}%
\begin{array}
[c]{rlcl}%
\text{DGP }1,\text{ }2,\text{ }5: & (1-\theta_{0,N})\sqrt{N}\underset
{N\rightarrow\infty}{\rightarrow}0\text{ } & \text{for which it is sufficient
that} & \theta_{0,N}=1-\frac{e}{N^{\frac{1}{2}(1+\epsilon)}}\\
\text{DGP }3: & (1-\theta_{0,N}^{2})N\underset{N\rightarrow\infty}%
{\rightarrow}0 & \text{for which it is sufficient that} & \theta_{0,N}%
=1-\frac{e}{N^{1+\epsilon}}\\
\text{DGP }4: & \frac{N}{g}\underset{N\rightarrow\infty,\text{ }%
g\rightarrow\infty}{\rightarrow}0, &  &
\end{array}
\label{startup}%
\end{equation}
with $e$ a constant and $\epsilon$ some real number larger than zero. In the
case of DGP 4, (\ref{startup}) implies that the process has been running
longer than the sample size $N.$ Kruiniger (2009)\nocite{krui09} uses the
above specification of DGP 3 with $\epsilon=0$ and DGP\ 4 with $N/g$
converging to a constant to construct local to unity asymptotic approximations
of the distributions of two step GMM estimators that use the Dif, Lev or Sys
moment conditions.

We do not confine ourselves to a specific DGP for the initial observations so
we obtain results that apply more generally. While the (non-) identification
conditions for identifying $\theta$ that result from the above data generating
processes might be (in)plausible, it is the arbitrariness of them which is
problematic. Additionally, the identification condition might hold but it can
still lead to large size distortions of Wald test statistics, like, the t-test.

To analyze the identification of $\theta$ by the different moment conditions
for a general number of time periods $T$, we start out with a representation
theorem. For the different moment conditions, it states the behavior of the
sample moments and their derivatives under Assumptions 1 and 2a.

\paragraph{Theorem 1 (Representation Theorem).}

\textit{Under Assumptions 1 and 2a, we can characterize the large sample
behavior of the Dif, Lev, NL, AS and Sys sample moments for }$T$ \textit{time
series observations and their derivatives by:}%
\begin{equation}%
\begin{array}
[c]{ll}%
\left(
\begin{array}
[c]{c}%
f_{N}^{j}(\theta)\\
q_{N}^{j}(\theta)
\end{array}
\right)  = & \left(
\begin{array}
[c]{c}%
A_{f}^{j}(\theta)\\
A_{q}^{j}(\theta)
\end{array}
\right)  \left[  \frac{1}{h_{N}(\theta_{0,N})\sqrt{N}}\left(  \psi
-h_{N}(\theta_{0,N})\sigma_{1,N}\iota_{T-1}\psi_{c}\right)  -\iota_{T-1}%
d_{2}\right]  +\\
& \left(
\begin{array}
[c]{c}%
\mu_{f}^{j}(\theta,\bar{\sigma}^{2})\\
\mu_{q}^{j}(\theta,\bar{\sigma}^{2})
\end{array}
\right)  +o_{p}(1),
\end{array}
\label{theo1}%
\end{equation}
\textit{with }$j=Dif,$ $Lev,$ $NL,$ $AS,$ $Sys$. \textit{The specifications of
the }$k_{j}$\textit{-dimensional sample moments }$f_{N}^{j}(\theta
)$\textit{\ and derivatives }$q_{N}^{j}(\theta)$\textit{\ are given in the
Appendix. Furthermore,} $A_{f}^{j}(\theta),$ $A_{q}^{j}(\theta),$ $\mu_{f}%
^{j}(\theta,\bar{\sigma}^{2})$\textit{\ and }$\mu_{q}^{j}(\theta,\bar{\sigma
}^{2})$ \textit{are constant }$k_{j}\times(T-1),$ $k_{j}\times(T-1),$
$k_{j}\times1$ \textit{and }$k_{j}\times1$ \textit{dimensional matrices,
}$\bar{\sigma}^{2}=(\sigma_{2}^{2}\ldots\sigma_{T}^{2}),$
\begin{equation}%
\begin{array}
[c]{rl}%
\frac{h_{N}(\theta_{0,N})}{\sqrt{N}}\sum_{i=1}^{N}\left(
\begin{array}
[c]{c}%
y_{i1}u_{i2}\\
\vdots\\
y_{i1}u_{iT}%
\end{array}
\right)  \underset{d}{\rightarrow} & \psi\\
\frac{1}{\sqrt{N}}\sum_{i=1}^{N}\frac{u_{i1}}{\sigma_{1,N}}c_{i}\underset
{d}{\rightarrow} & \psi_{c},
\end{array}
\label{psidef}%
\end{equation}
\textit{so }$\psi$\textit{\ is a }$(T-1)$\textit{-dimensional normal random
vector, }$\psi\sim N(0,$\textit{diag(}$\sigma_{2}^{2}\ldots\sigma_{T}^{2}))$,
$\psi_{c}\sim N(0,$\textit{var(}$c_{i}))$ \textit{and} \textit{independent
from }$\psi,$ \textit{and }$\iota_{T-1}$\textit{\ is a }$(T-1)$%
\textit{-dimensional vector of ones. }

\noindent\textit{The specifications of }$A_{f}^{j}(\theta),$\textit{\ }%
$A_{q}^{j}(\theta),\mathit{\ }\mu_{f}^{j}(\theta,\bar{\sigma}^{2}%
),\mathit{\ }\mu_{q}^{j}(\theta,\bar{\sigma}^{2})$\textit{\ for values of }$T$
\textit{equal to 3-5 are all stated in the Appendix.}\medskip

\begin{proof}
see the Appendix.\medskip
\end{proof}

The representation theorem in Theorem 1 is reminiscent of the cointegration
representation theorem, see $e.g.$ Engle and Granger (1987)\nocite{eg87} and
Johansen (1991).\nocite{JS91} Identical to that representation theorem,
Theorem 1 shows that the behavior of the moment series changes over different directions.

Theorem 1 implies that the sample moment and its derivative diverge in the
direction of $\binom{A_{f}^{j}(\theta)}{A_{q}^{j}(\theta)}$ since the latter
components get multiplied by $\frac{1}{h(\theta_{0,N})\sqrt{N}},$ which under
Assumption 2a goes off to infinity when the sample size increases. The only
identifying information for $\theta$ then results from that part of the sample
moment which does not depend on $\psi$. Since $\psi$ only affects the part of
the sample moments spanned by $A_{f}^{j}(\theta), $ the sample moments are
independent of $\psi$ in the direction of the maximal non-degenerate space
spanned by vectors orthogonal to $A_{f}^{j}(\theta)$ to which we refer as the
orthogonal complement of $A_{f}^{j}(\theta)$. We construct the orthogonal
complement, which we denote by $A_{f}^{j}(\theta)_{\perp}$, as the full rank
matrix projecting on the orthogonal complement of the range space of
$A_{f}^{j}(\theta)$. It consists of the minimal set of vectors spanning the
null space of the columns of $A_{f}^{j}(\theta)$. In the case the null space
has dimension zero, a full rank specification of $A_{f}^{j}(\theta)_{\perp}$
can not be constructed.

When we pre-multiply the sample moments by the orthogonal complement of
$A_{f}^{j}(\theta),$ we obtain%
\begin{equation}%
\begin{array}
[c]{cl}%
A_{f}^{j}(\theta)_{\perp}^{\prime}f_{N}^{j}(\theta)= & A_{f}^{j}%
(\theta)_{\perp}^{\prime}\mu_{f}^{j}(\theta,\bar{\sigma}^{2})+o_{p}(1).
\end{array}
\label{theo2}%
\end{equation}
Compared with expression (\ref{theo1}) in\ Theorem 1, the elements multiplied
by $A_{f}^{j}(\theta)$ have dropped out since $A_{f}^{j}(\theta)_{\perp
}^{\prime}A_{f}^{j}(\theta)\equiv0.$ The right hand side of (\ref{theo2}) now
contains all remaining identifying elements of the original moment conditions.
From expression (\ref{theo2}), it is seen that identification results only
when (1) $A_{f}^{j}(\theta)_{\perp}$ is a full rank matrix; (2) $A_{f}%
^{j}(\theta)_{\perp}^{\prime}\mu_{f}^{j}(\theta,\bar{\sigma}^{2})\neq0$ for
all $\theta\neq\theta_{0,N}$.

For an illustrative example of Theorem 1, consider the large sample behavior
for $T=3$ of the Lev sample moment, $\frac{1}{N}\sum_{i=1}^{N}\Delta
y_{i2}(y_{i3}-\theta y_{i2}),$ and its derivative, $-\frac{1}{N}\sum_{i=1}%
^{N}y_{i2}\Delta y_{i2},$ when $\theta_{0,N}$ converges to one according to
(\ref{limseq}) and mean stationarity (\ref{init3})-(\ref{init2}) applies. The
Lev moment condition has been proposed by Arellano and Bover (1995) and
Blundell and Bond (1998) to overcome the identification problems of the Dif
moment condition near the unit root. Under Assumption 1, the relevant elements
for the large sample behavior are:%
\begin{equation}%
\begin{array}
[c]{rl}%
f_{N}^{Lev}(\theta)= & \frac{1}{N}\sum_{i=1}^{N}\Delta y_{i2}(y_{i3}-\theta
y_{i2})\\
= & (1-\theta)\left\{  \frac{1}{N}\sum_{i=1}^{N}u_{i2}^{2}+\frac{1}{N}%
\sum_{i=1}^{N}u_{i2}y_{i1}+\right. \\
& \left.  \frac{1}{N}\sum_{i=1}^{N}(\theta_{0,N}-1)u_{i1}y_{i1}\right\}
+o_{p}(1),\\
q_{N}^{Lev}(\theta)= & -\frac{1}{N}\sum_{i=1}^{N}y_{i2}\Delta y_{i2}\\
= & -\frac{1}{N}\sum_{i=1}^{N}u_{i2}^{2}-\frac{1}{N}\sum_{i=1}^{N}u_{i2}%
y_{i1}\\
& -\frac{1}{N}\sum_{i=1}^{N}(1-\theta_{0,N})u_{i1}y_{i1}+o_{p}(1),
\end{array}
\label{limlevavg}%
\end{equation}
see the proof of Theorem 1 in the Appendix for a derivation. The $o_{p}%
(1)$\ remainder terms contain all elements in (\ref{limlevavg}) that can not
dominate the large sample behavior when $\theta_{0,N}$\ goes to one according
to the drifting parameter sequences defined in Assumption 1. The components
explicitly specified in (\ref{limlevavg}) either have a non-zero mean or
depend on the initial observations $y_{i1}$. Under Assumption 1, we have that%
\begin{equation}%
\begin{array}
[c]{c}%
h_{N}(\theta_{0,N})\frac{1}{\sqrt{N}}\sum_{i=1}^{N}u_{i2}y_{i1}\underset
{d}{\rightarrow}\psi_{2},\text{ }\frac{1}{\sqrt{N}}\sum_{i=1}^{N}\frac{u_{i1}%
}{\sigma_{1,N}}c_{i}\underset{d}{\rightarrow}\psi_{c},
\end{array}
\label{limh}%
\end{equation}
which is proven in Lemma 1 in the Appendix and where $\psi_{2}$\ and $\psi
_{c}$ are independent normal random variables with mean zero and variance
$\sigma_{2}^{2}$ and $\sigma_{c}^{2},$ $\sigma_{c}^{2}=$var($c_{i})$. It
explains why $\frac{1}{N}\sum_{i=1}^{N}u_{i2}y_{i1}$\ and $\frac{1}{N}%
\sum_{i=1}^{N}(\theta_{0,N}-1)u_{i1}y_{i1}=\frac{1}{N}\sum_{i=1}^{N}%
(\theta_{0,N}-1)u_{i1}^{2}+\frac{1}{N}\sum_{i=1}^{N}u_{i1}c_{i}$ explicitly
appear in (\ref{limlevavg}). When $d_{1}$\ in (\ref{limseq}) equals zero, the
rate at which $h_{N}(\theta_{0,N})$\ goes to zero, or the variance of the
initial observation goes to infinity, determines the behavior of the sample
moments in (\ref{limlevavg}). For example, when $d_{1}=0$ and these sequences
are as in Assumption 2b, it holds that%
\begin{equation}%
\begin{array}
[c]{rcl}%
\frac{1}{N}\sum_{i=1}^{N}y_{i2}\Delta y_{i2} & \underset{p}{\rightarrow} &
\sigma_{2}^{2}-d_{2}.
\end{array}
\label{convjaclev}%
\end{equation}
Although Assumption 1 does not fully pin down $d_{2}$, which value depends on
the particular DGP for the initial observations, it is clear that the
probability limit of the sample Jacobian typically differs from zero. Hence,
the Lev moment condition seems to identify $\theta$ irrespective of its true
value, see Arellano and Bover (1995) and Blundell and Bond (1998). There is a
caveat though since, under Assumption 2a, Theorem 1 shows that:
\begin{equation}%
\begin{array}
[c]{rl}%
f_{N}^{Lev}(\theta)= & \frac{1}{h_{N}(\theta_{0,N})\sqrt{N}}\frac{h_{N}%
(\theta_{0,N})}{\sqrt{N}}\sum_{i=1}^{N}\Delta y_{i2}(y_{i3}-\theta y_{i2})\\
= & (1-\theta)\left\{  \frac{1}{h_{N}(\theta_{0,N})\sqrt{N}}(\psi_{2}%
-h_{N}(\theta_{0,N})\sigma_{1,N}\psi_{c})+(\sigma_{2}^{2}-d_{2})\right\}
+o_{p}(1),\\
q_{N}^{Lev}(\theta)= & -\frac{1}{h_{N}(\theta_{0,N})\sqrt{N}}\frac
{h_{N}(\theta_{0,N})}{\sqrt{N}}\sum_{i=1}^{N}y_{i2}\Delta y_{i2},\\
= & -\frac{1}{h_{N}(\theta_{0,N})\sqrt{N}}(\psi_{2}-h_{N}(\theta_{0,N}%
)\sigma_{1,N}\psi_{c})-(\sigma_{2}^{2}-d_{2})+o_{p}(1),
\end{array}
\label{convmom}%
\end{equation}
which implies that the sample moments of the Lev population moment and
Jacobian diverge when the sample size increases. The Lev sample moment then no
longer identifies $\theta$ since the components that would identify $\theta$
in the Jacobian identification condition, i.e. $\frac{1}{N}\sum_{i=1}%
^{N}u_{i2}^{2},$ gets dominated by the component $\frac{1}{N}\sum_{i=1}%
^{N}u_{i2}y_{i1}$ and possibly $\frac{1}{N}\sum_{i=1}^{N}(1-\theta
_{0,N})u_{i1}y_{i1}.$

We next discuss what Theorem 1 implies for the different sets of moment
conditions discussed previously and their respective orthogonal complements of
$A_{f}(\theta).$

\paragraph{Dif and Lev conditions}

When $T=3$ or $4,$ the specifications of $\mu_{f}^{j}(\theta,\bar{\sigma}%
^{2})$, $A_{f}^{j}(\theta)$ and $A_{f}^{j}(\theta)_{\perp}$ for the Dif and
Lev moment conditions, which are stated in the proof of Theorem 1 in the
Appendix, are:%
\begin{equation}%
\begin{array}
[c]{cll}%
\text{\textbf{Dif:}} & \mathbf{T=3} & \mu_{f}^{Dif}(\theta,\bar{\sigma}%
^{2})=0,\text{ }A_{f}^{Dif}(\theta)=(-\theta\text{ }1),\text{ }A_{f}%
^{Dif}(\theta)_{\perp}=(1\text{ }\theta)\text{ }\\
& \mathbf{T=4} & \mu_{f}^{Dif}(\theta,\bar{\sigma}^{2})=\left(
\begin{array}
[c]{c}%
0\\
0\\
0
\end{array}
\right)  ,\text{ }A_{f}^{Dif}(\theta)=\left(
\begin{array}
[c]{ccc}%
-\theta & 1 & 0\\
0 & -\theta & 1\\
0 & -\theta & 1
\end{array}
\right)  ,\text{ }A_{f}^{Dif}(\theta)_{\perp}=\left(
\begin{array}
[c]{c}%
0\\
-1\\
1
\end{array}
\right)  .\\
\text{\textbf{Lev:}} & \mathbf{T=3} & \mu_{f}^{Lev}(\theta,\bar{\sigma}%
^{2})=\left(  1-\theta\right)  \left(
\begin{array}
[c]{c}%
\sigma_{2}^{2}\\
0
\end{array}
\right)  ,\text{ }A_{f}^{Lev}(\theta)=(1-\theta\text{ }0),\text{ }A_{f}%
^{Lev}(\theta)_{\perp}\text{ does not exist}\\
& \mathbf{T=4} & \mu_{f}^{Lev}(\theta,\bar{\sigma}^{2})=\left(  1-\theta
\right)  \left(
\begin{array}
[c]{c}%
\sigma_{2}^{2}\\
\sigma_{3}^{2}\\
0
\end{array}
\right)  ,\text{ }A_{f}^{Lev}(\theta)=\left(
\begin{array}
[c]{ccc}%
1-\theta & 0 & 0\\
0 & 1-\theta & 0
\end{array}
\right)  ,\\
&  & A_{f}^{Lev}(\theta)_{\perp}\text{ does not exist.}%
\end{array}
\label{diflevt=4}%
\end{equation}
The expressions of $A_{f}^{Lev}(\theta)$ are all such that we cannot specify a
non-zero matrix $A_{f}^{Lev}(\theta)_{\perp}$ such that $A_{f}^{Lev}%
(\theta)_{\perp}^{\prime}A_{f}^{Lev}(\theta)=0.$ This remains so when $T$
exceeds four, see the Appendix. Hence, $A_{f}^{Lev}(\theta)_{\perp}$ does not
exist (as a non-zero matrix). Regarding the Dif moments, when $T>3$ the rank
of the orthogonal complement of $A_{f}^{Dif}(\theta),$ $A_{f}^{Dif}%
(\theta)_{\perp},$ is larger than zero. However, since $\mu_{f}^{Dif}%
(\theta,\bar{\sigma}^{2})$ equals zero for any value of $T$, $A_{f}%
^{Dif}(\theta)_{\perp}^{\prime}\mu_{f}^{Dif}(\theta,\bar{\sigma}^{2})=0$ so
the Dif moment conditions do not identify $\theta.$ Summarizing, we have:%
\begin{equation}%
\begin{array}
[c]{rll}%
\text{\textbf{Dif}:} & \mu_{f}^{Dif}(\theta,\bar{\sigma}^{2})\text{ is vector
of all zeros.} & \text{No identification when }T\geq3.\\
\text{\textbf{Lev:}} & A_{f}^{Lev}(\theta)_{\perp}\text{ does not exist.} &
\text{No identification when }T\geq3.
\end{array}
\end{equation}

\paragraph{NL condition}

The NL\ moment condition is not defined for $T=3.$ When $T=4,$ the expressions
of $\mu_{f}^{j}(\theta,\bar{\sigma}^{2})$, $A_{f}^{j}(\theta)$ and $A_{f}%
^{j}(\theta)_{\perp}$ read
\begin{equation}%
\begin{array}
[c]{cl}%
\text{\textbf{NL:}} & \mu_{f}^{NL}(\theta,\bar{\sigma}^{2})=\left(
1-\theta\right)  \left(  \sigma_{3}^{2}-\theta\sigma_{2}^{2}\right)  ,\text{
}A_{f}^{NL}(\theta)=\left(
\begin{array}
[c]{ccc}%
\theta(\theta-1) & 1-\theta & 0
\end{array}
\right)  ,\text{ }\\
& A_{f}^{NL}(\theta)_{\perp}\text{ does not exist.}%
\end{array}
\label{difnlt=4}%
\end{equation}
Since the orthogonal complement does not exist, the NL moment condition does
not identify $\theta.$ The expression of $A_{f}^{NL}(\theta)$ for a larger
number of time series observations (see the Appendix) is also such that the
orthogonal complement $A_{f}^{NL}(\theta)_{\perp}$ also does not exist. Hence
for larger values of $T,$ the NL moment conditions also do not identify
$\theta.$

\paragraph{AS and Sys conditions}

The expressions of $\mu_{f}^{j}(\theta,\bar{\sigma}^{2})$, $A_{f}^{j}(\theta)$
and $A_{f}^{j}(\theta)_{\perp}$ when $T=3,$ $4$ for the AS and Sys moment
conditions result from stacking those of the Dif and NL and Dif and Lev moment
conditions respectively:%
\[%
\begin{array}
[c]{cll}%
\text{\textbf{AS:}} & \mathbf{T=4} & \mu_{f}^{AS}(\theta,\bar{\sigma}%
^{2})=\left(
\begin{array}
[c]{c}%
0\\
0\\
0\\
\left(  1-\theta\right)  \left(  \sigma_{3}^{2}-\theta\sigma_{2}^{2}\right)
\end{array}
\right)  ,\text{ }A_{f}^{AS}(\theta)=\left(
\begin{array}
[c]{ccc}%
-\theta & 1 & 0\\
0 & -\theta & 1\\
0 & -\theta & 1\\
\theta(\theta-1) & 1-\theta & 0
\end{array}
\right)  ,\text{ }\\
&  & A_{f}^{AS}(\theta)_{\perp}=\left(
\begin{array}
[c]{cc}%
\theta-1 & 0\\
0 & -1\\
0 & 1\\
1 & 0
\end{array}
\right)  .\\
\text{\textbf{Sys:}} & \mathbf{T=3} & \mu_{f}^{Sys}(\theta,\bar{\sigma}%
^{2})=\left(  1-\theta\right)  \left(
\begin{array}
[c]{c}%
0\\
\sigma_{2}^{2}%
\end{array}
\right)  ,\text{ }A_{f}^{Sys}(\theta)=\left(
\begin{array}
[c]{cc}%
-\theta & 1\\
1-\theta & 0
\end{array}
\right)  ,\text{ }\\
&  & A_{f}^{Sys}(\theta)_{\perp}\text{ does not exist.}%
\end{array}
\]%
\begin{equation}%
\begin{array}
[c]{cll}%
\text{\textbf{Sys:}} & \mathbf{T=4} & \mu_{f}^{Sys}(\theta,\bar{\sigma}%
^{2})=\left(  1-\theta\right)  \left(
\begin{array}
[c]{c}%
0\\
0\\
0\\
\sigma_{2}^{2}\\
\sigma_{3}^{2}%
\end{array}
\right)  ,\text{ }A_{f}^{Sys}(\theta)=\left(
\begin{array}
[c]{ccc}%
-\theta & 1 & 0\\
0 & -\theta & 1\\
0 & -\theta & 1\\
1-\theta & 0 & 0\\
0 & 1-\theta & 0
\end{array}
\right)  ,\text{ }\\
&  & A_{f}^{Sys}(\theta)_{\perp}=\left(
\begin{array}
[c]{cc}%
\theta-1 & 0\\
0 & -1\\
0 & 1\\
-\theta & 0\\
1 & 0
\end{array}
\right)  .
\end{array}
\label{assyst=4}%
\end{equation}

When $T=3,$ $A_{f}^{Sys}(\theta)$ is a full rank square matrix so its
orthogonal complement does not exist. It implies that the Sys moment
conditions do not identify $\theta$ when $T=3.$ When $T=4$, the orthogonal
complement of $A_{f}^{j}(\theta),$ $A_{f}^{j}(\theta)_{\perp},$ has rank
larger than zero for both AS and Sys moments. Furthermore, the specification
of $\mu_{f}^{j}(\theta,\bar{\sigma}^{2})$ for the AS and Sys moment conditions
in (\ref{assyst=4}) is such that $A_{f}^{j}(\theta)_{\perp}^{\prime}\mu
_{f}^{j}(\theta,\bar{\sigma}^{2})\neq0$ for all $\theta\neq\theta_{0,N}$,
while it is not difficult to see that $\lim_{N\rightarrow\infty}A_{f}%
^{j}(\theta_{0,N})_{\perp}^{\prime}\mu_{f}^{j}(\theta_{0,N},\bar{\sigma}%
^{2})=0$ which just reflects that the moment conditions hold at the true
value. This implies that although the AS and Sys sample moments diverge in the
direction of $A_{f}^{j}(\theta),$ so that part cannot be used to identify
$\theta,$ the AS and Sys sample moments identify $\theta$ by their part which
is spanned by the orthogonal complement of $A_{f}^{j}(\theta).$ The
expressions of $\mu_{f}^{j}(\theta,\bar{\sigma}^{2})$ and $A_{f}^{j}(\theta)$
in the proof of Theorem 1 in the Appendix show that this argument extends to
all values of $T$ larger than three.

Our preceding analysis is summarized by Corollary 1:

\paragraph{Corollary 1 (Identification of $\theta$).}

\textit{Under Assumptions 1 and 2a}, $\theta$ \textit{is identified by the AS
and Sys moment conditions when }$T$ \textit{exceeds three. Furthermore,
}$\theta$\textit{\ is not identified by the Dif, Lev and NL moment conditions
separately\ for any value of T and the Sys moment conditions when T equals
three.}\medskip

Corollary 1 proves stylized facts 1-4 from Section 3, which are illustrated by
Panels 1-2. It also shows that the identification from the Lev moment
condition remains problematic for larger values of $T$ but the Sys and AS
moment conditions generally identify $\theta$ for values of $T$ larger than three.

Regarding the NL moments we find that they are not robust to all settings of
nuisance parameters like the variance of the initial observations. Alvarez and
Arellano (2004)\nocite{alar04} and Kruiniger (2013) have shown that, when the
data, including the initial observation, have finite second moments and the
autoregressive parameter equals one, $\theta$\ is identified by the NL and,
hence, the AS moment conditions if and only if $T\geq4.$\ Furthermore, if
$T\geq4$, $\theta$\ is only locally identified when the unconditional
variances of the errors change at a constant rate of growth between $t=2$\ and
$t=T-1$\ and only second-order but globally identified when the unconditional
variances between $t=2$\ and $t=T-1$\ are equal. Unlike Alvarez and Arellano
(2004) and Kruiniger (2013), our limiting sequence for the variance of the
initial observations allows for unbounded values. Theorem 1 then shows that
identification by the NL moment conditions is lost when its convergence rate
accords with (\ref{convinfzero}). The intuition is that the NL moment
conditions are a product of levels and first differences so they are unlikely
to identify the parameters in limit sequences where the variance of the
initial observations increases faster than the sample size.

Theorem 1 can be used to construct the non-standard limiting behavior of one
and two step GMM estimators that result from the different moment conditions.
These are similar to the non-standard results in \textit{e.g.} Madsen
(2003)\nocite{mad03} and Kruiniger (2009)\nocite{krui09} so we, for reasons of
brevity, refrain from stating them.

\paragraph{Robust sample moments}

Theorem 1 shows that the identification of $\theta$ when the variance of the
initial observations is large results from the part of the (AS or Sys) moment
conditions that lies in the direction of $A_{f}^{j}(\theta)_{\perp}$.
Expressions of the orthogonal complements of $A_{f}^{j}(\theta)$ for $T=4$ and
5 for the AS and Sys moment conditions are stated in (\ref{assyst=4}). They
can be specified (see the Appendix) as%
\begin{equation}
A_{f}^{j}(\theta)_{\perp}=(G_{f,T}^{j}(\theta)\text{ }\vdots\text{ }%
G_{2,T}^{j}), \label{OCdec}%
\end{equation}
where $T$\ indicates the number of time periods and $G_{2,T}^{j}$\ is such
that $G_{2,T}^{j\prime}\mu_{f}^{j}(\theta,\bar{\sigma}^{2})=0$ for all
$\theta.$\ Furthermore, $G_{f,T}^{j}(\theta)$\ is the only part of $A_{f}%
^{j}(\theta)_{\perp}$\ that depends on $\theta$. The orthogonal complements
are then such that the resulting, what we refer to as, robust moment
conditions are quadratic in $\theta:$%
\begin{equation}
g_{f,T}^{j}(\theta)=A_{f}(\theta)_{\perp}^{j\prime}f_{N}^{j}(\theta
)=a\theta^{2}+b\theta+d, \label{quadspec}%
\end{equation}
where the expressions for $a,$ $b$ and $d$ are constructed in the Appendix:

\begin{description}
\item[\textbf{T=4:}] 

\begin{description}
\item[\textbf{Sys}] $a=\frac{1}{N}\sum_{i=1}^{N}\binom{(\Delta y_{i2})^{2}}%
{0},$ $b=-\frac{1}{N}\sum_{i=1}^{N}\binom{(y_{i3}-y_{i1})^{2}}{\Delta
y_{i2}\Delta y_{i3}},$ $d=\frac{1}{N}\sum_{i=1}^{N}\binom{(y_{i4}%
-y_{i1})\Delta y_{i3}}{\Delta y_{i2}\Delta y_{i4}}.$

\item[\textbf{AS}] $a=\frac{1}{N}\sum_{i=1}^{N}\binom{(y_{i3}-y_{i1})\Delta
y_{i2}}{0},$ $b=-\frac{1}{N}\sum_{i=1}^{N}\binom{(y_{i3}-y_{i1})\Delta
y_{i3}+(y_{i4}-y_{i1})\Delta y_{i2}}{\Delta y_{i2}\Delta y_{i3}},\allowbreak$

$d=\frac{1}{N}\sum_{i=1}^{N}\binom{(y_{i4}-y_{i1})\Delta y_{i3}}{\Delta
y_{i2}\Delta y_{i4}}.$
\end{description}

\item[\textbf{T=5:}] 

\begin{description}
\item[\textbf{Sys}] $a=\frac{1}{N}\sum_{i=1}^{N}\left(
\begin{array}
[c]{c}%
(\Delta y_{i2})^{2}\\
(y_{i3}-y_{i1})\Delta y_{i3}\\
(\Delta y_{i3})^{2}\\
0\\
0
\end{array}
\right)  ,$ $b=-\frac{1}{N}\sum_{i=1}^{N}\left(
\begin{array}
[c]{c}%
(y_{i3}-y_{i1})^{2}\\
(y_{i4}-y_{i1})(y_{i4}-y_{i2})\\
(y_{i4}-y_{i2})^{2}\\
\Delta y_{i2}\Delta y_{i4}\\
\Delta y_{i3}\Delta y_{i4}%
\end{array}
\right)  ,$

$d=\frac{1}{N}\sum_{i=1}^{N}\left(
\begin{array}
[c]{c}%
(y_{i4}-y_{i1})\Delta y_{i3}\\
(y_{i5}-y_{i1})\Delta y_{i4}\\
(y_{i5}-y_{i2})\Delta y_{i4}\\
\Delta y_{i2}\Delta y_{i5}\\
\Delta y_{i3}\Delta y_{i5}%
\end{array}
\right)  .$

\item[\textbf{AS}] $a=\frac{1}{N}\sum_{i=1}^{N}\left(
\begin{array}
[c]{c}%
(y_{i3}-y_{i1})\Delta y_{i2}\\
(y_{i4}-y_{i1})\Delta y_{i3}\\
(y_{i4}-y_{i2})\Delta y_{i3}\\
0\\
0
\end{array}
\right)  ,$ $b=-\frac{1}{N}\sum_{i=1}^{N}\left(
\begin{array}
[c]{c}%
(y_{i4}-y_{i1})\Delta y_{i2}+(y_{i3}-y_{i1})\Delta y_{i3}\\
(y_{i4}-y_{i1})\Delta y_{i4}+(y_{i5}-y_{i1})\Delta y_{i3}\\
(y_{i4}-y_{i2})\Delta y_{i4}+(y_{i5}-y_{i2})\Delta y_{i3}\\
\Delta y_{i2}\Delta y_{i4}\\
\Delta y_{i3}\Delta y_{i4}%
\end{array}
\right)  ,$

$d=\frac{1}{N}\sum_{i=1}^{N}\left(
\begin{array}
[c]{c}%
(y_{i4}-y_{i1})\Delta y_{i3}\\
(y_{i5}-y_{i1})\Delta y_{i4}\\
(y_{i5}-y_{i2})\Delta y_{i4}\\
\Delta y_{i2}\Delta y_{i5}\\
\Delta y_{i3}\Delta y_{i5}%
\end{array}
\right)  ,$\newline
\end{description}
\end{description}

\noindent and similar specifications of $a,$ $b$ and $d$ result for larger
values of $T.$

It is interesting to see that these robust moments only depend on differences
of the data so the initial observations get differenced out. This explains why
these moments are robust to the variance of the initial observations. When the
autoregressive parameter equals one and in the case of iid normal errors\ and
time series homoskedasticity, Ahn and Thomas (2006) and Kruiniger (2013) show
that the maximum likelihood estimator of Hsiao et al.
(2002)\nocite{hsiaopestah02} and the random effects estimator of Anderson and
Hsiao (1982)\nocite{ah82} have the same limiting distributions. These results
show that, similar to our findings, moment conditions involving levels of the
data are redundant in this setting and only moment conditions using
differences of the data, like our robust moment conditions, are informative.

\paragraph{Large individual effect variance}

So far we have focused on highly persistent panel data resulting from a large
autoregressive parameter. However, the representation theorem for the moment
conditions and their derivatives in Theorem 1 applies to any setting where the
variance of the initial observations gets large. The expression of the initial
observation in (\ref{init1}) shows that its variance becomes large when either
the variance of the initial disturbance term, $u_{i1},$\ or the individual
specific effect, $\mu_{i},$\ becomes large. Theorem 1 focusses on a large
variance that results from the autoregressive parameter converging to one.
Theorem 1 does, however, extend to the case where jointly with the sample
size, the individual specific effect variance becomes large in such a manner
that Assumption 2a holds. This drifting sequence applies to any value of the
autoregressive parameter so the resulting identification issues are then no
longer confined to the unit root value. Hence, they also apply to the cases
with only moderate autoregressive dynamics, but a large variance of the
unobserved heterogeneity. The robust moments in (\ref{quadspec}) also apply to
this case. Kruiniger (2002) extensively analyzes the setting of a large
variance of the individual specific effects. He shows that only moment
conditions based on differences of the data yield a consistent estimator so
moment conditions involving levels are redundant. He also constructs the set
of optimal moment conditions assuming time series homoskedasticity. Our robust
moments (\ref{quadspec}) extend his set of optimal moment conditions since
they remain valid under a large variance of the individual specific effect and
also allow for time series heteroskedasticity.

\section{KLM test and robust sample moments}

Theorem 1 establishes identification results for the AS and Sys moment
conditions, which are based on the robust sample moments. It is not clear,
however, how an identification robust test procedure makes use of it. In this
section, we show that the KLM test based on the original AS or Sys moment
conditions just uses the robust sample moments when only the latter contain
identifying information on the autoregressive parameter. We show that, under
large variances of the initial observation and when the true value of $\theta$
is close to one, the KLM test based on either the AS or Sys moment conditions
exploits the identifying information from the robust moment conditions in an
optimal manner. For practical purposes, this implies that we do not have to
explicitly use the robust sample moments since they are implicitly used when
conducting a KLM\ test using AS or Sys moment conditions.

We obtain the above result in four steps. First, we characterize the limit
behavior of the robust sample moments. Second, we use it to determine
asymptotic sequences for the true and hypothesized values so the power
properties of the corresponding identification robust test statistics when
using the robust moments are not trivial and stay informative. Third, we
construct the largest (infeasible) discriminatory power that can be obtained
from combining the robust moments. Fourth, we show that it coincides with the
rejection frequency of KLM\ tests using either AS or Sys moment conditions.
Summarizing, the KLM test based on original AS\ or Sys moment conditions
implicitly resorts to using the robust sample moments in an optimal manner
when only these contain information on $\theta.$

\subsection{Large sample behavior of robust sample moments}

To construct the limiting behavior of the robust sample moments for settings
where only they contain information on $\theta,$ we first state the
probability limits of the quantities $a,$ $b$ and $d$ in (\ref{quadspec})
under Assumption 1. The components that comprise the robust sample moments do
not depend on the variance of the initial observations so they are not
affected by Assumption 2. Since we analyze the behavior when the true value
$\theta_{0,N}$ is converging to one, we specify this convergence behavior of
$\theta_{0,N}$ so it is dominated by the random components present in the
limit behavior of $a,$ $b$ and $d$ which are of order $O_{p}(N^{-\frac{1}{2}%
}).$ This then implies that $\theta_{0,N}$ converges rather rapidly to one
with a convergence rate that is faster than $N^{-\frac{1}{2}}.$ Hence,
$\theta_{0,N}$ is considered to be in the close neighborhood of one.

\paragraph{Theorem 2.}

\textit{Under Assumption 1,} \textit{the limit behavior of the different
components of }$g_{f,T}^{j}(\theta)$\textit{, }$j=AS,$ $Sys,$ \textit{for
}$\theta_{0,N}=1+\frac{l}{N^{\tau}}$\textit{\ with l a fixed constant, }$l<0,
$\textit{\ and }$\tau>\frac{1}{2},$ \textit{is characterized by:}

\paragraph{T=4:}

$a=\binom{\sigma_{2}^{2}}{0}+O_{p}(N^{-\frac{1}{2}}),$ $b=-\binom{\sigma
_{2}^{2}+\sigma_{3}^{2}}{0}+O_{p}(N^{-\frac{1}{2}}),$ $d=\binom{\sigma_{3}%
^{2}}{0}+O_{p}(N^{-\frac{1}{2}}).$

\paragraph{T=5:}

$%
\begin{array}
[c]{c}%
a=\left(
\begin{array}
[c]{c}%
\sigma_{2}^{2}\\
\sigma_{3}^{2}\\
\sigma_{3}^{2}\\
0\\
0
\end{array}
\right)  +O_{p}(N^{-\frac{1}{2}}),\text{ }b=\left(
\begin{array}
[c]{c}%
\sigma_{2}^{2}+\sigma_{3}^{2}\\
\sigma_{3}^{2}+\sigma_{4}^{2}\\
\sigma_{3}^{2}+\sigma_{4}^{2}\\
0\\
0
\end{array}
\right)  +O_{p}(N^{-\frac{1}{2}}),\text{ }d=\left(
\begin{array}
[c]{c}%
\sigma_{3}^{2}\\
\sigma_{4}^{2}\\
\sigma_{4}^{2}\\
0\\
0
\end{array}
\right)  +O_{p}(N^{-\frac{1}{2}}).
\end{array}
$\medskip

\begin{proof}
see the Appendix.\medskip
\end{proof}

Although AS and Sys robust moments are different, Theorem 2 implies that under
Assumption 1 the probability limits of $a$, $b$ and $d$ are identical.
Furthermore, Theorem 2 implies that the Jacobian of the robust moment equation
(\ref{quadspec}) is of full column rank when $\sigma_{t}^{2}\neq\sigma^{2}$
for at least one value of $t=2,...,T$. This fulfills one of the sufficient
conditions for standard asymptotic theory for GMM\ inference based on the
robust sample moments, which since the other sufficient conditions can be
shown to hold as well, applies for these settings.

\subsection{Asymptotic sequence for the hypothesized value}

We want to compare tests of H$_{0}:\theta=\theta^{\ast}$ using the robust
sample moments to KLM\ tests of H$_{0}$ using the original AS and Sys moments
for settings where the identification can be problematic, which occurred for
true values of $\theta$ close to one and large variances of the initial
observations. Because we want to analyze local asymptotic power while the true
value $\theta_{0,N}$ is converging to one according to $\theta_{0,N}%
=1+\frac{l}{N^{\tau}}$, we also consider a local to unity drifting sequence
for the hypothesized value $\theta^{\ast},$ which we denote by $\theta(e)$
with $e<0$ the localizing parameter. Although less common in asymptotic power
analysis, the advantage of a drifting hypothesized value is that our results
hold for a range of hypothesized values.

The asymptotic sequence $\theta(e)$ is such that the behavior of the
identification robust tests is not diverging and informative about $\theta,$
when the true value $\theta_{0,N}$ is converging to one. Theorem 3 establishes
the particular rate at which $\theta(e)$ converges to one which makes these
conditions hold. Note that there is a slight abuse of notation as from now on
we suppress the superscript $j$ in $g_{f,T}^{j}(\theta(e))$, $j=AS,$ $Sys,$
which is inconsequential for the results to follow.

\paragraph{Theorem 3.}

\textit{Under Assumption 1, }$\theta_{0,N}=1+\frac{l}{N^{\tau}}$\textit{\ with
l a fixed constant, }$l<0,$\textit{\ and }$\tau>\frac{1}{2},$ \textit{the
robust moments }$\sqrt{N}g_{f,T}(\theta(e))$\textit{\ are informative about
}$\theta$ \textit{and converge to a bounded in probability, non-degenerate
random variable under the following local to unity drifting sequence }%
$\theta(e)$\textit{:}

\noindent\textbf{1. }$\theta(e)=1+\frac{e}{\sqrt[4]{N}}$\textit{\ in the case
of }$\sigma_{t}^{2}=\sigma^{2},$\textit{\ }$t=2,\ldots T,$

\noindent\textbf{2.} $\theta(e)=1+\frac{e}{\sqrt{N}}$ \textit{when }%
$\sigma_{t}^{2}\neq\sigma^{2},$\textit{\ for at least one value of} $t,$
$t=2,\ldots T-1,$

\noindent\textit{with }$e<0$ \textit{a finite constant.}\medskip

\begin{proof}
see the Appendix.\medskip
\end{proof}

The quartic root convergence rate in Theorem 3.1 results since the Jacobian of
the robust moment equation (\ref{quadspec}) is then equal to zero but the
Hessian is not. It is thus a setting of so-called second order identification
with first order underidentification. Estimators then generally have quartic
root convergence rates, see $e.g.$ Dovonon and Renault (2013),\nocite{dr13}
Dovonon and Hall (2018)\nocite{dh16} and Dovonon $et$ $al.$
(2020).\nocite{dhk17} A quartic root convergence rate for estimators in
dynamic panel data models is also found by Ahn and Thomas (2006) and Kruiniger (2013).

The quartic root convergence rate for the robust sample moments results from
specifying $\theta(e)=1+\frac{e}{N^{1/4}}$ and $\sigma_{t}^{2}=\sigma^{2},
$\textit{\ }$t=2,\ldots T.$ All elements of the robust sample moments which
are linear in $e$ then cancel out in the limit. We are then left with a
quadratic term in $e$ and components that converge at the rate $\frac{1}%
{\sqrt{N}}$. A quartic root convergence rate makes all these components of the
same order of magnitude. Theorem 3 shows that error variances which are
constant over time, $\sigma_{t}^{2}=\sigma^{2},$\textit{\ }$t=2,\ldots T,$
lead to this slow convergence rate.

\subsection{Largest rejection frequencies of robust sample moments}

To show that the KLM test of H$_{0}$ using AS\ and Sys moment conditions just
uses the robust sample moments when only these contain information on
$\theta,$ we use the largest rejection frequencies that result in such
instances from the robust sample moments. To obtain these largest rejection
frequencies, we first consider the GMM-AR test of H$_{p}:\theta(e)=1+\frac
{e}{\sqrt[4]{N}}$ using the robust sample moments, which is specified as:
\begin{equation}%
\begin{array}
[c]{c}%
\text{GMM-AR(}\theta(e))=Ng_{f,T}(\theta(e))^{\prime}\hat{V}_{gg}%
(\theta(e))^{-1}g_{f,T}(\theta(e)),
\end{array}
\label{gmmare}%
\end{equation}
with $g_{f,T}(\theta(e))$\ the moments in (\ref{quadspec}) evaluated at
$\theta(e)=1+\frac{e}{\sqrt[4]{N}}$\ and $\hat{V}_{gg}(\theta(e))$\ the
(Eicker-White) covariance matrix estimator of the covariance matrix of
$g_{f,T}(\theta(e)).$ For $T=4$ and 5:\footnote{We thank an anonymous referee
for showing this.}%
\[%
\begin{array}
[c]{ccl}%
\mathbf{T=4:} & g_{f,T=4}^{AS}(\theta(e))= & \left(
\begin{array}
[c]{lc}%
1 & -\theta(e)\\
0 & 1
\end{array}
\right)  g_{f,T=4}^{Sys}(\theta(e))\\
\mathbf{T=5:} & g_{f,T=5}^{AS}(\theta(e))= & \left(
\begin{array}
[c]{ccccc}%
1 & -\theta(e)/(1-\theta(e)) & \theta(e)/(1-\theta(e)) & 0 & 0\\
0 & 1 & 0 & 0 & -\theta(e)\\
0 & 0 & 1 & 0 & -\theta(e)\\
0 & 0 & 0 & 1 & 0\\
0 & 0 & 0 & 0 & 1
\end{array}
\right)  g_{f,T=5}^{Sys}(\theta(e))
\end{array}
\]
so GMM-AR($\theta(e))$ is equivalent for the AS and Sys moment conditions
since the invertible matrix by which $g_{f,T}^{Sys}(\theta(e))$ has to be
pre-multiplied to obtain $g_{f,T}^{AS}(\theta(e))$ cancels out in
GMM-AR($\theta(e)).$ This result can be extended to larger values of $T.$

\paragraph{\textbf{Theorem 4.}}

\textit{Under Assumption 1, }$\theta_{0,N}=1+\frac{l}{N^{\tau}}$\textit{\ with
l a fixed constant, }$l<0,$\textit{\ and }$\tau>\frac{1}{2},$ $\sigma_{t}%
^{2}=\sigma^{2},$\textit{\ }$t=2,\ldots T,$ \textit{the large sample
distribution of the GMM-AR statistic (\ref{gmmare}) for testing H}$_{p}%
:\theta(e)=1+\frac{e}{\sqrt[4]{N}},$\textit{\ in a sample of size }$N$
\textit{is characterized by }%
\begin{equation}%
\begin{array}
[c]{c}%
\chi^{2}(\delta(N),p_{\max}),
\end{array}
\label{limgmmare}%
\end{equation}
\textit{with }$\delta(N)=(e\sigma)^{4}\binom{\iota_{p}}{0}^{\prime
}(B(N)^{\prime}V_{abd}B(N))^{-1}\binom{\iota_{p}}{0},$\textit{\ }%
$p$\textit{\ the number of columns G}$_{f,T}(\theta),$\textit{\ so when
}$T=4,$\textit{\ }$p=1$\textit{\ and when }$T=5,$\textit{\ }$p=3,$%
\textit{\ and }$p_{\max}$\textit{\ the number of elements of }$g_{f,T}%
(\theta(e))$\textit{, so, when }$T=4,$\textit{\ }$p_{\max}=2,$\textit{\ while
}$p_{\max}=5$\textit{\ for }$T=5,$
\begin{equation}%
\begin{array}
[c]{c}%
B(N)=(\iota_{3}\otimes I_{p_{\max}})+\frac{e}{\sqrt[4]{N}}\left[  (2+\frac
{e}{\sqrt[4]{N}})(e_{1,3}\otimes I_{p_{\max}})+(e_{2,3}\otimes I_{p_{\max}%
})\right]  ,
\end{array}
\label{bnspec}%
\end{equation}
$V_{abd}$\textit{\ the covariance matrix of }$a,$\textit{\ }$b$\textit{\ and
}$d$\textit{, }$I_{p_{\max}}$\textit{\ the }$p_{\max}\times p_{\max}%
$\textit{\ dimensional identity matrix, }$e_{1,3}$\textit{\ and }$e_{2,3}%
$\textit{\ the first and second }$3\times1$\textit{\ dimensional unity vectors
and }$\chi^{2}(\delta,p_{\max})$\textit{\ a non-central }$\chi^{2}%
$\textit{\ distribution with non-centrality parameter }$\delta$\textit{\ and
}$p_{\max}$ \textit{degrees of freedom.}$\medskip\medskip$

\begin{proof}
see the Appendix.$\medskip$
\end{proof}

The expression of the large sample distribution in Theorem 4 depends on the
sample size. Given the quartic root convergence rate, convergence to the
limiting distribution is very slow so it is important for the accuracy of the
approximation of the finite sample distribution to incorporate higher order
components. The proof of Theorem 4 in the Appendix therefore from the outset
considers all higher order components of $g_{f,T}(\theta(e))$ in order to
construct a large sample approximation of the distribution of GMM-AR($\theta
(e)).$

To obtain the maximal rejection frequencies using the robust sample moments,
we use a (infeasible) weighted average of the moment equations in
$g_{f,T}(\theta(e))$\ where the weights are chosen such that the
non-centrality parameter equals the one of the non-central $\chi^{2}%
$\ limiting distribution of the GMM-AR statistic while the degrees of freedom
is equal to one ($i.e.$ the number of elements of $\theta$). This value of the
non-centrality parameter is also the maximal one that can be obtained using a
weighted average of the robust sample moments.

\paragraph{Theorem 5.}

\textit{Under Assumption 1, }$\theta_{0,N}=1+\frac{l}{N^{\tau}}$\textit{\ with
l a fixed constant, }$l<0,$\textit{\ and }$\tau>\frac{1}{2},$ $\sigma_{t}%
^{2}=\sigma^{2},$\textit{\ }$t=2,\ldots T,$\textit{\ an optimal (infeasible)
GMM-AR test of H}$_{p}:\theta(e)=1+\frac{e}{\sqrt[4]{N}}$ \textit{that uses a
weighted average of the robust sample moments can be constructed that has
approximately a }%
\begin{equation}%
\begin{array}
[c]{c}%
\chi^{2}(\delta(N),1),
\end{array}
\label{noncentral}%
\end{equation}
\textit{distribution in large samples of size }$N$\textit{. }$\medskip$

\begin{proof}
see the Appendix.$\medskip$
\end{proof}

The GMM-AR statistics in Theorems 4 and 5 both have non-central $\chi^{2}$
distributions with the same non-centrality parameter so the one with the
smallest number of degrees of freedom, i.e. the statistic in Theorem 5, has
the largest power.

Figure 4 illustrates Theorem 5 and shows the maximal rejection frequencies
based on combining the robust sample moments based on either AS or Sys moment
condition in a GMM-AR test\footnote{We use the covariance matrix estimator for
each simulated data set to compute the GMM-AR statistics.} for $T=4$\ and $5.$
It uses DGP 1 from Section 3 with a true value of $\theta$ which is very close
to one (0.99) and a large value of $\sigma_{c}^{2}$ (ten) compared to
$\sigma^{2}$ (one), which amplifies the variance of the initial conditions.
The DGP\ thus satisfies mean stationarity (\ref{init1})-(\ref{init2}) and also
time series homoskedasticity, $i.e.$ $\sigma_{t}^{2}=\sigma^{2}$ for
$t=2,\ldots,T$. We use $N=2000$, a relatively large value and test for a wide
range of values for $\theta$, which together with $N$ provides a mapping to
the constant $e$ $(=\sqrt[4]{N}(\theta-1))$ in Figure 4 (horizontal axis). The
usual power curve, as shown earlier in the Figures in Panels 1 and 2, reports
the rejection frequencies of tests of the hypothesized parameter value as a
function of the parameter value used in the DGP\ where the data is simulated
from. Figure 4, however, reports for a fixed parameter value equal to one in
the DGP used to simulate the data, the rejection frequencies as a function of
a varying localizing parameter $e$ and, hence, autoregressive parameter
$\theta(e),$ under the tested null hypothesis. The rejection frequencies in
Figure 4 thus report those observed at one for a range of the usual power
curves where the tested parameter values correspond with those on the
horizontal axes in Figure 4.

Because of the equivalence of the GMM-AR test for the AS and Sys robust
moments, the rejection frequencies are identical for the AS\ and Sys based
robust sample moments and only differ over $T.$ Any remaining differences in
Figure 4 are due to sampling noise.\bigskip

\begin{center}
$%
\begin{array}
[c]{c}%
\text{Figure 4. Rejection frequencies of GMM-AR tests of H}_{p}:\theta
(e)=1+\frac{e}{\sqrt[4]{N}}\\
\text{using weighted robust sample moments}\\%
\raisebox{-0pt}{\includegraphics[
height=2.2139in,
width=2.9386in
]%
{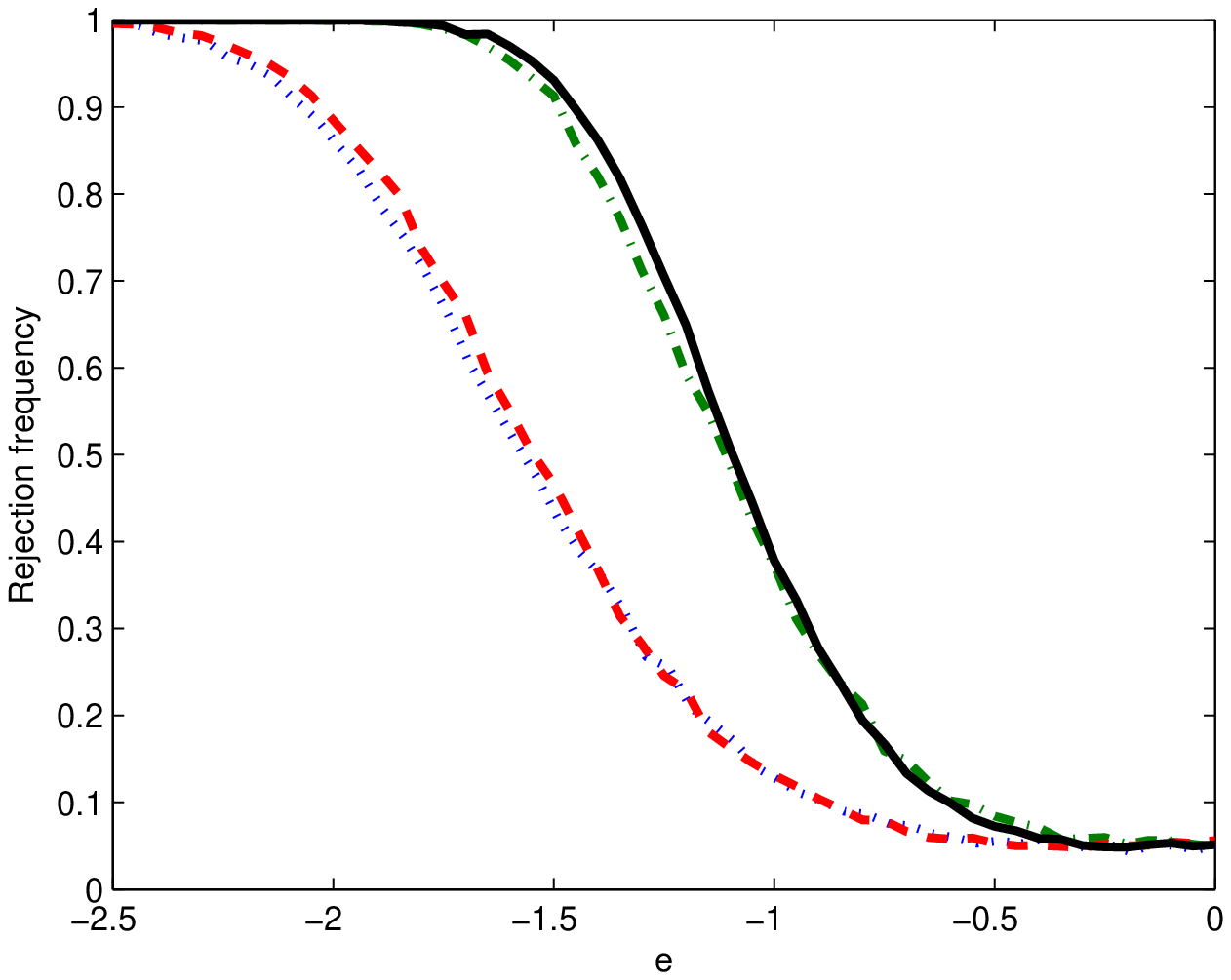}%
}
\\
\multicolumn{1}{l}{\text{Note: 5\% significance level, true value of }%
\theta\text{\ is 0.99, }N=2000\text{, Sys \& }T=4\text{ (dashed), AS \&}}\\
\multicolumn{1}{l}{\text{ }T=4\text{ (dotted), Sys \& }T=5\text{ (solid), AS
\& }T=5\text{ (dash-dotted).}}%
\end{array}
$\bigskip
\end{center}

\subsection{Large sample behavior of the KLM test}

Finally, we construct the large sample distribution of KLM\ tests of
H$_{p}:\theta(e)=1+\frac{e}{\sqrt[4]{N}}$ using AS and Sys moment conditions
when $\theta_{0,N}$ accords with the drifting sequences in Assumptions 1 and
2a so only the robust sample moments contain information on $\theta.$

\paragraph{Theorem 6.}

\textit{Under Assumptions 1 and 2a, }$\theta_{0,N}=1+\frac{l}{N^{\tau}}%
$\textit{\ with l a fixed constant, }$l<0,$\textit{\ and }$\tau>\frac{1}{2},$
$\sigma_{t}^{2}=\sigma^{2},\mathit{\ }t=2,\ldots T,$ \textit{the large sample
distribution of the KLM statistic using the AS or Sys moments for}
\textit{testing the hypothesis H}$_{p}:\theta(e)=1+\frac{e}{\sqrt[4]{N}}%
$\textit{\ is characterized by}
\begin{equation}%
\begin{array}
[c]{rl}%
\text{KLM(}\theta(e))\sim & \chi^{2}(\delta(N),1)
\end{array}
\label{worstcasedis}%
\end{equation}
\textit{with }$\delta(N)$\textit{\ defined in Theorem 4}$.$

\begin{proof}
see the Appendix.\medskip
\end{proof}

Under Assumptions 1 and 2a, Theorem 1 implies that the GMM sample moments
diverge in one direction and converge in another one. Identical to tests for
cointegration, Theorem 6 shows that the diverging parts of the GMM sample
moments cancel out in the large sample distribution of the KLM\ test so it
only contains elements from the converging part of the GMM sample moments. The
proof of the large sample distribution of the KLM test is therefore rather
elaborate since this has to be shown for each of the different components of
the KLM\ test.

Theorem 6 shows that the large sample distribution of the KLM\ test using AS
or Sys moment conditions when only the robust sample moments contain
information on $\theta$ is identical to the limiting distribution of the
GMM-AR test that optimally combines the robust sample moments for these
settings. It proves that KLM\ tests using the AS and Sys moment conditions
then only use the robust sample moments. It is similar to what happens in
cointegration where since the cointegrating vector and stochastic trends
operate orthogonally, a likelihood ratio test on the cointegration vector also
does not depend on the stochastic trends, see e.g. Johansen (1991).

Theorem 6 is illustrated by the Figures in Panel 5, which show the rejection
frequencies of 5\% significance tests using a KLM test of H$_{p}%
:\theta(e)=1+\frac{e}{\sqrt[4]{N}}$ with AS and Sys moment conditions when $T$
equals four, Figure 5.1, and five, Figure 5.2, respectively. It uses the same
DGP as for Figure 4. Also identical to Figure 4, the rejection frequencies in
Panel 5 report the rejection frequencies when using a fixed parameter value in
the DGP\ where we simulate the data from, as a function of a varying parameter
value under the tested hypothesis.

Panel 5 shows, for both $T=4$ and $T=5$, that the rejection frequencies that
result from using the KLM test with either AS or Sys moment conditions are
equal to the largest rejection frequencies, that can be obtained with the
robust moments when only they contain information on $\theta$. It illustrates
that the robust sample moments are (implicitly) used when you conduct
KLM\ tests with AS or Sys moment conditions. Hence, in practice one can just
use AS or Sys moment conditions in the construction of the KLM test, i.e.
there is no need to switch to the robust sample moments.

Panel 5 also provides a visual proof of stylized fact 5 from\ Section 3, i.e.
rejection frequencies for the KLM test using AS or Sys moment conditions are
almost identical when the true value of $\theta$ is close to one and for large
variances of the initial observations, and that it is not specific for the
tested values used there but holds generally for different tested values of
$\theta.$%

\[%
\begin{array}
[c]{c}%
\text{Panel 5: Rejection frequencies of KLM tests of H}_{p}:\theta
(e)=1+\frac{e}{\sqrt[4]{N}}\text{ using AS (dashed) }\\
\text{and Sys (dash-dotted) and GMM-AR tests using (infeasible) optimal
weighted average }\\
\text{of robust sample moments (solid line)}\\
\\%
\begin{array}
[c]{ccc}%
\raisebox{-0pt}{\includegraphics[
height=2.2139in,
width=2.9386in
]%
{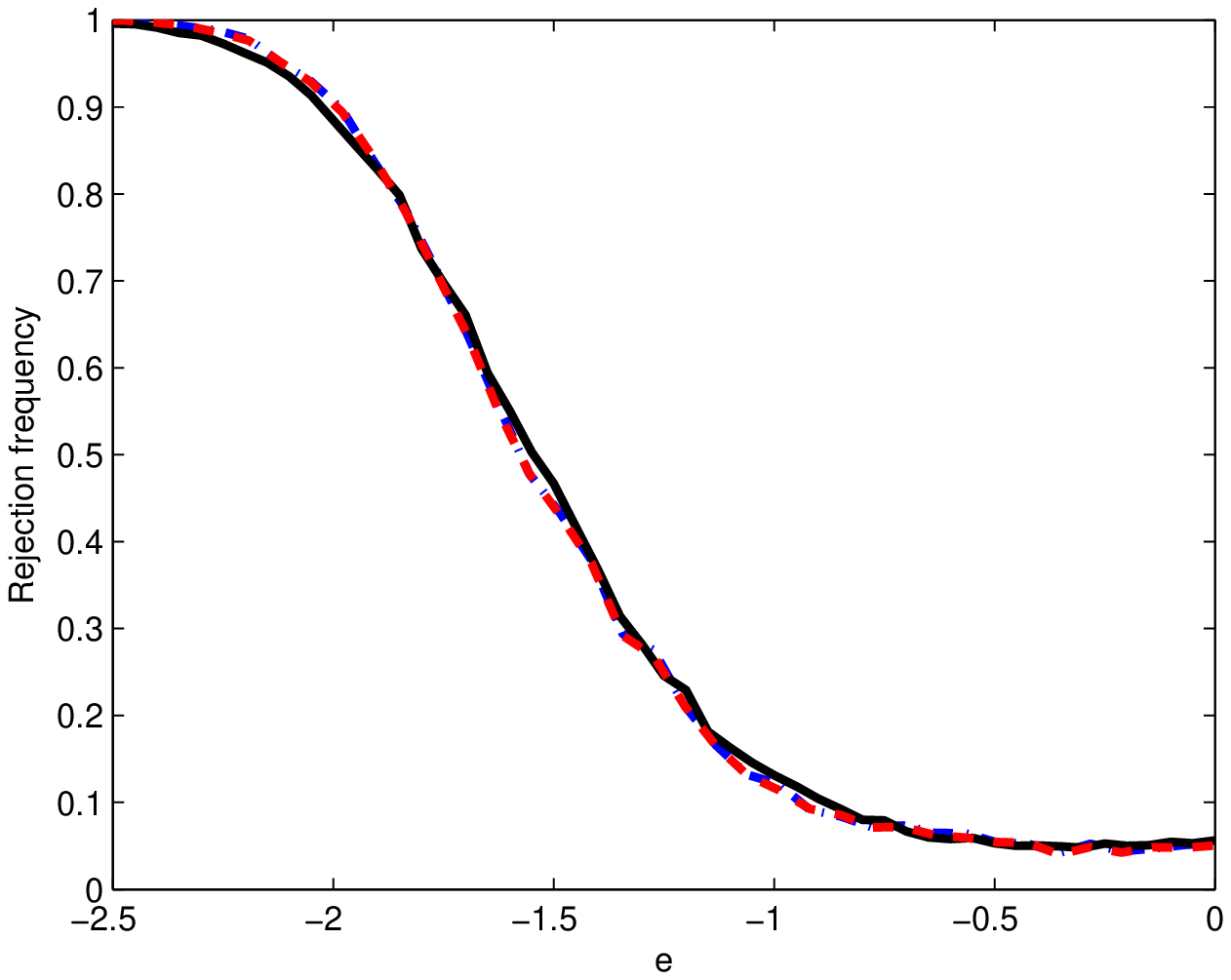}%
}
&  &
\raisebox{-0pt}{\includegraphics[
height=2.2139in,
width=2.9386in
]%
{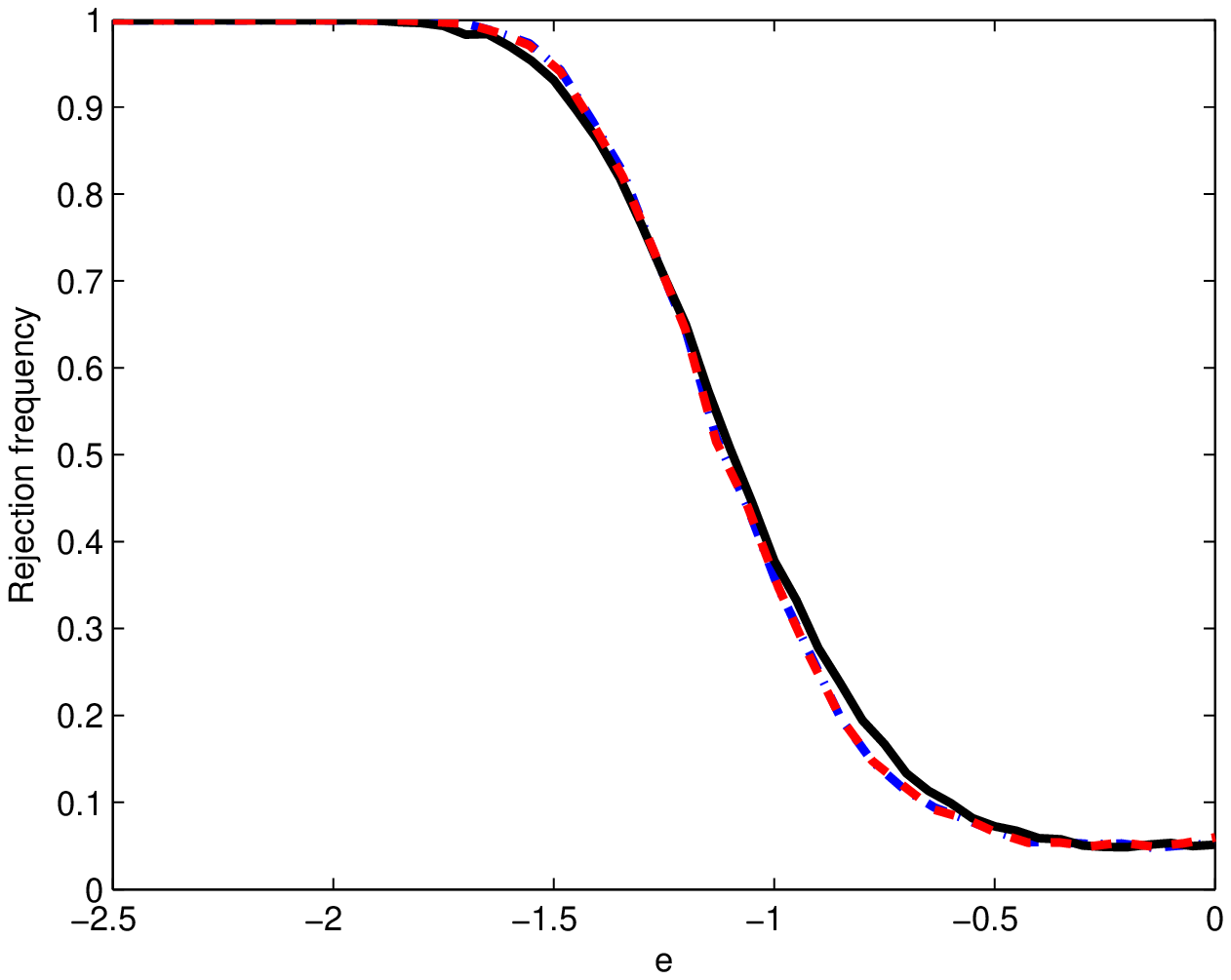}%
}
\\
\text{Figure 5.1: }T=4 &  & \text{Figure 5.2: }T=5
\end{array}
\\
\\
\multicolumn{1}{l}{\text{Note: 5\% significance level, true value of }%
\theta\text{\ is 0.99, N=2000.}}%
\end{array}
\]

\section{Conclusions}

We have analyzed GMM inference for dynamic panel data models involving highly
persistent panel data. We show that the Dif, Lev and NL moment conditions
separately do not identify the parameters in dynamic panel data models for a
general number of time periods. This results from the divergence of the
initial observations for some plausible data generating processes involving
highly persistent panel data. When there are more than three time periods, the
AS and Sys moment conditions, however, do lead to identification. The
identification based on the AS and Sys moment conditions for the problematic
cases of divergent initial observations results from so-called robust sample
moments. They are combinations of either the AS or Sys sample moments and do
not depend on the initial observations.

Despite the positive identification results for AS and Sys moment conditions,
conventional inference based on two step GMM estimators is not valid since
these estimators have non-standard limiting distributions near the unit root.
Similar results hold for two step GMM\ estimators based on our robust sample
moments. We have therefore analyzed the large sample properties of
identification robust GMM test procedures. These test statistics are size
correct, easy to implement and have been used in a variety of models analyzed
using GMM. We show that the identification robust KLM statistic based on the
AS and Sys sample moments implicitly resorts to using the robust sample
moments when only the latter contain identifying information.

Based on the theoretical analysis and numerical results a number of remarks
can be made regarding the implementation of GMM inference for applied linear
dynamic panel data analysis. First, statistical inference, i.e. hypothesis
testing and confidence intervals, should be based on identification robust
tests, like, for example, the KLM\ or GMM-AR test. The non-standard limiting
behavior of the two step GMM coefficient estimator makes the use of
conventional GMM inference hazardous in applied research when there are
identification issues. Second, one should always use either AS or Sys moment
conditions since these deliver identification under more general conditions
when $T>3.$ An advantage of the AS moments is that they are valid under less
restrictive assumptions than the Sys moments. Third, when mean stationarity
applies, the Sys moments are preferred. Although AS and Sys moments contain
the same amount of identifying information when $\theta$ is close to one and
the variance of the initial observations is large, in practice the opposite
may well be the case if one is not close to the unit root (or if time series
heteroskedasticity is present). This is shown, for example, by our simulated
KLM power curves in Section 2. Fourth, the original AS or Sys moments should
be used in an identification robust GMM test statistic and not the implied
robust sample moments. Although only the latter preserve identification when
the variance of the initial observations is large, we have shown that the
identification robust KLM test based on the AS or Sys moments implicitly uses
the robust sample moments.

Finally, for expository purposes we have only analyzed the first-order
autoregressive panel data model. The extension to panel data models with
multiple endogenous regressors, e.g. dynamic models with additional endogenous
regressors, is an important area for future research.\newpage

\noindent{\LARGE Appendix. Specification of GMM sample moments and
proofs}\bigskip

\paragraph{Specification of sample moment functions}

For the Dif moment conditions in (\ref{difmom}), $k_{Dif}$ equals $\frac{1}%
{2}(T-2)(T-1)$ while $f_{i}^{Dif}(\theta)$ and $q_{i}^{Dif}(\theta)$ read%
\[%
\begin{array}
[c]{rl}%
f_{i}^{Dif}(\theta)= & Z_{i}^{Dif}\varphi_{i}^{Dif}(\theta)\\
q_{i}^{Dif}(\theta)= & -Z_{i}^{Dif}\Delta y_{-1,i},
\end{array}
\]
with $\varphi_{i}^{Dif}(\theta)=(\Delta y_{i3}-\theta\Delta y_{i2}\ldots\Delta
y_{iT}-\theta\Delta y_{iT-1})^{\prime},$ $\Delta y_{-1,i}=(\Delta y_{i2}%
\ldots\Delta y_{iT-1})^{\prime}$ and%
\[%
\begin{array}
[c]{c}%
Z_{i}^{Dif}=\left(
\begin{array}
[c]{ccc}%
y_{i1} & 0\ldots0 & 0\\
0 & \ddots & 0\\
0 & 0\ldots0 & \left(
\begin{array}
[c]{c}%
y_{i1}\\
\vdots\\
y_{iT-2}%
\end{array}
\right)
\end{array}
\right)  :\frac{1}{2}(T-1)(T-2)\times(T-2).
\end{array}
\]
For the Lev moment conditions in (\ref{levmom}), $k_{Lev}$ equals $T-2$ while
the sample moment functions are%
\[%
\begin{array}
[c]{rl}%
f_{i}^{Lev}(\theta)= & Z_{i}^{Lev}\varphi_{i}^{Lev}(\theta)\\
q_{i}^{Lev}(\theta)= & -Z_{i}^{Lev}y_{-1,i},
\end{array}
\]
with $\varphi_{i}^{Lev}(\theta)=(y_{i3}-\theta y_{i2}\ldots y_{iT}-\theta
y_{iT-1})^{\prime},$ $y_{-1,i}=(y_{i2}\ldots y_{iT-1})^{\prime},$ and%
\[%
\begin{array}
[c]{c}%
Z_{i}^{Lev}=\left(
\begin{array}
[c]{ccc}%
\Delta y_{i2} & 0\ldots0 & 0\\
0 & \ddots & 0\\
0 & 0\ldots0 & \Delta y_{iT-1}%
\end{array}
\right)  :(T-2)\times(T-2).
\end{array}
\]
For the NL moment conditions in (\ref{ahnschmidt}), $k_{NL}$ equals $T-3$
while the sample moment functions can be specified as%
\[%
\begin{array}
[c]{rl}%
f_{i}^{NL}(\theta)= & Z_{i}^{NL}(\theta)\varphi_{i}^{NL}(\theta)\\
q_{i}^{NL}(\theta)= & \left(  \frac{\partial}{\partial\theta}Z_{i}^{NL}%
(\theta)\right)  \varphi_{i}^{NL}(\theta)+Z_{i}^{NL}(\theta)\left(
\frac{\partial}{\partial\theta}\varphi_{i}^{NL}(\theta)\right)  ,
\end{array}
\]
with $\varphi_{i}^{NL}(\theta)=((y_{i4}-\theta y_{i3})\ldots(y_{iT}-\theta
y_{iT-1}))^{\prime}$ and
\[%
\begin{array}
[c]{c}%
Z_{i}^{NL}(\theta)=\left(
\begin{array}
[c]{ccc}%
(\Delta y_{i3}-\theta\Delta y_{i2}) & 0\ldots0 & 0\\
0 & \ddots & 0\\
0 & 0\ldots0 & (\Delta y_{iT-1}-\theta\Delta y_{iT-2})
\end{array}
\right)  :(T-3)\times(T-3).
\end{array}
\]
The sample moments for the AS moment conditions result by just stacking the
appropriate sample moments stated above so $k_{AS}$ equals $\frac{1}%
{2}(T-1)(T-2)+T-3$. In a similar manner, the Sys sample moments result so
$k_{Sys}$ equals $\frac{1}{2}(T+1)(T-2)$.

\paragraph{Lemma 1.}

We state some intermediate results, which involve the different terms in the
sample moments and their derivatives. Assumption 1 implies the following:%
\[%
\begin{array}
[c]{rrll}%
\mathbf{i.} & \frac{1}{N}\sum_{i=1}^{N}(\theta_{0,N}-1)y_{i1}u_{i1}= &
-d_{2}-\frac{\sigma_{1,N}}{\sqrt{N}}\psi_{c}+o_{p}(1), & \\
\mathbf{ii.} & \frac{1}{N}\sum_{i=1}^{N}(1-\theta_{0,N})u_{i1}u_{it}%
\underset{p}{\rightarrow} & 0, & t>1,\\
\mathbf{iii.} & \frac{1}{N}\sum_{i=1}^{N}u_{it}^{2}\underset{p}{\rightarrow} &
\sigma_{t}^{2}, & t>1,\\
\mathbf{iv.} & \frac{1}{N}\sum_{i=1}^{N}\Delta y_{it}\Delta y_{it}\underset
{p}{\rightarrow} & \sigma_{t}^{2}, & t>1,\\
\mathbf{v.} & \frac{1}{N}\sum_{i=1}^{N}\Delta y_{it}\Delta y_{is}\underset
{p}{\rightarrow} & 0, & t,s>1,t\neq s.\\
\mathbf{vi.} & \frac{h_{N}(\theta_{0,N})}{\sqrt{N}}\sum_{i=1}^{N}\left(
\begin{array}
[c]{c}%
y_{i1}u_{i2}\\
\vdots\\
y_{i1}u_{iT}%
\end{array}
\right)  \underset{d}{\rightarrow} & \psi, &
\end{array}
\]
with $\psi=(\psi_{2}\ldots\psi_{T})^{\prime}\sim N(0,$diag($\sigma_{2}%
^{2},\ldots,\sigma_{T}^{2}))$ independent from $\psi_{c}\sim N(0,\sigma
_{c}^{2}),$ $\sigma_{c}^{2}=$var($c_{i}$)$.$

\paragraph{Proof of Lemma 1.}

\textbf{i. }Under mean stationarity, we have:
\[%
\begin{array}
[c]{c}%
\frac{1}{N}\sum_{i=1}^{N}(\theta_{0,N}-1)y_{i1}u_{i1}=\frac{1}{N}\sum
_{i=1}^{N}(\theta_{0,N}-1)u_{i1}^{2}+\frac{1}{N}\sum_{i=1}^{N}(\theta
_{0,N}-1)u_{i1}\mu_{i}.
\end{array}
\]
Assumption 1c implies that $(1-\theta_{0,N})^{\frac{1}{2}}u_{i1}$ is a random
variable with finite fourth moments so a law of large numbers applies:%
\[%
\begin{array}
[c]{c}%
\frac{1}{N}\sum_{i=1}^{N}(\theta_{0,N}-1)u_{i1}^{2}\underset{p}{\rightarrow
}-d_{2}.
\end{array}
\]
Since $c_{i}=(1-\theta_{0,N})\mu_{i},$ we can specify:
\[%
\begin{array}
[c]{c}%
\frac{1}{N}\sum_{i=1}^{N}(\theta_{0,N}-1)u_{i1}\mu_{i}=-\frac{1}{N}\sum
_{i=1}^{N}u_{i1}c_{i}=-\frac{\sigma_{1,N}}{\sqrt{N}}\frac{1}{\sqrt{N}}%
\sum_{i=1}^{N}\frac{u_{i1}}{\sigma_{1,N}}c_{i},
\end{array}
\]
because
\[%
\begin{array}
[c]{c}%
\frac{1}{\sqrt{N}}\sum_{i=1}^{N}\frac{u_{i1}}{\sigma_{1,N}}c_{i}\underset
{d}{\rightarrow}\psi_{c},
\end{array}
\]
with $\psi_{c}$ independent of $\psi_{j},$ $j=2,\ldots,T,$ as $c_{i}$ is
independent from $u_{ij},$ $j=2,,\ldots,T.$ Upon combining, we obtain:%
\[%
\begin{array}
[c]{c}%
\frac{1}{N}\sum_{i=1}^{N}(\theta_{0,N}-1)y_{i1}\mu_{i}=-d_{2}-\frac
{\sigma_{1,N}}{\sqrt{N}}\psi_{c}+o_{p}(1).
\end{array}
\]

\textbf{ii. }Since $u_{it}$ are independently distributed, $t=1,\ldots,T,$ and
$(1-\theta_{0,N})^{\frac{1}{2}}u_{i1}$ is a random variable with finite fourth
moments, a law of large numbers applies:%
\[%
\begin{array}
[c]{rll}%
\frac{1}{N}\sum_{i=1}^{N}(1-\theta_{0,N})u_{i1}u_{it}\underset{p}{\rightarrow}
& 0, & t>1.
\end{array}
\]

\textbf{iii. }Finite fourth moments of $u_{it}$ implies that a law of large
numbers applies:%
\[%
\begin{array}
[c]{rll}%
\frac{1}{N}\sum_{i=1}^{N}u_{it}^{2}\underset{p}{\rightarrow} & \sigma_{t}%
^{2}, & t>1.
\end{array}
\]

\textbf{iv. }Mean stationarity implies $\Delta y_{i2}=u_{i2}+\left(
\theta_{0,N}-1\right)  u_{i1},$ so%
\[%
\begin{array}
[c]{c}%
\frac{1}{N}\sum_{i=1}^{N}\Delta y_{i2}\Delta y_{i2}=\frac{1}{N}\sum_{i=1}%
^{N}u_{i2}^{2}+\left(  \theta_{0,N}-1\right)  \frac{1}{N}\sum_{i=1}^{N}\left(
\theta_{0,N}-1\right)  u_{i1}^{2}+\frac{2}{N}\sum_{i=1}^{N}\left(
\theta_{0,N}-1\right)  u_{i1}u_{i2}.
\end{array}
\]
Because $\frac{1}{N}\sum_{i=1}^{N}(1-\theta_{0,N})u_{i1}^{2}\underset
{p}{\rightarrow}d_{2}$\ and $(1-\theta_{0,N})\underset{N\rightarrow\infty
}{\rightarrow}0$, we have%
\[%
\begin{array}
[c]{c}%
\left(  \theta_{0,N}-1\right)  \frac{1}{N}\sum_{i=1}^{N}\left(  \theta
_{0,N}-1\right)  u_{i1}^{2}\underset{p}{\rightarrow}0,
\end{array}
\]
which shows that $\left(  \theta_{0,N}-1\right)  \frac{1}{N}\sum_{i=1}%
^{N}\left(  \theta_{0,N}-1\right)  u_{i1}^{2}=o_{p}(1)$. Furthermore, since
both $\left(  \theta_{0,N}-1\right)  ^{\frac{1}{2}}u_{i1}$ and $u_{i2}$ have
finite fourth moments and are independent, $\frac{2}{N}\sum_{i=1}^{N}\left(
\theta_{0,N}-1\right)  u_{i1}u_{i2}=o_{p}(1)$, which implies that
\[%
\begin{array}
[c]{c}%
\frac{1}{N}\sum_{i=1}^{N}\Delta y_{i2}\Delta y_{i2}=\frac{1}{N}\sum_{i=1}%
^{N}u_{i2}^{2}+o_{p}(1).
\end{array}
\]
Finally, we have $E\left(  u_{i2}^{2}\right)  =\sigma_{2}^{2}$\ and finite
fourth moment, hence%
\[%
\begin{array}
[c]{c}%
\frac{1}{N}\sum_{i=1}^{N}\Delta y_{i2}^{2}\underset{p}{\rightarrow}\sigma
_{2}^{2}.
\end{array}
\]
Along the same lines as the above this can be shown to hold for other values
of $t$ as well.

\textbf{v. }Similar to the above, when substituting for $\Delta y_{i2}$\ and
$\Delta y_{i3}$\ we have%
\[%
\begin{array}
[c]{c}%
\frac{1}{N}\sum_{i=1}^{N}\Delta y_{i2}\Delta y_{i3}=\frac{1}{N}\sum_{i=1}%
^{N}u_{i2}u_{i3}+\left(  \theta_{0,N}-1\right)  \frac{1}{N}\sum_{i=1}%
^{N}u_{i2}^{2}+\theta_{0,N}\left(  \theta_{0,N}-1\right)  \frac{1}{N}%
\sum_{i=1}^{N}u_{i1}u_{i2}+\\
+(\theta_{0,N}-1)\frac{1}{N}\sum_{i=1}^{N}u_{i1}u_{i3}+(\theta_{0,N}%
-1)^{2}\frac{1}{N}\sum_{i=1}^{N}u_{i1}u_{i2}+\theta_{0,N}(\theta_{0,N}%
-1)^{2}\frac{1}{N}\sum_{i=1}^{N}u_{i1}^{2}.
\end{array}
\]
Similar derivations as before show that $\frac{1}{N}\sum_{i=1}^{N}\theta
_{0,N}\left(  \theta_{0,N}-1\right)  ^{2}u_{i1}^{2}\underset{p}{\rightarrow
}0,$\ $\frac{1}{N}\sum_{i=1}^{N}(\theta_{0,N}-1)^{2}u_{i1}u_{i2}\underset
{p}{\rightarrow}0,$\ $\frac{1}{N}\sum_{i=1}^{N}(\theta_{0,N}-1)u_{i1}%
u_{i3}\underset{p}{\rightarrow}0,$\ $\frac{1}{N}\sum_{i=1}^{N}\theta
_{0,N}\left(  \theta_{0,N}-1\right)  u_{i1}u_{i2}\underset{p}{\rightarrow}%
0,$\ $\frac{1}{N}\sum_{i=1}^{N}(\theta_{0,N}-1)u_{i2}^{2}\underset
{p}{\rightarrow}0$, $\frac{1}{N}\sum_{i=1}^{N}u_{i2}u_{i3}\underset
{p}{\rightarrow}0$, so all these terms are $o_{p}(1)$ and have probability
limit 0, implying that
\[%
\begin{array}
[c]{rl}%
\frac{1}{N}\sum_{i=1}^{N}\Delta y_{i2}\Delta y_{i3}\underset{p}{\rightarrow} &
0.
\end{array}
\]
Along similar lines this can be proven to extend to the first differences at
other time periods.

\textbf{vi. }Since $h_{N}(\theta_{0,N})^{-2}=var(y_{i1}),$ the random variable
$h_{N}(\theta_{0,N})y_{i1}$ has variance equal to one. Since $y_{i1}$ and
$u_{it},$ $t>1,$ are independent, because of Assumption 1e, $E(h_{N}%
(\theta_{0,N})y_{i1}u_{it})=0$. Furthermore, Assumption 1d implies that
$Var\left(  h_{N}(\theta_{0,N})y_{i1}u_{it}\right)  =\sigma_{t}^{2}$, which is
finite. A central limit theorem therefore applies:%
\[%
\begin{array}
[c]{rl}%
\frac{h_{N}(\theta_{0,N})}{\sqrt{N}}\sum_{i=1}^{N}\left(
\begin{array}
[c]{c}%
y_{i1}u_{i2}\\
\vdots\\
y_{i1}u_{iT}%
\end{array}
\right)  =\frac{1}{\sqrt{N}}\sum_{i=1}^{N}\left(
\begin{array}
[c]{c}%
h_{N}(\theta_{0,N})y_{i1}u_{i2}\\
\vdots\\
h_{N}(\theta_{0,N})y_{i1}u_{iT}%
\end{array}
\right)  \underset{d}{\rightarrow} & \psi,
\end{array}
\]
with $\psi=(\psi_{y_{1i}u_{i2}}\ldots\psi_{y_{1i}u_{iT}})^{\prime}$ a $T-1$
dimensional, mean zero normal random vector. Assumption 1e states that
$u_{i1}/\sigma_{1,N},$ $u_{i2}\ldots,u_{iT}$\textit{\ }and $c_{i}$\ are
independently distributed within individuals and over the different
individuals. It implies that $u_{i1}c_{i}$\textit{\ }and $y_{i1}u_{it}$ are
uncorrelated. Since $\psi$ and $\psi_{c}$ are the limits of the scaled sums of
$y_{i1}u_{it}$ and $u_{i1}c_{i}$, they are uncorrelated normal random
variables and therefore independent. As a result of this, the $T\times T$
covariance matrix of $\psi$ and $\psi_{c}$ is diagonal:%
\[%
\begin{array}
[c]{rl}%
V_{\binom{\psi}{\psi_{c}}\binom{\psi}{\psi_{c}}}= & var(\psi,\psi_{c})=\left(
\begin{array}
[c]{cc}%
V_{\psi\psi} & V_{\psi\psi_{c}}\\
V_{\psi_{c}\psi} & V_{\psi_{c}\psi_{c}}%
\end{array}
\right) \\
= & E\left[  \lim_{N\rightarrow\infty}\frac{1}{N}\sum_{i=1}^{N}\left(
\begin{array}
[c]{c}%
h_{N}(\theta_{0,N})y_{i1}u_{i2}\\
\vdots\\
h_{N}(\theta_{0,N})y_{i1}u_{iT}\\
\frac{u_{i1}}{\sigma_{1,N}}c_{i}%
\end{array}
\right)  \left(
\begin{array}
[c]{c}%
h_{N}(\theta_{0,N})y_{i1}u_{i2}\\
\vdots\\
h_{N}(\theta_{0,N})y_{i1}u_{iT}\\
\frac{u_{i1}}{\sigma_{1,N}}c_{i}%
\end{array}
\right)  ^{\prime}\right] \\
= & E\left[  \lim_{N\rightarrow\infty}\left(
\begin{array}
[c]{c}%
h_{N}(\theta_{0,N})y_{i1}u_{i2}\\
\vdots\\
h_{N}(\theta_{0,N})y_{i1}u_{iT}\\
\frac{u_{i1}}{\sigma_{1,N}}c_{i}%
\end{array}
\right)  \left(
\begin{array}
[c]{c}%
h_{N}(\theta_{0,N})y_{i1}u_{i2}\\
\vdots\\
h_{N}(\theta_{0,N})y_{i1}u_{iT}\\
\frac{u_{i1}}{\sigma_{1,N}}c_{i}%
\end{array}
\right)  ^{\prime}\right] \\
= & \text{diag(}\sigma_{2}^{2}\ldots\sigma_{T}^{2}\text{ }\sigma_{c}^{2}).
\end{array}
\]

\paragraph{Proof of Theorem 1.}

\noindent\textbf{T=3. }Under mean stationarity we have%
\[%
\begin{array}
[c]{l}%
\Delta y_{i2}=u_{i2}+\left(  \theta_{0,N}-1\right)  u_{i1}\\
\Delta y_{i3}=u_{i3}+\left(  \theta_{0,N}-1\right)  u_{i2}+\theta_{0,N}\left(
\theta_{0,N}-1\right)  u_{i1}.
\end{array}
\]
Substituting these expressions, we can specify the Dif sample moment and its
derivative as%
\[%
\begin{array}
[c]{rl}%
f_{N}^{Dif}(\theta)= & \frac{1}{N}\sum_{i=1}^{N}\left(  y_{i1}\Delta
y_{i3}-\theta y_{i1}\Delta y_{i2}\right) \\
= & \frac{1}{N}\sum_{i=1}^{N}y_{i1}u_{i3}+\left(  \theta_{0,N}-1-\theta
\right)  \frac{1}{N}\sum_{i=1}^{N}y_{i1}u_{i2}+(\theta_{0,N}-\theta)\frac
{1}{N}\sum_{i=1}^{N}(\theta_{0,N}-1)y_{i1}u_{i1},\\
q_{N}^{Dif}(\theta)= & -\frac{1}{N}\sum_{i=1}^{N}y_{i1}\Delta y_{i2}\\
= & -\frac{1}{N}\sum_{i=1}^{N}y_{i1}u_{i2}-\frac{1}{N}\sum_{i=1}^{N}%
(\theta_{0,N}-1)y_{i1}u_{i1}.
\end{array}
\]
Combining convergence results stated in Lemma 1, the large sample behavior of
the Dif sample moment and derivative can thus be characterized by%
\[%
\begin{array}
[c]{l}%
f_{N}^{Dif}(\theta)=\frac{1}{h_{N}(\theta_{0,N})\sqrt{N}}\left[  \left(
\psi_{3}-\theta\psi_{2}\right)  -(1-\theta)h_{N}(\theta_{0,N})\sigma_{1,N}%
\psi_{c}\right]  -\left(  1-\theta\right)  d_{2}+o_{p}\left(  1\right)  ,\\
q_{N}^{Dif}(\theta)=-\frac{1}{h_{N}(\theta_{0,N})\sqrt{N}}\left[  \psi
_{2}-h_{N}(\theta_{0,N})\sigma_{1,N}\psi_{c}\right]  +d_{2}+o_{p}\left(
1\right)  ,
\end{array}
\]
where we note that $h_{N}(\theta_{0,N})\sigma_{1,N}\leq1,$ since
var($y_{i1})\geq$var($u_{i1}),$ from which it is readily seen that%
\[%
\begin{array}
[c]{l}%
A_{f}^{Dif}(\theta)=\left(
\begin{array}
[c]{cc}%
-\theta & 1
\end{array}
\right)  ,\text{ }\mu_{f}^{Dif}(\theta,\bar{\sigma}^{2})=0,\\
A_{q}^{Dif}(\theta)=\left(
\begin{array}
[c]{cc}%
-1 & 0
\end{array}
\right)  ,\text{ }\mu_{q}^{Dif}(\theta,\bar{\sigma}^{2})=0.
\end{array}
\]

Regarding the Lev moment, using
\[%
\begin{array}
[c]{l}%
y_{i2}=\Delta y_{i2}+y_{i1}\\
y_{i3}=\Delta y_{i3}+\Delta y_{i2}+y_{i1},
\end{array}
\]
we have%
\[%
\begin{array}
[c]{rl}%
f_{N}^{Lev}(\theta)= & \frac{1}{N}\sum_{i=1}^{N}(y_{i3}-\theta y_{i2})\Delta
y_{i2}\\
= & \frac{1}{N}\sum_{i=1}^{N}\left(  \Delta y_{i3}+(1-\theta)\Delta
y_{i2}\right)  \Delta y_{i2}+(1-\theta)\frac{1}{N}\sum_{i=1}^{N}y_{i1}\Delta
y_{i2}.
\end{array}
\]
Exploiting mean stationarity and substituting for $\Delta y_{i2}$ and $\Delta
y_{i3},$ we write%
\[%
\begin{array}
[c]{c}%
(1-\theta)\frac{1}{N}\sum_{i=1}^{N}y_{i1}\Delta y_{i2}=(1-\theta)\frac{1}%
{N}\sum_{i=1}^{N}y_{i1}u_{i2}+(1-\theta)\left(  \theta_{0,N}-1\right)
\frac{1}{N}\sum_{i=1}^{N}y_{i1}u_{i1},
\end{array}
\]
and using Lemma 1, we have%
\[%
\begin{array}
[c]{c}%
\frac{1}{N}\sum_{i=1}^{N}\left(  \Delta y_{i3}+(1-\theta)\Delta y_{i2}\right)
\Delta y_{i2}=\left(  1-\theta\right)  \frac{1}{N}\sum_{i=1}^{N}u_{i2}%
^{2}+o_{p}\left(  1\right)  .
\end{array}
\]
Regarding the Lev derivative, we have%
\[%
\begin{array}
[c]{rl}%
q_{N}^{Lev}(\theta)= & -\frac{1}{N}\sum_{i=1}^{N}y_{i2}\Delta y_{i2}\\
= & -\frac{1}{N}\sum_{i=1}^{N}\Delta y_{i2}\Delta y_{i2}-\frac{1}{N}\sum
_{i=1}^{N}y_{i1}\Delta y_{i2},
\end{array}
\]
where
\[%
\begin{array}
[c]{c}%
\frac{1}{N}\sum_{i=1}^{N}y_{i1}\Delta y_{i2}=\frac{1}{N}\sum_{i=1}^{N}%
y_{i1}u_{i2}+\left(  \theta_{0,N}-1\right)  \frac{1}{N}\sum_{i=1}^{N}%
y_{i1}u_{i1},
\end{array}
\]
and
\[%
\begin{array}
[c]{c}%
\frac{1}{N}\sum_{i=1}^{N}\Delta y_{i2}\Delta y_{i2}=\frac{1}{N}\sum_{i=1}%
^{N}u_{i2}^{2}+o_{p}\left(  1\right)  .
\end{array}
\]
Therefore, we can write the Lev moment condition and derivative as%
\[%
\begin{array}
[c]{rl}%
f_{N}^{Lev}(\theta)= & (1-\theta)\left\{  \frac{1}{N}\sum_{i=1}^{N}u_{i2}%
^{2}+\frac{1}{N}\sum_{i=1}^{N}y_{i1}u_{i2}+\right. \\
& \left.  \frac{1}{N}\sum_{i=1}^{N}(\theta_{0,N}-1)y_{i1}u_{i1}\right\}
+o_{p}(1).\\
q_{N}^{Lev}(\theta)= & -\frac{1}{N}\sum_{i=1}^{N}u_{i2}^{2}-\frac{1}{N}%
\sum_{i=1}^{N}y_{i1}u_{i2}\\
& -\frac{1}{N}\sum_{i=1}^{N}(\theta_{0,N}-1)y_{i1}u_{i1}+o_{p}(1).
\end{array}
\]
Combining this and other convergence results from Lemma 1, the large sample
behavior of the Lev sample moment and derivative can thus be characterized by%
\[%
\begin{array}
[c]{rl}%
f_{N}^{Lev}(\theta)= & (1-\theta)\left\{  \frac{1}{h_{N}(\theta_{0,N})\sqrt
{N}}\left[  \psi_{2}-h_{N}(\theta_{0,N})\sigma_{1,N}\psi_{c}\right]  +\left(
\sigma_{2}^{2}-d_{2}\right)  \right\}  +o_{p}\left(  1\right) \\
q_{N}^{Lev}(\theta)= & -\frac{1}{h_{N}(\theta_{0,N})\sqrt{N}}\left[  \psi
_{2}-h_{N}(\theta_{0,N})\sigma_{1,N}\psi_{c}\right]  -\left(  \sigma_{2}%
^{2}-d_{2}\right)  +o_{p}\left(  1\right)  ,
\end{array}
\]
so this implies that%
\[%
\begin{array}
[c]{l}%
A_{f}^{Lev}(\theta)=(1-\theta\text{ }0),\text{ }\mu_{f}^{Lev}(\theta
,\bar{\sigma}^{2})=\left(  1-\theta\right)  \sigma_{2}^{2},\\
A_{q}^{Lev}(\theta)=(-1\text{ }0),\text{ }\mu_{q}^{Lev}(\theta,\bar{\sigma
}^{2})=-\sigma_{2}^{2}.
\end{array}
\]
From this last result, it is not difficult to see that, under Assumption 2b,
we have%
\[%
\begin{array}
[c]{c}%
\frac{1}{N}\sum_{i=1}^{N}y_{i2}\Delta y_{i2}\underset{p}{\rightarrow}%
\sigma_{2}^{2}-d_{2}.
\end{array}
\]
The reason for this is that Assumption 2b amounts to $h_{N}(\theta_{0,N}%
)\sqrt{N}=\frac{\sqrt{N}}{\sqrt{var(y_{i1})}}\underset{N\rightarrow\infty
}{\rightarrow}\infty$ and, since var($y_{i1})\geq$var($u_{i1})$, it implies
that $\sigma_{1,N}^{2}/N\underset{N\rightarrow\infty}{\rightarrow}0.$ Finally,
the Sys sample moment and derivative simply result from stacking the Dif and
Lev sample moments and derivatives:%
\[%
\begin{array}
[c]{rl}%
f_{N}^{Sys}(\theta)= & \frac{1}{N}\sum_{i=1}^{N}\left(
\begin{array}
[c]{c}%
y_{i1}\Delta y_{i3}-\theta y_{i1}\Delta y_{i2}\\
y_{i3}\Delta y_{i2}-\theta y_{i2}\Delta y_{i2}%
\end{array}
\right)  ,\\
q_{N}^{Sys}(\theta)= & -\frac{1}{N}\sum_{i=1}^{N}\left(
\begin{array}
[c]{c}%
y_{i1}\Delta y_{i2}\\
y_{i2}\Delta y_{i2}%
\end{array}
\right)  .
\end{array}
\]
Combining earlier convergence results, the large sample behavior of the Sys
sample moment and derivative can thus be characterized by%
\[%
\begin{array}
[c]{rl}%
f_{N}^{Sys}(\theta)= & \left(
\begin{array}
[c]{cc}%
-\theta & 1\\
1-\theta & 0
\end{array}
\right)  \left[  \frac{1}{h_{N}(\theta_{0,N})\sqrt{N}}\left\{  \left(
\begin{array}
[c]{c}%
\psi_{2}\\
\psi_{3}%
\end{array}
\right)  -h_{N}(\theta_{0,N})\sigma_{1,N}\psi_{c}\iota_{2}\right\}  -\iota
_{2}d_{2}\right]  +\\
& \left(  1-\theta\right)  \left(
\begin{array}
[c]{c}%
0\\
\sigma_{2}^{2}%
\end{array}
\right)  +o_{p}(1),\\
q_{N}^{Sys}(\theta)= & -\left(
\begin{array}
[c]{cc}%
1 & 0\\
1 & 0
\end{array}
\right)  \left[  \frac{1}{h_{N}(\theta_{0,N})\sqrt{N}}\left\{  \left(
\begin{array}
[c]{c}%
\psi_{2}\\
\psi_{3}%
\end{array}
\right)  -h_{N}(\theta_{0,N})\sigma_{1,N}\psi_{c}\iota_{2}\right\}  -\iota
_{2}d_{2}\right]  -\\
& \left(
\begin{array}
[c]{c}%
0\\
\sigma_{2}^{2}%
\end{array}
\right)  +o_{p}(1),
\end{array}
\]
from which it is readily seen that%
\[%
\begin{array}
[c]{l}%
A_{f}^{Sys}(\theta)=\left(
\begin{array}
[c]{cc}%
-\theta & 1\\
1-\theta & 0
\end{array}
\right)  ,\text{ }\mu_{f}^{Sys}(\theta,\bar{\sigma}^{2})=\left(
1-\theta\right)  \left(
\begin{array}
[c]{c}%
0\\
\sigma_{2}^{2}%
\end{array}
\right)  ,\\
A_{q}^{Sys}(\theta)=\left(
\begin{array}
[c]{cc}%
-1 & 0\\
-1 & 0
\end{array}
\right)  ,\text{ }\mu_{q}^{Sys}(\theta,\bar{\sigma}^{2})=\left(
\begin{array}
[c]{c}%
0\\
-\sigma_{2}^{2}%
\end{array}
\right)  .
\end{array}
\]

\noindent\textbf{T=4. }Under mean stationarity, we have%
\[%
\begin{array}
[c]{l}%
\Delta y_{i2}=u_{i2}+\left(  \theta_{0,N}-1\right)  u_{i1}\\
\Delta y_{i3}=u_{i3}+\left(  \theta_{0,N}-1\right)  u_{i2}+\theta_{0,N}\left(
\theta_{0,N}-1\right)  u_{i1}\\
\Delta y_{i4}=u_{i4}+\left(  \theta_{0,N}-1\right)  u_{i3}+\theta_{0,N}\left(
\theta_{0,N}-1\right)  u_{i2}+\theta_{0,N}^{2}\left(  \theta_{0,N}-1\right)
u_{i1}.
\end{array}
\]
Substituting these expressions and $y_{i2}=\Delta y_{i2}+y_{i1}$, we can
specify the Dif sample moments and their derivatives as%
\[%
\begin{array}
[c]{l}%
f_{N}^{Dif}(\theta)=\frac{1}{N}\sum_{i=1}^{N}\left(
\begin{array}
[c]{c}%
y_{i1}\Delta y_{i3}-\theta y_{i1}\Delta y_{i2}\\
y_{i1}\Delta y_{i4}-\theta y_{i1}\Delta y_{i3}\\
y_{i2}\Delta y_{i4}-\theta y_{i2}\Delta y_{i3}%
\end{array}
\right) \\
=\left(
\begin{array}
[c]{ccc}%
\theta_{0,N}-1-\theta & 1 & 0\\
\left(  \theta_{0,N}-\theta\right)  (\theta_{0,N}-1) & \theta_{0,N}-1-\theta &
1\\
\left(  \theta_{0,N}-\theta\right)  (\theta_{0,N}-1) & \theta_{0,N}-1-\theta &
1
\end{array}
\right)  \frac{1}{N}\sum_{i=1}^{N}\left(
\begin{array}
[c]{c}%
y_{i1}u_{i2}\\
y_{i1}u_{i3}\\
y_{i1}u_{i4}%
\end{array}
\right)  +\\
(\theta_{0,N}-\theta)(\theta_{0,N}-1)\left(
\begin{array}
[c]{c}%
1\\
\theta_{0,N}\\
\theta_{0,N}%
\end{array}
\right)  \frac{1}{N}\sum_{i=1}^{N}y_{i1}u_{i1}+\frac{1}{N}\sum_{i=1}%
^{N}\left(
\begin{array}
[c]{c}%
0\\
0\\
\Delta y_{i2}(\Delta y_{i4}-\theta\Delta y_{i3})
\end{array}
\right)  ,
\end{array}
\]

\[%
\begin{array}
[c]{l}%
q_{N}^{Dif}(\theta)=-\frac{1}{N}\sum_{i=1}^{N}\left(
\begin{array}
[c]{c}%
y_{i1}\Delta y_{i2}\\
y_{i1}\Delta y_{i3}\\
y_{i2}\Delta y_{i3}%
\end{array}
\right) \\
=-\left(
\begin{array}
[c]{ccc}%
1 & 0 & 0\\
\theta_{0,N}-1 & 1 & 0\\
\theta_{0,N}-1 & 1 & 0
\end{array}
\right)  \frac{1}{N}\sum_{i=1}^{N}\left(
\begin{array}
[c]{c}%
y_{i1}u_{i2}\\
y_{i1}u_{i3}\\
y_{i1}u_{i4}%
\end{array}
\right)  -(\theta_{0,N}-1)\left(
\begin{array}
[c]{c}%
1\\
\theta_{0,N}\\
\theta_{0,N}%
\end{array}
\right)  \frac{1}{N}\sum_{i=1}^{N}y_{i1}u_{i1}\\
-\frac{1}{N}\sum_{i=1}^{N}\left(
\begin{array}
[c]{c}%
0\\
0\\
\Delta y_{i2}\Delta y_{i3}%
\end{array}
\right)  .
\end{array}
\]
The limit behavior of the first two terms in each expression has been
established before. Furthermore, Lemma 1 shows that the last term in each
expression is $o_{p}(1)$. Therefore, the large Dif sample moment and
derivative can be expressed as:%
\begin{align*}
f_{N}^{Dif}(\theta)  &  =\left(
\begin{array}
[c]{ccc}%
-\theta & 1 & 0\\
0 & -\theta & 1\\
0 & -\theta & 1
\end{array}
\right)  \left[  \frac{1}{h_{N}(\theta_{0,N})\sqrt{N}}\left\{  \left(
\begin{array}
[c]{c}%
\psi_{2}\\
\psi_{3}\\
\psi_{4}%
\end{array}
\right)  -h_{N}(\theta_{0,N})\sigma_{1,N}\psi_{c}\iota_{3}\right\}  -\iota
_{3}d_{2}\right]  +o_{p}(1),\\
q_{N}^{Dif}(\theta)  &  =-\left(
\begin{array}
[c]{ccc}%
1 & 0 & 0\\
0 & 1 & 0\\
0 & 1 & 0
\end{array}
\right)  \left[  \frac{1}{h_{N}(\theta_{0,N})\sqrt{N}}\left\{  \left(
\begin{array}
[c]{c}%
\psi_{2}\\
\psi_{3}\\
\psi_{4}%
\end{array}
\right)  -h_{N}(\theta_{0,N})\sigma_{1,N}\psi_{c}\iota_{3}\right\}  -\iota
_{3}d_{2}\right]  +o_{p}(1),
\end{align*}
from which it is readily seen that%
\[%
\begin{array}
[c]{l}%
A_{f}^{Dif}(\theta)=\left(
\begin{array}
[c]{ccc}%
-\theta & 1 & 0\\
0 & -\theta & 1\\
0 & -\theta & 1
\end{array}
\right)  ,\text{ }\mu_{f}^{Dif}(\theta,\bar{\sigma}^{2})=\left(
\begin{array}
[c]{c}%
0\\
0\\
0
\end{array}
\right)  ,\\
A_{q}^{Dif}(\theta)=-\left(
\begin{array}
[c]{ccc}%
1 & 0 & 0\\
0 & 1 & 0\\
0 & 1 & 0
\end{array}
\right)  ,\text{ }\mu_{q}^{Dif}(\theta,\bar{\sigma}^{2})=\left(
\begin{array}
[c]{c}%
0\\
0\\
0
\end{array}
\right)  .
\end{array}
\]

After some algebra, we can specify the Lev sample moments and their
derivatives as%
\[%
\begin{array}
[c]{l}%
f_{N}^{Lev}(\theta)=\frac{1}{N}\sum_{i=1}^{N}\left(
\begin{array}
[c]{c}%
y_{i3}\Delta y_{i2}-\theta y_{i2}\Delta y_{i2}\\
y_{i4}\Delta y_{i3}-\theta y_{i3}\Delta y_{i3}%
\end{array}
\right) \\
=\frac{1}{N}\sum_{i=1}^{N}\left(
\begin{array}
[c]{ccc}%
1-\theta & 0 & 0\\
(1-\theta)(\theta_{0,N}-1) & 1-\theta & 0
\end{array}
\right)  \left(
\begin{array}
[c]{c}%
y_{i1}u_{i2}\\
y_{i1}u_{i3}\\
y_{i1}u_{i4}%
\end{array}
\right)  +(1-\theta)(\theta_{0,N}-1)\left(
\begin{array}
[c]{c}%
1\\
\theta_{0,N}%
\end{array}
\right)  \frac{1}{N}\sum_{i=1}^{N}y_{i1}u_{i1}\\
+(1-\theta)\frac{1}{N}\sum_{i=1}^{N}\left(
\begin{array}
[c]{c}%
\Delta y_{i2}\Delta y_{i2}\\
\Delta y_{i3}\Delta y_{i3}%
\end{array}
\right)  +\frac{1}{N}\sum_{i=1}^{N}\left(
\begin{array}
[c]{c}%
\Delta y_{i3}\Delta y_{i2}\\
\left(  \Delta y_{i4}+(1-\theta)\Delta y_{i2}\right)  \Delta y_{i3}%
\end{array}
\right)  ,
\end{array}
\]%
\[%
\begin{array}
[c]{l}%
q_{N}^{Lev}(\theta)=-\frac{1}{N}\sum_{i=1}^{N}\left(
\begin{array}
[c]{c}%
y_{i2}\Delta y_{i2}\\
y_{i3}\Delta y_{i3}%
\end{array}
\right) \\
=-\frac{1}{N}\sum_{i=1}^{N}\left(
\begin{array}
[c]{ccc}%
1 & 0 & 0\\
\theta_{0,N}-1 & 1 & 0
\end{array}
\right)  \left(
\begin{array}
[c]{c}%
y_{i1}u_{i2}\\
y_{i1}u_{i3}\\
y_{i1}u_{i4}%
\end{array}
\right)  -(\theta_{0,N}-1)\left(
\begin{array}
[c]{c}%
1\\
\theta_{0,N}%
\end{array}
\right)  \frac{1}{N}\sum_{i=1}^{N}y_{i1}u_{i1}\\
-\frac{1}{N}\sum_{i=1}^{N}\left(
\begin{array}
[c]{c}%
\Delta y_{i2}\Delta y_{i2}\\
\Delta y_{i3}\Delta y_{i3}%
\end{array}
\right)  -\frac{1}{N}\sum_{i=1}^{N}\left(
\begin{array}
[c]{c}%
0\\
\Delta y_{i2}\Delta y_{i3}%
\end{array}
\right)  .
\end{array}
\]
Using Lemma 1, the large sample behavior of these expressions is equal to:%
\begin{align*}
f_{N}^{Lev}(\theta)  &  =\left(
\begin{array}
[c]{ccc}%
1-\theta & 0 & 0\\
0 & 1-\theta & 0
\end{array}
\right)  \left[  \frac{1}{h_{N}(\theta_{0,N})\sqrt{N}}\left\{  \left(
\begin{array}
[c]{c}%
\psi_{2}\\
\psi_{3}\\
\psi_{4}%
\end{array}
\right)  -h_{N}(\theta_{0,N})\sigma_{1,N}\psi_{c}\iota_{2}\right\}  -\iota
_{2}d_{2}\right]  +\\
&  \left(  1-\theta\right)  \left(
\begin{array}
[c]{c}%
\sigma_{2}^{2}\\
\sigma_{3}^{2}%
\end{array}
\right)  +o_{p}(1),
\end{align*}

\begin{align*}
q_{N}^{Lev}(\theta)  &  =-\left(
\begin{array}
[c]{ccc}%
1 & 0 & 0\\
0 & 1 & 0
\end{array}
\right)  \left[  \frac{1}{h_{N}(\theta_{0,N})\sqrt{N}}\left\{  \left(
\begin{array}
[c]{c}%
\psi_{2}\\
\psi_{3}\\
\psi_{4}%
\end{array}
\right)  -h_{N}(\theta_{0,N})\sigma_{1,N}\psi_{c}\iota_{2}\right\}  -\iota
_{2}d_{2}\right]  -\\
&  \left(
\begin{array}
[c]{c}%
\sigma_{2}^{2}\\
\sigma_{3}^{2}%
\end{array}
\right)  +o_{p}(1),
\end{align*}
so this implies that%
\[%
\begin{array}
[c]{l}%
A_{f}^{Lev}(\theta)=\left(
\begin{array}
[c]{ccc}%
1-\theta & 0 & 0\\
0 & 1-\theta & 0
\end{array}
\right)  ,\text{ }\mu_{f}^{Lev}(\theta,\bar{\sigma}^{2})=\left(
1-\theta\right)  \left(
\begin{array}
[c]{c}%
\sigma_{2}^{2}\\
\sigma_{3}^{2}%
\end{array}
\right) \\
A_{q}^{Lev}(\theta)=-\left(
\begin{array}
[c]{ccc}%
1 & 0 & 0\\
0 & 1 & 0
\end{array}
\right)  ,\text{ }\mu_{q}^{Lev}(\theta,\bar{\sigma}^{2})=-\left(
\begin{array}
[c]{c}%
\sigma_{2}^{2}\\
\sigma_{3}^{2}%
\end{array}
\right)  .
\end{array}
\]

We can specify the NL sample moment and its derivative as%
\begin{align*}
&
\begin{array}
[c]{l}%
f_{N}^{NL}(\theta)=\frac{1}{N}\sum_{i=1}^{N}\left(  y_{i4}-\theta
y_{i3}\right)  \left(  \Delta y_{i3}-\theta\Delta y_{i2}\right) \\
=\frac{1}{N}\sum_{i=1}^{N}\left(
\begin{array}
[c]{ccc}%
(1-\theta)(\theta_{0,N}-\theta-1) & (1-\theta) & 0
\end{array}
\right)  \left(
\begin{array}
[c]{c}%
y_{i1}u_{i2}\\
y_{i1}u_{i3}\\
y_{i1}u_{i4}%
\end{array}
\right)  +\\
\frac{1}{N}\sum_{i=1}^{N}(\theta_{0,N}-1)(\theta_{0,N}-\theta)(1-\theta
)y_{i1}u_{i1}+\left(  1-\theta\right)  \frac{1}{N}\sum_{i=1}^{N}\left(  \Delta
y_{i3}\Delta y_{i3}-\theta\Delta y_{i2}\Delta y_{i2}\right)  +\\
\frac{1}{N}\sum_{i=1}^{N}\left(  (\Delta y_{i4}+\left(  1-\theta\right)
\Delta y_{i2})\Delta y_{i3}-(\Delta y_{i4}+\left(  1-\theta\right)  \Delta
y_{i3})\theta\Delta y_{i2}\right)  ,
\end{array}
\\
&
\begin{array}
[c]{l}%
q_{N}^{NL}(\theta)=-\frac{1}{N}\sum_{i=1}^{N}\left(
\begin{array}
[c]{ccc}%
\theta_{0,N}-2\theta & -1 & 0
\end{array}
\right)  \left(
\begin{array}
[c]{c}%
y_{i1}u_{i2}\\
y_{i1}u_{i3}\\
y_{i1}u_{i4}%
\end{array}
\right)  +\frac{1}{N}\sum_{i=1}^{N}(\theta_{0,N}-1)(1+\theta_{0,N}%
-2\theta)y_{i1}u_{i1}-\\
\frac{1}{N}\sum_{i=1}^{N}\left(  \Delta y_{i3}\Delta y_{i3}+\left(
1-2\theta\right)  \Delta y_{i2}\Delta y_{i2}\right)  -\frac{1}{N}\sum
_{i=1}^{N}\left(  \Delta y_{i2}\Delta y_{i3}+\left(  \Delta y_{i4}+\left(
1-2\theta\right)  \Delta y_{i3}\right)  \Delta y_{i2}\right)  .
\end{array}
\end{align*}
Using Lemma 1, the large sample behavior of these expressions is equal to:%
\[%
\begin{array}
[c]{l}%
f_{N}^{NL}(\theta)=\frac{1}{N}\sum_{i=1}^{N}\left(
\begin{array}
[c]{ccc}%
\theta(\theta-1) & 1-\theta & 0
\end{array}
\right)  \left[  \frac{1}{h_{N}(\theta_{0,N})\sqrt{N}}\left\{  \left(
\begin{array}
[c]{c}%
\psi_{2}\\
\psi_{3}\\
\psi_{4}%
\end{array}
\right)  -h_{N}(\theta_{0,N})\sigma_{1,N}\psi_{c}\iota_{2}\right\}  -\iota
_{2}d_{2}\right]  +\\
\left(  1-\theta\right)  \left(  \sigma_{3}^{2}-\theta\sigma_{2}^{2}\right)
+o_{p}(1),
\end{array}
\]%
\[%
\begin{array}
[c]{l}%
q_{N}^{NL}(\theta)=-\frac{1}{N}\sum_{i=1}^{N}\left(
\begin{array}
[c]{ccc}%
1-2\theta & 1 & 0
\end{array}
\right)  \left[  \frac{1}{h_{N}(\theta_{0,N})\sqrt{N}}\left\{  \left(
\begin{array}
[c]{c}%
\psi_{2}\\
\psi_{3}\\
\psi_{4}%
\end{array}
\right)  -h_{N}(\theta_{0,N})\sigma_{1,N}\psi_{c}\iota_{2}\right\}  -\iota
_{2}d_{2}\right]  -\\
\sigma_{3}^{2}-(1-2\theta)\sigma_{2}^{2}+o_{p}(1),
\end{array}
\]
so this implies that:%
\[%
\begin{array}
[c]{l}%
A_{f}^{NL}(\theta)=\left(
\begin{array}
[c]{ccc}%
\theta(\theta-1) & 1-\theta & 0
\end{array}
\right)  ,\text{ }\mu_{f}^{NL}(\theta,\bar{\sigma}^{2})=\left(  1-\theta
\right)  \left(  \sigma_{3}^{2}-\theta\sigma_{2}^{2}\right) \\
A_{q}^{NL}(\theta)=\left(
\begin{array}
[c]{ccc}%
2\theta-1 & -1 & 0
\end{array}
\right)  ,\text{ }\mu_{q}^{NL}(\theta,\bar{\sigma}^{2})=\left(  2\theta
-1\right)  \sigma_{2}^{2}-\sigma_{3}^{2}.
\end{array}
\]

Finally, regarding AS and Sys moment conditions, we simply have%
\[%
\begin{array}
[c]{l}%
A_{f}^{Sys}(\theta)=\left(
\begin{array}
[c]{c}%
A_{f}^{Dif}(\theta)\\
A_{f}^{Lev}(\theta)\text{ }\vdots\text{ }0
\end{array}
\right)  ,\text{ }\mu_{f}^{Sys}(\theta,\bar{\sigma}^{2})=\left(
\begin{tabular}
[c]{c}%
$\mu_{f}^{Dif}(\theta,\bar{\sigma}^{2})$\\
$\mu_{f}^{Lev}(\theta,\bar{\sigma}^{2})$%
\end{tabular}
\right)  ,\\
A_{q}^{Sys}(\theta)=\left(
\begin{array}
[c]{c}%
A_{q}^{Dif}(\theta)\\
A_{q}^{Lev}(\theta)\text{ }\vdots\text{ }0
\end{array}
\right)  ,\text{ }\mu_{q}^{Sys}(\theta,\bar{\sigma}^{2})=\left(
\begin{tabular}
[c]{c}%
$\mu_{q}^{Dif}(\theta,\bar{\sigma}^{2})$\\
$\mu_{q}^{Lev}(\theta,\bar{\sigma}^{2})$%
\end{tabular}
\right)  .
\end{array}
\]%
\[%
\begin{array}
[c]{l}%
A_{f}^{AS}(\theta)=\left(
\begin{array}
[c]{c}%
A_{f}^{Dif}(\theta)\\
A_{f}^{NL}(\theta)\text{ }\vdots\text{ }0
\end{array}
\right)  ,\text{ }\mu_{f}^{AS}(\theta,\bar{\sigma}^{2})=\left(
\begin{tabular}
[c]{c}%
$\mu_{f}^{Dif}(\theta,\bar{\sigma}^{2})$\\
$\mu_{f}^{NL}(\theta,\bar{\sigma}^{2})$%
\end{tabular}
\right)  ,\\
A_{q}^{AS}(\theta)=\left(
\begin{array}
[c]{c}%
A_{q}^{Dif}(\theta)\\
A_{q}^{NL}(\theta)\text{ }\vdots\text{ }0
\end{array}
\right)  ,\text{ }\mu_{q}^{AS}(\theta,\bar{\sigma}^{2})=\left(
\begin{tabular}
[c]{c}%
$\mu_{q}^{Dif}(\theta,\bar{\sigma}^{2})$\\
$\mu_{q}^{NL}(\theta,\bar{\sigma}^{2})$%
\end{tabular}
\right)  .
\end{array}
\]

\noindent\textbf{T=5. }Using similar calculations, we obtain:%
\[
A_{f}^{Dif}(\theta)=\left(
\begin{array}
[c]{cccc}%
-\theta & 1 & 0 & 0\\
0 & -\theta & 1 & 0\\
0 & -\theta & 1 & 0\\
0 & 0 & -\theta & 1\\
0 & 0 & -\theta & 1\\
0 & 0 & -\theta & 1
\end{array}
\right)  ,\text{ }\mu_{f}^{Dif}(\theta,\bar{\sigma}^{2})=\left(
\begin{array}
[c]{c}%
0\\
0\\
0\\
0\\
0\\
0
\end{array}
\right)  ,
\]%
\[
A_{f}^{Lev}(\theta)=\left(
\begin{array}
[c]{cccc}%
1-\theta & 0 & 0 & 0\\
0 & 1-\theta & 0 & 0\\
0 & 0 & 1-\theta & 0
\end{array}
\right)  ,\text{ }\mu_{f}^{Lev}(\theta,\bar{\sigma}^{2})=\left(
1-\theta\right)  \left(
\begin{array}
[c]{c}%
\sigma_{2}^{2}\\
\sigma_{3}^{2}\\
\sigma_{4}^{2}%
\end{array}
\right)  ,
\]%
\[
A_{f}^{NL}(\theta)=\left(
\begin{array}
[c]{cccc}%
\theta(\theta-1) & 1-\theta & 0 & 0\\
0 & \theta(\theta-1) & 1-\theta & 0
\end{array}
\right)  ,\text{ }\mu_{f}^{NL}(\theta,\bar{\sigma}^{2})=\left(  1-\theta
\right)  \left(
\begin{array}
[c]{c}%
\sigma_{3}^{2}-\theta\sigma_{2}^{2}\\
\sigma_{4}^{2}-\theta\sigma_{3}^{2}%
\end{array}
\right)  .
\]

\noindent\textbf{General T. }Along the lines of the above, it is also possible
to construct the expressions of $A_{f}^{j}(\theta),A_{q}^{j}(\theta),$
$\mu_{f}^{j}(\theta,\bar{\sigma}^{2})$ and $\mu_{q}^{j}(\theta,\bar{\sigma
}^{2})$ for larger values of $T$ which we, for reasons of brevity, refrain
from.\bigskip

\paragraph{Orthogonal complements of $A_{f}^{AS}(\theta)$ and $A_{f}%
^{Sys}(\theta)$ for $T=4$ and $5$ and the specification of the robust sample
moments}

We specify the orthogonal complements as in (\ref{OCdec}), which we repeat
here for convenience:%
\[
A_{f}^{j}(\theta)_{\perp}=(G_{f,T}^{j}(\theta)\text{ }\vdots\text{ }%
G_{2,T}^{j}),
\]
where $T$ indicates the number of time periods and $G_{2,T}^{j}$ is such that
$G_{2,T}^{j\prime}\mu_{f}^{j}(\theta,\bar{\sigma}^{2})=0.$ This notation is
used in the proofs of subsequent theorems.

\noindent\textbf{T=4. }From the expressions of $A_{f}^{j}(\theta)$ and
$\mu_{f}^{j}(\theta,\bar{\sigma}^{2})$ in (\ref{assyst=4}), $G_{f,T=4}%
^{j}(\theta)$ and $G_{2,T=4}^{j}$ for $j=AS,$ $Sys$ result as:%
\begin{align*}
G_{f,T=4}^{AS}(\theta)  &  =\left(
\begin{array}
[c]{c}%
-(1-\theta)\\
0\\
0\\
1
\end{array}
\right)  ,\text{ }G_{2,T=4}^{AS}=\left(
\begin{array}
[c]{c}%
0\\
-1\\
1\\
0
\end{array}
\right)  ,\\
G_{f,T=4}^{Sys}(\theta)  &  =\left(
\begin{array}
[c]{c}%
-\left(  1-\theta\right) \\
0\\
0\\
-\theta\\
1
\end{array}
\right)  ,\text{ }G_{2,T=4}^{Sys}=\left(
\begin{array}
[c]{c}%
0\\
-1\\
1\\
0\\
0
\end{array}
\right)  .
\end{align*}
From these expressions and (\ref{assyst=4}), it is easily seen that%
\begin{align*}
A_{f}^{AS}(\theta)_{\perp}^{\prime}\mu_{f}^{AS}(\theta,\bar{\sigma}^{2})  &
=\left(
\begin{array}
[c]{c}%
\left(  1-\theta\right)  \left(  \sigma_{3}^{2}-\theta\sigma_{2}^{2}\right) \\
0
\end{array}
\right)  ,\\
A_{f}^{Sys}(\theta)_{\perp}^{\prime}\mu_{f}^{Sys}(\theta,\bar{\sigma}^{2})  &
=\left(
\begin{array}
[c]{c}%
\sigma_{3}^{2}-\theta\sigma_{2}^{2}\\
0
\end{array}
\right)  ,
\end{align*}
from which follows that $A_{f}^{j}(\theta)_{\perp}^{\prime}\mu_{f}^{j}%
(\theta,\bar{\sigma}^{2})\neq0$ for all $\theta\neq\theta_{0,N}, $ $j=AS,$
$Sys.$

\noindent\textbf{T=5. }The expressions for $A_{f}^{j}(\theta),$ $\mu_{f}%
^{j}(\theta,\bar{\sigma}^{2}),$ $G_{f,T=5}^{j}(\theta)$ and $G_{2,T=5}^{j}$
for $j=AS,$ $Sys$ are:%
\begin{align*}
A_{f}^{AS}(\theta)  &  =\left(
\begin{array}
[c]{cccc}%
-\theta & 1 & 0 & 0\\
0 & -\theta & 1 & 0\\
0 & -\theta & 1 & 0\\
0 & 0 & -\theta & 1\\
0 & 0 & -\theta & 1\\
0 & 0 & -\theta & 1\\
\theta(\theta-1) & 1-\theta & 0 & 0\\
0 & \theta(\theta-1) & 1-\theta & 0
\end{array}
\right)  \text{, }\mu_{f}^{AS}(\theta,\bar{\sigma}^{2})=\left(  1-\theta
\right)  \left(
\begin{array}
[c]{c}%
0\\
0\\
0\\
0\\
0\\
0\\
\sigma_{3}^{2}-\theta\sigma_{2}^{2}\\
\sigma_{4}^{2}-\theta\sigma_{3}^{2}%
\end{array}
\right)  ,\\
G_{f,T=5}^{AS}(\theta)  &  =\left(
\begin{array}
[c]{ccc}%
-(1-\theta) & 0 & 0\\
0 & -(1-\theta) & 0\\
0 & 0 & -(1-\theta)\\
0 & 0 & 0\\
0 & 0 & 0\\
0 & 0 & 0\\
1 & 0 & 0\\
0 & 1 & 1
\end{array}
\right)  ,\text{ }G_{2,T=5}^{AS}=\left(
\begin{array}
[c]{cc}%
0 & 0\\
0 & 0\\
0 & 0\\
-1 & 0\\
1 & -1\\
0 & 1\\
0 & 0\\
0 & 0
\end{array}
\right)  ,
\end{align*}%
\begin{align*}
A_{f}^{Sys}(\theta)  &  =\left(
\begin{array}
[c]{cccc}%
-\theta & 1 & 0 & 0\\
0 & -\theta & 1 & 0\\
0 & -\theta & 1 & 0\\
0 & 0 & -\theta & 1\\
0 & 0 & -\theta & 1\\
0 & 0 & -\theta & 1\\
1-\theta & 0 & 0 & 0\\
0 & 1-\theta & 0 & 0\\
0 & 0 & 1-\theta & 0
\end{array}
\right)  \text{, }\mu_{f}^{Sys}(\theta,\bar{\sigma}^{2})=\left(
1-\theta\right)  \left(
\begin{array}
[c]{c}%
0\\
0\\
0\\
0\\
0\\
0\\
\sigma_{2}^{2}\\
\sigma_{3}^{2}\\
\sigma_{4}^{2}%
\end{array}
\right)  ,\\
G_{f,T=5}^{Sys}(\theta)  &  =\left(
\begin{array}
[c]{ccc}%
-(1-\theta) & 0 & 0\\
0 & -(1-\theta) & 0\\
0 & 0 & -(1-\theta)\\
0 & 0 & 0\\
0 & 0 & 0\\
0 & 0 & 0\\
-\theta & 0 & 0\\
1 & -\theta & -\theta\\
0 & 1 & 1
\end{array}
\right)  ,\text{ }G_{2,T=5}^{Sys}=\left(
\begin{array}
[c]{cc}%
0 & 0\\
0 & 0\\
0 & 0\\
-1 & 0\\
1 & -1\\
0 & 1\\
0 & 0\\
0 & 0\\
0 & 0
\end{array}
\right)  .
\end{align*}
Straightforward algebra shows that $A_{f}^{j}(\theta)_{\perp}^{\prime}\mu
_{f}^{j}(\theta,\bar{\sigma}^{2})\neq0$ for all $\theta\neq\theta_{0,N}, $
$j=AS,$ $Sys.$

The robust sample moments are defined as
\[
g_{f,T}^{j}(\theta)=A_{f}(\theta)_{\perp}^{j\prime}f_{N}^{j}(\theta),
\]
with $A_{f}(\theta)_{\perp}^{j}=(G_{f,T}^{j}(\theta)$ $\vdots$ $G_{2,T}^{j}).$
For the Sys moment conditions, $G_{f,T}^{j}(\theta)$ is a linear function of
$\theta$ and $G_{2,T}^{j}$ does not depend on $\theta.$ Since $f_{N}%
^{j}(\theta)$ is linear in $\theta$ as well for the Sys sample moments, the
part of $g_{f,T}^{j}(\theta)$ resulting from $G_{f,T}^{j}(\theta)^{\prime
}f_{N}^{j}(\theta)$ is quadratic in $\theta$ while the part that results from
$G_{2,T}^{j\prime}f_{N}^{j}(\theta)$ is linear in $\theta.$ Given the
specification of $G_{f,T}^{j}(\theta),$ $G_{2,T}^{j}$ and $f_{N}^{j}(\theta),$
it is then straightforward to compute the specification of $a,$ $b$ and $d.$

For the AS moment conditions, $G_{f,T}^{j}(\theta)$ is a linear function of
$\theta$ and $G_{2,T}^{j}$ does not depend on $\theta.$ For the AS sample
moments, $f_{N}^{j}(\theta)$ is quadratic in $\theta$ but the part of
$g_{f,T}^{j}(\theta)$ resulting from $G_{f,T}^{j}(\theta)^{\prime}f_{N}%
^{j}(\theta)$ is not of third order in $\theta$ as expected but just a
quadratic function of $\theta.$ The part of $g_{f,T}^{j}(\theta)$ that results
from $G_{2,T}^{j\prime}f_{N}^{j}(\theta)$ is linear in $\theta.$ Given the
specification of $G_{f,T}^{j}(\theta),$ $G_{2,T}^{j}$ and $f_{N}^{j}(\theta),$
it is then again straightforward to compute the specification of $a,$ $b$ and
$d.$

\paragraph{Proof of Theorem 2.}

Under mean stationarity, we can write
\[%
\begin{array}
[c]{rl}%
\Delta y_{i2}= & (\theta_{0,N}-1)u_{i1}+u_{i2}\\
\Delta y_{i3}= & \theta_{0,N}(\theta_{0,N}-1)u_{i1}+(\theta_{0,N}%
-1)u_{i2}+u_{i3}\\
\Delta y_{i4}= & \theta_{0,N}^{2}(\theta_{0,N}-1)u_{i1}+\theta_{0,N}%
(\theta_{0,N}-1)u_{i2}+(\theta_{0,N}-1)u_{i3}+u_{i4}\\
\Delta y_{i5}= & \theta_{0,N}^{3}(\theta_{0,N}-1)u_{i1}+\theta_{0,N}%
^{2}(\theta_{0,N}-1)u_{i2}+\theta_{0,N}(\theta_{0,N}-1)u_{i3}+(\theta
_{0,N}-1)u_{i4}+u_{i5}\\
y_{i3}-y_{i1}= & (1+\theta_{0,N})(\theta_{0,N}-1)u_{i1}+\theta_{0,N}%
u_{i2}+u_{i3}\\
y_{i4}-y_{i1}= & (1+\theta_{0,N}+\theta_{0,N}^{2})(\theta_{0,N}-1)u_{i1}%
+\theta_{0,N}^{2}u_{i2}+\theta_{0,N}u_{i3}+u_{i4}\\
y_{i4}-y_{i2}= & (\theta_{0,N}+\theta_{0,N}^{2})(\theta_{0,N}-1)u_{i1}%
+(\theta_{0,N}^{2}-1)u_{i2}+\theta_{0,N}u_{i3}+u_{i4}\\
y_{i5}-y_{i1}= & (1+\theta_{0,N}+\theta_{0,N}^{2}+\theta_{0,N}^{3}%
)(\theta_{0,N}-1)u_{i1}+\theta_{0,N}^{3}u_{i2}+\theta_{0,N}^{2}u_{i3}%
+\theta_{0,N}u_{i4}+u_{i5}\\
y_{i5}-y_{i2}= & (\theta_{0,N}+\theta_{0,N}^{2}+\theta_{0,N}^{3})(\theta
_{0,N}-1)u_{i1}+(\theta_{0,N}^{3}-1)u_{i2}+\theta_{0,N}^{2}u_{i3}+\theta
_{0,N}u_{i4}+u_{i5}.
\end{array}
\]

The robust sample moments consist of products of the above expressions. To
obtain the probability limits in Theorem 2 of the elements comprising the
robust sample moments, we use that
\[%
\begin{array}
[c]{lc}%
\frac{1}{N}\sum_{i=1}^{N}(\theta_{0,N}-1)u_{it}^{2} & \underset{p}%
{\rightarrow}0,\\
\frac{1}{N}\sum_{i=1}^{N}(\theta_{0,N}-1)u_{it}u_{is} & \underset
{p}{\rightarrow}0,
\end{array}
\]
for all $s$ and $t$, $t>1,$ $t\neq s,$ which is implied by Assumption 1.
Therefore, the $a,$ $b$ and $d$ components of the robust sample moments
simplify to:

\paragraph{T=4, Sys:}%

\begin{align*}
a  &  =\frac{1}{N}\sum_{i=1}^{N}\binom{(\Delta y_{i2})^{2}}{0}=\frac{1}{N}%
\sum_{i=1}^{N}\binom{(\theta_{0,N}-1)^{2}u_{i1}^{2}+u_{i2}^{2}}{0}%
+O_{p}(N^{-1/2}),\\
b  &  =-\frac{1}{N}\sum_{i=1}^{N}\binom{(y_{i3}-y_{i1})^{2}}{\Delta
y_{i2}\Delta y_{i3}}\\
&  =-\frac{1}{N}\sum_{i=1}^{N}\binom{((1+\theta_{0,N})^{2}(\theta_{0,N}%
-1)^{2}u_{i1}^{2}+\theta_{0,N}^{2}u_{i2}^{2}+u_{i3}^{2}}{\theta_{0,N}%
(\theta_{0,N}-1)^{2}u_{i1}^{2}+(\theta_{0,N}-1)u_{i2}^{2}}+O_{p}(N^{-1/2}),\\
d  &  =\frac{1}{N}\sum_{i=1}^{N}\binom{(y_{i4}-y_{i1})\Delta y_{i3}}{\Delta
y_{i2}\Delta y_{i4}}\\
&  =\frac{1}{N}\sum_{i=1}^{N}\binom{\theta_{0,N}(1+\theta_{0}+\theta_{0,N}%
^{2})(\theta_{0,N}-1)^{2}u_{i1}^{2}+\theta_{0,N}^{2}(\theta_{0,N}-1)u_{i2}%
^{2}+\theta_{0,N}u_{i3}^{2}}{\theta_{0,N}^{2}(\theta_{0,N}-1)^{2}u_{i1}%
^{2}+\theta_{0,N}(\theta_{0,N}-1)u_{i2}^{2}}+O_{p}(N^{-1/2}),
\end{align*}
where the $O_{p}(N^{-1/2})$ remainder terms result from the interaction terms
between the different errors, like, for example $\frac{1}{N}\sum_{i=1}%
^{N}u_{i2}u_{i3},$ which converge at rate $N^{-\frac{1}{2}},$ since their
correlation equals zero.

Using next that, because of Assumption 1c, $\frac{1}{N}\sum_{i=1}^{N}%
(1-\theta_{0,N})^{2}u_{i1}^{2}\underset{p}{\rightarrow}0,$ and $\theta
_{0,N}=1+\frac{l}{N^{\tau}},$ with $l$ a fixed constant,\textit{\ }$l<0,$ we
have that%
\[%
\begin{array}
[c]{cl}%
a= & \binom{\sigma_{2}^{2}}{0}+O_{p}(N^{-1/2})\\
b= & -\binom{(1+2\frac{l}{N^{\tau}}+\frac{l^{2}}{N^{2\tau}})\sigma_{2}%
^{2}+\sigma_{3}^{2}}{\frac{l}{N^{\tau}}\sigma_{2}^{2}}+O_{p}(N^{-1/2})\\
d= & \binom{(\frac{l}{N^{\tau}}+2\frac{l^{2}}{N^{2\tau}}+\frac{l^{3}}%
{N^{3\tau}})^{2}\sigma_{2}^{2}+(1+\frac{l}{N^{\tau}})\sigma_{3}^{2}}{(\frac
{l}{N^{\tau}}+\frac{l^{2}}{N^{2\tau}})\sigma_{2}^{2}}+O_{p}(N^{-1/2}),
\end{array}
\]
so, if $\tau>\frac{1}{2},$
\[%
\begin{array}
[c]{cl}%
a= & \binom{\sigma_{2}^{2}}{0}+O_{p}(N^{-1/2})\\
b= & -\binom{\sigma_{2}^{2}+\sigma_{3}^{2}}{0}+O_{p}(N^{-1/2})\\
d= & \binom{\sigma_{3}^{2}}{0}+O_{p}(N^{-1/2}).
\end{array}
\]

\paragraph{T=4, AS:}%

\begin{align*}
a  &  =\frac{1}{N}\sum_{i=1}^{N}\binom{(y_{i3}-y_{i1})\Delta y_{i2}}{0}%
=\frac{1}{N}\sum_{i=1}^{N}\binom{(1+\theta_{0,N})(1-\theta_{0,N})^{2}%
u_{i1}^{2}+\theta_{0,N}u_{i2}^{2}}{0}+O_{p}(N^{-1/2}),\\
b  &  =-\frac{1}{N}\sum_{i=1}^{N}\binom{(y_{i3}-y_{i1})\Delta y_{i3}%
+(y_{i4}-y_{i1})\Delta y_{i2}}{\Delta y_{i2}\Delta y_{i3}}\\
&  =-\frac{1}{N}\sum_{i=1}^{N}\binom{(1-\theta_{0,N})^{2}[(1+2\theta
_{0,N}(1+\theta_{0,N})]u_{i,1}^{2}+(2\theta_{0,N}^{2}-\theta_{0,N})u_{i,2}%
^{2}+u_{i,3}^{2}}{\theta_{0,N}(\theta_{0,N}-1)^{2}u_{i1}^{2}+(\theta
_{0,N}-1)u_{i2}^{2}}+O_{p}(N^{-1/2}),\\
d  &  =\frac{1}{N}\sum_{i=1}^{N}\binom{(y_{i4}-y_{i1})\Delta y_{i3}}{\Delta
y_{i2}\Delta y_{i4}}\\
&  =\frac{1}{N}\sum_{i=1}^{N}\binom{\theta_{0,N}(1+\theta_{0}+\theta_{0,N}%
^{2})(\theta_{0,N}-1)^{2}u_{i1}^{2}+\theta_{0,N}^{2}(\theta_{0,N}-1)u_{i2}%
^{2}+\theta_{0,N}u_{i3}^{2}}{\theta_{0,N}^{2}(\theta_{0,N}-1)^{2}u_{i1}%
^{2}+\theta_{0,N}(\theta_{0,N}-1)u_{i2}^{2}}+O_{p}(N^{-1/2}),
\end{align*}
so also,
\[%
\begin{array}
[c]{cl}%
a= & \binom{\sigma_{2}^{2}}{0}+O_{p}(N^{-1/2})\\
b= & -\binom{\sigma_{2}^{2}+\sigma_{3}^{2}}{0}+O_{p}(N^{-1/2})\\
d= & \binom{\sigma_{3}^{2}}{0}+O_{p}(N^{-1/2}).
\end{array}
\]
We use similar calculations for $T=5$ to obtain that:

\paragraph{T=5, Sys:}%

\begin{align*}
a  &  =\frac{1}{N}\sum_{i=1}^{N}\left(
\begin{array}
[c]{c}%
(\Delta y_{i2})^{2}\\
(y_{i3}-y_{i1})\Delta y_{i3}\\
(\Delta y_{i3})^{2}\\
0\\
0
\end{array}
\right)  =\left(
\begin{array}
[c]{c}%
\sigma_{2}^{2}\\
\sigma_{3}^{2}\\
\sigma_{3}^{2}\\
0\\
0
\end{array}
\right)  +O_{p}(N^{-\frac{1}{2}}),\\
b  &  =-\frac{1}{N}\sum_{i=1}^{N}\left(
\begin{array}
[c]{c}%
(y_{i3}-y_{i1})^{2}\\
(y_{i4}-y_{i1})(y_{i4}-y_{i2})\\
(y_{i4}-y_{i2})^{2}\\
\Delta y_{i2}\Delta y_{i4}\\
\Delta y_{i3}\Delta y_{i4}%
\end{array}
\right)  =-\left(
\begin{array}
[c]{c}%
\sigma_{2}^{2}+\sigma_{3}^{2}\\
\sigma_{3}^{2}+\sigma_{4}^{2}\\
\sigma_{3}^{2}+\sigma_{4}^{2}\\
0\\
0
\end{array}
\right)  +O_{p}(N^{-\frac{1}{2}}),\\
d  &  =\frac{1}{N}\sum_{i=1}^{N}\left(
\begin{array}
[c]{c}%
(y_{i4}-y_{i1})\Delta y_{i3}\\
(y_{i5}-y_{i1})\Delta y_{i4}\\
(y_{i5}-y_{i2})\Delta y_{i4}\\
\Delta y_{i2}\Delta y_{i5}\\
\Delta y_{i3}\Delta y_{i5}%
\end{array}
\right)  =\left(
\begin{array}
[c]{c}%
\sigma_{3}^{2}\\
\sigma_{4}^{2}\\
\sigma_{4}^{2}\\
0\\
0
\end{array}
\right)  +O_{p}(N^{-\frac{1}{2}}).
\end{align*}

\paragraph{T=5, AS:}%

\begin{align*}
a  &  =\frac{1}{N}\sum_{i=1}^{N}\left(
\begin{array}
[c]{c}%
(y_{i3}-y_{i1})\Delta y_{i2}\\
(y_{i4}-y_{i1})\Delta y_{i3}\\
(y_{i4}-y_{i2})\Delta y_{i3}\\
0\\
0
\end{array}
\right)  =\left(
\begin{array}
[c]{c}%
\sigma_{2}^{2}\\
\sigma_{3}^{2}\\
\sigma_{3}^{2}\\
0\\
0
\end{array}
\right)  +O_{p}(N^{-\frac{1}{2}}),\\
b  &  =-\frac{1}{N}\sum_{i=1}^{N}\left(
\begin{array}
[c]{c}%
(y_{i4}-y_{i1})\Delta y_{i2}+(y_{i3}-y_{i1})\Delta y_{i3}\\
(y_{i4}-y_{i1})\Delta y_{i4}+(y_{i5}-y_{i1})\Delta y_{i3}\\
(y_{i4}-y_{i2})\Delta y_{i4}+(y_{i5}-y_{i2})\Delta y_{i3}\\
\Delta y_{i2}\Delta y_{i4}\\
\Delta y_{i3}\Delta y_{i4}%
\end{array}
\right)  =-\left(
\begin{array}
[c]{c}%
\sigma_{2}^{2}+\sigma_{3}^{2}\\
\sigma_{3}^{2}+\sigma_{4}^{2}\\
\sigma_{3}^{2}+\sigma_{4}^{2}\\
0\\
0
\end{array}
\right)  +O_{p}(N^{-\frac{1}{2}}),\\
d  &  =\frac{1}{N}\sum_{i=1}^{N}\left(
\begin{array}
[c]{c}%
(y_{i4}-y_{i1})\Delta y_{i3}\\
(y_{i5}-y_{i1})\Delta y_{i4}\\
(y_{i5}-y_{i2})\Delta y_{i4}\\
\Delta y_{i2}\Delta y_{i5}\\
\Delta y_{i3}\Delta y_{i5}%
\end{array}
\right)  =\left(
\begin{array}
[c]{c}%
\sigma_{3}^{2}\\
\sigma_{4}^{2}\\
\sigma_{4}^{2}\\
0\\
0
\end{array}
\right)  +O_{p}(N^{-\frac{1}{2}}).
\end{align*}

\paragraph{Proof of Theorem 3.}

The proof of Theorem 3 establishes the probability limits of $a,$ $b$ and $d$
for $\theta_{0,N}=1+\frac{l}{N^{\tau}},$ $l<0,$ and $\tau>\frac{1}{2}.$
Denoting these probability limits by, $a_{p},$ $b_{p}$ and $d_{p},$ the large
sample behavior of $a,$ $b,$ and $d$ is characterized by, for $\theta
_{0,N}=1+\frac{l}{N^{\tau}}$ with $\tau>\frac{1}{2}:$%
\[%
\begin{array}
[c]{c}%
\sqrt{N}(a-a_{p})\underset{d}{\rightarrow}\varepsilon_{a},\text{ }\sqrt
{N}(b-b_{p})\underset{d}{\rightarrow}\varepsilon_{b},\text{ }\sqrt{N}%
(d-d_{p})\underset{d}{\rightarrow}\varepsilon_{d},
\end{array}
\]
with $(\varepsilon_{a},$ $\varepsilon_{b},$ $\varepsilon_{d})$ jointly normal,
mean zero random variables, which follows straightforwardly from an
appropriate CLT applied to the highest order remainder terms in the proof of
Theorem 2 which are all sample averages over iid mean zero random variables.
We want to determine the appropriate rate for $\xi$ in $g_{f,T}(\theta(e)), $
so we can analyze its behavior in a neighborhood of the true value
$\theta_{0,N}=1+\frac{l}{N^{\tau}},$ \textit{\ }$l<0,$ with $\tau>\frac{1}{2}$
while $N$ goes to infinity, with
\[%
\begin{array}
[c]{c}%
\theta(e)=1+\frac{e}{N^{\xi}}.
\end{array}
\]
Substituting $\theta(e)$ and the above large sample characterizations of $a, $
$b$ and $d$ in (\ref{quadspec}), we can write$:$%
\[%
\begin{array}
[c]{c}%
g_{f,T}(\theta(e))=(1+\frac{e}{N^{\xi}})^{2}(a_{p}+\frac{\varepsilon_{a}%
}{\sqrt{N}})+(1+\frac{e}{N^{\xi}})(b_{p}+\frac{\varepsilon_{b}}{\sqrt{N}%
})+d_{p}+\frac{\varepsilon_{d}}{\sqrt{N}}+o_{p}(N^{-1/2}).
\end{array}
\]
To determine $\xi$ we impose two conditions: (1) $\sqrt{N}g_{f,T}(\theta(e))$
converges to a non-degenerate bounded random variable of order $O_{p}(1)$; (2)
$g_{f,T}(\theta(e))$ is informative about the value of $e$ when $N$ gets
large. We discriminate between two different cases for $\sigma_{t}^{2}:$

\textbf{1. }For $\sigma_{t}^{2}=\sigma^{2},$ $t=2,\ldots,T:$%
\begin{align*}
&
\begin{array}
[c]{c}%
g_{f,T}(\theta(e))=
\end{array}
\\
&
\begin{array}
[c]{lc}%
(1+\frac{e}{N^{\xi}})^{2}(a_{p}+\frac{\varepsilon_{a}}{\sqrt{N}})+(1+\frac
{e}{N^{\xi}})(b_{p}+\frac{\varepsilon_{b}}{\sqrt{N}})+d_{p}+\frac
{\varepsilon_{d}}{\sqrt{N}}+o_{p}(N^{-1/2}) & =\\
a_{p}+b_{p}+d_{p}+\frac{1}{\sqrt{N}}(\varepsilon_{a}+\varepsilon
_{b}+\varepsilon_{d})+\left(  \frac{e}{N^{\xi}}\right)  ^{2}a_{p}+ & \\
\frac{e}{N^{\xi}}(b_{p}+2a_{p})+\frac{e}{N^{\xi}\sqrt{N}}(\varepsilon
_{b}+2\varepsilon_{a})+\frac{e^{2}}{N^{2\xi}N^{1/2}}\varepsilon_{a}%
+o_{p}(N^{-1/2}) &
\end{array}
\end{align*}
since $a_{p}+b_{p}+d_{p}=0$ and $b_{p}+2a_{p}=0,$ we distinguish three settings:

\begin{description}
\item[$\xi<1/4:$]
\[%
\begin{array}
[c]{c}%
g_{f,T}(\theta(e))=\frac{e^{2}}{N^{2\xi}}a_{p}+o_{p}(N^{-2\xi})
\end{array}
\]

\item[$\xi=1/4:$]
\[%
\begin{array}
[c]{rl}%
g_{f,T}(\theta(e))= & \frac{1}{\sqrt{N}}(\varepsilon_{a}+\varepsilon
_{b}+\varepsilon_{d}+e^{2}a_{p})+\frac{e}{\sqrt{N}\sqrt[4]{N}}(\varepsilon
_{b}+2\varepsilon_{a})+\frac{e^{2}\varepsilon_{a}}{N}+o_{p}(N^{-1/2})\\
= & \frac{1}{\sqrt{N}}(\varepsilon_{a}+\varepsilon_{b}+\varepsilon_{d}%
+e^{2}a_{p})+o_{p}(N^{-1/2})
\end{array}
\]

\item[$\xi>1/4:$]
\[%
\begin{array}
[c]{c}%
g_{f,T}(\theta(e))=\frac{1}{\sqrt{N}}(\varepsilon_{a}+\varepsilon
_{b}+\varepsilon_{d})+o_{p}(N^{-1/2}).
\end{array}
\]

\end{description}

\noindent This shows that the appropriate rate corresponds with $\xi=1/4$. For
a smaller value of $\xi,$ $\sqrt{N}g_{f,T}(\theta(e))$ diverges. For a larger
value, $\sqrt{N}g_{f,T}(\theta(e))$ converges to a mean zero normal random
variable unaffected by the choice of $e$. Although in this case $\sqrt
{N}g_{f,T}(\theta(e))$ is not informative about $e$, we do not need to worry
about $e$ because standard asymptotics apply$.$

\textbf{2. }When $\sigma_{t}^{2}\neq\sigma_{s}^{2},$ for at least one $t\neq
s,$ $a_{p}+b_{p}+d_{p}=0$ but $b_{p}+2a_{p}\neq0,$ we can establish along the
lines of the above that the appropriate rate corresponds with $\xi=1/2:$%
\begin{align*}
&
\begin{array}
[c]{c}%
g_{f,T}(\theta(e))=
\end{array}
\\
&
\begin{array}
[c]{lc}%
(1+\frac{e}{\sqrt{N}})^{2}(a_{p}+\frac{\varepsilon_{a}}{\sqrt{N}})+(1+\frac
{e}{\sqrt{N}})(b_{p}+\frac{\varepsilon_{b}}{\sqrt{N}})+d_{p}+\frac
{\varepsilon_{d}}{\sqrt{N}}+o_{p}(N^{-1/2}) & =\\
a_{p}+b_{p}+d_{p}+\frac{1}{\sqrt{N}}(\varepsilon_{a}+\varepsilon
_{b}+\varepsilon_{d}+e(b_{p}+a_{p}))+ & \\
\frac{e}{N}(2\varepsilon_{a}+\varepsilon_{b}+eE(a))+\frac{e^{2}\varepsilon
_{a}}{N\sqrt{N}}+o_{p}(N^{-1/2}) & =\\
\frac{1}{\sqrt{N}}(\varepsilon_{a}+\varepsilon_{b}+\varepsilon_{d}%
+e(b_{p}+2a_{p}))+\frac{e}{N}(2\varepsilon_{a}+\varepsilon_{b}+ea_{p}%
)+\frac{e^{2}\varepsilon_{a}}{N\sqrt{N}}+o_{p}(N^{-1/2}) & =\\
\frac{1}{\sqrt{N}}(\varepsilon_{a}+\varepsilon_{b}+\varepsilon_{d}%
+e(b_{p}+2a_{p}))+o_{p}(N^{-1/2}). &
\end{array}
\end{align*}

\paragraph{Proof of Theorem 4.}

Denote with $g_{f,T}(\theta(e))$ the moments in (\ref{quadspec}) evaluated at
$\theta(e)=1+\frac{e}{\sqrt[4]{N}}$. When $\sigma_{t}^{2}=\sigma^{2}$ and
substituting the large sample characterization of $a$, $b$ and $d,$ $\sqrt
{N}g_{f,T}(\theta(e))$ can be expressed as:%
\[%
\begin{array}
[c]{c}%
\sqrt{N}g_{f,T}(\theta(e))=e^{2}a_{p}+\varepsilon_{a}(1+\frac{2e}{\sqrt[4]{N}%
}+\frac{e^{2}}{\sqrt{N}})+\varepsilon_{b}(1+\frac{e}{\sqrt[4]{N}}%
)+\varepsilon_{d}+o_{p}(1).
\end{array}
\]
Define%
\[%
\begin{array}
[c]{c}%
\phi(N)=e^{2}a_{p}+\varepsilon_{a}\left(  1+\frac{2e}{\sqrt[4]{N}}+\frac
{e^{2}}{\sqrt{N}}\right)  +\varepsilon_{b}\left(  1+\frac{e}{\sqrt[4]{N}%
}\right)  +\varepsilon_{d}.
\end{array}
\]
Since ($\varepsilon_{a},$ $\varepsilon_{b},$ $\varepsilon_{d}$) are jointly
normal distributed,
\[
\phi(N)\sim N(e^{2}a_{p},\ B(N)^{\prime}V_{abd}B(N))
\]
with
\[%
\begin{array}
[c]{c}%
B(N)=(\iota_{3}\otimes I_{p_{\max}})+\frac{e}{\sqrt[4]{N}}\left[  (2+\frac
{e}{\sqrt[4]{N}})(e_{1,3}\otimes I_{p_{\max}})+(e_{2,3}\otimes I_{p_{\max}%
})\right]  ,
\end{array}
\]
and $V_{abd}$ the covariance matrix of $(\varepsilon_{a}^{\prime}$ $\vdots$
$\varepsilon_{b}^{\prime}$ $\vdots$ $\varepsilon_{d}^{\prime})^{\prime},$
$\iota_{3}$ a $3\times1$ dimensional vector of ones, $I_{p_{\max}}$ the
$p_{\max}\times p_{\max}$ dimensional identity matrix, $p_{\max}$ equals the
number of elements of $a$ and $e_{1,3}$ and $e_{2,3}$ the first and second
$3\times1$ dimensional unity vectors.

Hence,
\[
\sqrt{N}g_{f,T}(\theta(e))=\phi(N)+o_{p}(1),
\]
so in a sample of size $N,$ $\sqrt{N}g_{f,T}(\theta(e))$ is normally
distributed up to a $o_{p}(1)$ term. While some of the components in $\phi(N)$
are essentially also $o_{p}(1),$ it is important to incorporate them for an
accurate approximation of the distribution of $\sqrt{N}g_{f,T}(\theta(e))$ for
a given sample of size $N$ since the low order components, of order $N^{-1/4}%
$, converge very slowly to zero.

The individual moments $g_{f,n}(\theta(e))$ in the sample average
$g_{f,T}(\theta(e))=\frac{1}{N}\sum\limits_{n=1}^{N}g_{f,n}(\theta(e))$ can be
specified as:%
\[%
\begin{array}
[c]{rl}%
g_{f,n}(\theta(e))= & (1+\frac{e}{\sqrt[4]{N}})^{2}a_{n}+(1+\frac{e}%
{\sqrt[4]{N}})b_{n}+d_{n}\\
= & (1+\frac{e}{\sqrt[4]{N}})^{2}[a_{p}+\varepsilon_{a_{n}}]+(1+\frac
{e}{\sqrt[4]{N}})[b_{p}+\varepsilon_{b_{n}}]+[d_{p}+\varepsilon_{d_{n}}]\\
= & (a_{p}+b_{p}+d_{p})+\frac{e}{\sqrt[4]{N}}(2a_{p}+b_{p})+\frac{e^{2}}%
{\sqrt{N}}a_{p}+\\
& \varepsilon_{a_{n}}+\varepsilon_{b_{n}}+\varepsilon_{d_{n}}+\frac
{e}{\sqrt[4]{N}}(2\varepsilon_{a_{n}}+\varepsilon_{b_{n}})+\frac{e^{2}}%
{\sqrt{N}}\varepsilon_{a_{n}}\\
= & \frac{e^{2}}{\sqrt{N}}a_{p}+\varepsilon_{a_{n}}+\varepsilon_{b_{n}%
}+\varepsilon_{d_{n}}+\frac{e}{\sqrt[4]{N}}(2\varepsilon_{a_{n}}%
+\varepsilon_{b_{n}})+\frac{e^{2}}{\sqrt{N}}\varepsilon_{a_{n}},
\end{array}
\]
with $a=\frac{1}{N}\sum\limits_{n=1}^{N}a_{n},$ $b=\frac{1}{N}\sum
\limits_{n=1}^{N}b_{n},$ $d=\frac{1}{N}\sum\limits_{n=1}^{N}d_{n},$
$\varepsilon_{a_{n}}=a_{n}-a_{p},$ $\varepsilon_{b_{n}}=b_{n}-b_{p},$
$\varepsilon_{d_{n}}=d_{n}-d_{p},$ so taking $g_{f,n}(\theta(e))$ in deviation
from its sample average $g_{f,T}(\theta(e))$ results in
\[%
\begin{array}
[c]{rl}%
g_{f,n}(\theta(e))-g_{f,T}(\theta(e))= & \varepsilon_{a_{n}}-\varepsilon
_{a}+\varepsilon_{b_{n}}-\varepsilon_{b}+\varepsilon_{d_{n}}-\varepsilon
_{d}+\\
& \frac{e}{\sqrt[4]{N}}(2(\varepsilon_{a_{n}}-\varepsilon_{a})+\varepsilon
_{b_{n}}-\varepsilon_{b})+\frac{e^{2}}{\sqrt{N}}(\varepsilon_{a_{n}%
}-\varepsilon_{a})+o_{p}(N^{-1/2})
\end{array}
\]
From the above, it then straightforwardly follows that%
\[%
\begin{array}
[c]{c}%
\hat{V}_{gg}(e)=\frac{1}{N}\sum_{i=1}^{N}\left(  g_{f,n}(\theta(e))-g_{f,T}%
(\theta(e))\right)  \left(  g_{f,n}(\theta(e))-g_{f,T}(\theta(e))\right)
^{\prime}=B(N)^{\prime}V_{abd}B(N)+o_{p}(1),
\end{array}
\]
so the distribution of the GMM-AR statistic testing H$_{p}$ for a sample of
size $N$ is characterized by%
\[
\chi^{2}(\delta(N),p_{\max})+o_{p}(1),
\]
with $\delta(N)=e^{4}a_{p}^{\prime}\left[  B(N)^{\prime}V_{abd}B(N)\right]
^{-1}a_{p}.$

\paragraph{Proof of Theorem 5.}

When we instead of the full vector $g_{f,T}(\theta(e))$ use a linear
combination of it, say $w^{\prime}g_{f,T}(\theta(e))$ with $w$ an orthonormal
$p_{\max}\times1$ vector, the approximating distribution of the GMM-AR
statistic for testing H$_{p}:\theta(e)=1+\frac{e}{\sqrt[4]{N}}$ that uses
$w^{\prime}g_{f,T}(\theta(e))$ as the moment vector reads%
\[%
\begin{array}
[c]{c}%
\chi^{2}(e^{4}(w^{\prime}a_{p})^{\prime}\left[  w^{\prime}B(N)^{\prime}%
V_{abd}B(N)w\right]  ^{-1}(w^{\prime}a_{p}),1).
\end{array}
\]
The optimal combination $w$ is the one that leads to the largest value of the
non-centrality parameter. The non-centrality parameter can be specified as%
\[%
\begin{array}
[c]{lc}%
e^{4}(w^{\prime}a_{p})^{\prime}\left[  w^{\prime}B(N)^{\prime}V_{abd}%
B(N)w\right]  ^{-1}(w^{\prime}a_{p}) & =e^{4}\frac{(w^{\prime}a_{p})^{2}%
}{w^{\prime}B(N)^{\prime}V_{abd}B(N)w}.
\end{array}
\]
The maximal value of $\frac{(w^{\prime}a_{p})^{2}}{w^{\prime}B(N)^{\prime
}V_{abd}B(N)w}$ results from the largest root of the generalized eigenvalue
problem%
\[
\left\vert \lambda B(N)^{\prime}V_{abd}B(N)-a_{p}a_{p}^{\prime}\right\vert =0
\]
and the optimal value of $w$ equals the eigenvector associated with the
largest root. Since $a_{p}$ is only a vector, just one root of the generalized
eigenvalue problem is non-zero so it is also the largest one. This root
results from using%
\[%
\begin{array}
[c]{cc}%
w= & (B(N)^{\prime}V_{abd}B(N))^{-1}a_{p}%
\end{array}
\]
and the largest root then equals
\[%
\begin{array}
[c]{cc}%
\lambda_{\max}= & a_{p}^{\prime}(B(N)^{\prime}V_{abd}B(N))^{-1}a_{p}%
\end{array}
\]
so the maximal value of the non-centrality parameter is%
\[%
\begin{array}
[c]{c}%
\delta(N)=e^{4}a_{p}^{\prime}(B(N)^{\prime}V_{abd}B(N))^{-1}a_{p}%
=(e\sigma)^{4}\binom{\iota_{p}}{0}^{\prime}(B(N)^{\prime}V_{abd}%
B(N))^{-1}\binom{\iota_{p}}{0}%
\end{array}
\]
since $a_{p}=\sigma^{2}\binom{\iota_{p}}{0}$ with $\iota_{p}$ a $p\times1 $
dimensional vector of ones and $p$ the number of columns of $G_{f,T}(\theta).$

\paragraph{Proof of Theorem 6.}

Before we start out to prove Theorem 6, we first state an addendum to Theorem
1, which incorporates some higher order components of order $O_{p}(N^{-1/2})$
that are needed for some of the intermediate results.

\paragraph{Addendum to Theorem 1: Theorem 1$^{\ast}$ (Representation
Theorem).}

\textit{Under Assumptions 1 and 2a, we can characterize the large sample
behavior of the Dif, Lev, NL, AS and Sys sample moments and their derivatives
by:}%
\[%
\begin{array}
[c]{ll}%
\left(
\begin{array}
[c]{c}%
f_{N}^{j}(\theta)\\
q_{N}^{j}(\theta)
\end{array}
\right)  = & \left(
\begin{array}
[c]{c}%
A_{f}^{j}(\theta)\\
A_{q}^{j}(\theta)
\end{array}
\right)  \left[  \frac{1}{h_{N}(\theta_{0,N})\sqrt{N}}(\psi-h_{N}(\theta
_{0,N})\sigma_{1,n}\iota_{T-1}\psi_{c})+\iota_{T-1}d_{2}\right]  +\\
& \left(
\begin{array}
[c]{c}%
\mu_{f}^{j}(\theta,\bar{\sigma}^{2})\\
\mu_{q}^{j}(\theta,\bar{\sigma}^{2})
\end{array}
\right)  +\frac{1}{\sqrt{N}}\left(
\begin{array}
[c]{c}%
B_{f}^{j}(\theta)\\
B_{q}^{j}(\theta)
\end{array}
\right)  \psi_{uu}+o_{p}(N^{-1/2}),
\end{array}
\]
\textit{with }$j=Dif,$ $Lev,$ $NL,$ $AS,$ $Sys$ \textit{and }$B_{f}^{j}%
(\theta),$ $B_{q}^{j}(\theta):$ $k_{j}\times m_{j}$ \textit{and} $k_{j}\times
m_{j},$ $k_{j}\times1$ \textit{dimensional matrices and }$\psi_{uu}%
$\textit{\ is a mean zero, finite variance, normal random vector that is
possibly dependent on }$\psi.$

\paragraph{Proof of large sample distribution KLM\ statistic.}

For the construction of the large sample distribution of the KLM\ statistic
under Assumptions 1 and 2a, we use that the part of the sample moments spanned
by $A_{f}^{j}(\theta(e))$ and the part spanned by $A_{f}^{j}(\theta
(e))_{\perp}$ converge at different rates. We use the normalized large sample
behavior of each of these parts to construct it. This amounts to
pre-multiplying the sample moments in the expression of the KLM statistic by
$(A_{f}^{j}(\theta(e))$ $\vdots$ $A_{f}^{j}(\theta(e))_{\perp})$ to which it
is invariant if $(A_{f}^{j}(\theta(e))$ $\vdots$ $A_{f}^{j}(\theta(e))_{\perp
})$ is invertible. The specification of $A_{f}^{j}(\theta(e))_{\perp}$ as
equal to $(G_{f,T}^{j}(\theta(e))$ $\vdots$ $G_{2,T}^{j})$, see (\ref{OCdec}),
is such that $(A_{f}^{j}(\theta(e))$ $\vdots$ $A_{f}^{j}(\theta(e))_{\perp})$
is invertible for the Sys moment conditions but not for the AS moment
conditions both when $T=4$ and $5 $ since $A_{f}^{j}(\theta(e))$ does not have
full column rank. To have an invertible specification of $(A_{f}^{j}%
(\theta(e))$ $\vdots$ $A_{f}^{j}(\theta(e))_{\perp}),$ we use that we can
specify $A_{f}^{j}(\theta(e))$ for the AS\ moments as:%
\[%
\begin{array}
[c]{crl}%
\mathbf{T=4:} & \mathbf{\ }A_{f}^{AS}(\theta)= & \left(
\begin{array}
[c]{ccc}%
-\theta & 1 & 0\\
0 & -\theta & 1\\
0 & -\theta & 1\\
\theta(\theta-1) & 1-\theta & 0
\end{array}
\right) \\
& = & A_{f,T=4}^{AS}(\theta)_{1}A_{f,T=4}^{AS}(\theta)_{2}\\
\mathbf{T=5:} & A_{f}^{AS}(\theta)= & \left(
\begin{array}
[c]{cccc}%
-\theta & 1 & 0 & 0\\
0 & -\theta & 1 & 0\\
0 & -\theta & 1 & 0\\
0 & 0 & -\theta & 1\\
0 & 0 & -\theta & 1\\
0 & 0 & -\theta & 1\\
\theta(\theta-1) & 1-\theta & 0 & 0\\
0 & \theta(\theta-1) & 1-\theta & 0
\end{array}
\right) \\
& = & A_{f,T=5}^{AS}(\theta)_{1}A_{f,T=5}^{AS}(\theta)_{2}%
\end{array}
\]
where
\[%
\begin{array}
[c]{cll}%
\mathbf{T=4:} & A_{f,T=4}^{AS}(\theta)_{1}=\left(
\begin{array}
[c]{cc}%
-\theta & 0\\
0 & 1\\
0 & 1\\
\theta(\theta-1) & 0
\end{array}
\right)  ,\mathbf{\ } & A_{f,T=4}^{AS}(\theta)_{2},\left(
\begin{array}
[c]{ccc}%
1 & -\theta^{-1} & 0\\
0 & -\theta & 1
\end{array}
\right) \\
\mathbf{T=5:} & A_{f,T=5}^{AS}(\theta)_{1}=\left(
\begin{array}
[c]{ccc}%
-\theta & 1 & 0\\
0 & -\theta & 0\\
0 & -\theta & 0\\
0 & 0 & 1\\
0 & 0 & 1\\
0 & 0 & 1\\
\theta(\theta-1) & 1-\theta & 0\\
0 & \theta(\theta-1) & 0
\end{array}
\right)  & A_{f,T=5}^{AS}(\theta)_{2}=\left(
\begin{array}
[c]{cccc}%
1 & 0 & -\theta^{-2} & 0\\
0 & 1 & -\theta^{-1} & 0\\
0 & 0 & -\theta & 1
\end{array}
\right)
\end{array}
\]
so unlike $A_{f}^{AS}(\theta),$ $A_{f}^{AS}(\theta)_{1}$ has full column rank.
For the Sys moments, for which $A_{f}^{Sys}(\theta)$ has full column rank, we
use $A_{f}^{Sys}(\theta)_{1}=A_{f}^{Sys}(\theta).$ The matrix $(A_{f}%
^{j}(\theta(e))_{1}$ $\vdots$ $A_{f}^{j}(\theta(e))_{\perp})$ is now
invertible for both $j=AS,$ $Sys,$ so we use it to construct the large sample
behavior of the KLM\ statistic to test H$_{p}:\theta(e)=1+\frac{e}{\sqrt[4]%
{N}}$ whilst the true value of $\theta$ is drifting to one in line with
Assumption 2a. We separately construct the behavior of the following four components:

\begin{enumerate}
\item[\textbf{1.}] $\sqrt{N}\hat{V}_{ff}(\theta(e))^{-1}f_{N}(\theta(e))$

\item[\textbf{2.}] $q_{N}(\theta(e))$

\item[\textbf{3.}] $\hat{V}_{\theta f}(\theta(e))$

\item[\textbf{4.}] $\hat{D}_{N}(\theta(e))$
\end{enumerate}

\noindent which provide the building blocks for the large sample distribution
of the KLM statistic. For each of these components, we determine their limit
behavior when multiplied by $(h_{N}(\theta_{0,N})A_{f}(\theta(e))_{1}$
$\vdots$ $A_{f}(e)_{\perp})$ for the last three components and its inverse for
the first one. Taken all together this implies that $(h_{N}(\theta_{0,N}%
)A_{f}(\theta(e))_{1}$ $\vdots$ $A_{f}(e)_{\perp})$ cancels out of the overall
expression of the KLM statistic.

\textbf{1. }To determine the limit behavior of $\sqrt{N}\hat{V}_{ff}%
(\theta(e))^{-1}f_{N}(\theta(e)),$ we disentangle the components with
different convergence rates which we do by pre-multiplying it by
$(h_{N}(\theta_{0,N})A_{f}(\theta(e))_{1}$ $\vdots$ $A_{f}(e)_{\perp})^{-1}:$
\[%
\begin{array}
[c]{l}%
(h_{N}(\theta_{0,N})A_{f}(\theta(e))_{1}\text{ }\vdots\text{ }A_{f}(e)_{\perp
})^{-1}\sqrt{N}\hat{V}_{ff}(\theta(e))^{-1}f_{N}(\theta(e))=\\
\left[  (h_{N}(\theta_{0,N})A_{f}(\theta(e))_{1}\text{ }\vdots\text{ }%
A_{f}(e)_{\perp})^{\prime}\hat{V}_{ff}(e)(h_{N}(\theta_{0,N})A_{f}%
(e)_{1}\text{ }\vdots\text{ }A_{f}(e)_{\perp})\right]  ^{-1}\\
\left[  \sqrt{N}(h_{N}(\theta_{0,N})A_{f}(e)_{1}\text{ }\vdots\text{ }%
A_{f}(e)_{\perp})^{\prime}f_{N}(e)\right]  .
\end{array}
\]
We next determine the large sample behavior of the different components under
Assumptions 1 and 2a. Our specification of $A_{f}(\theta(e))_{\perp}$ is such
that:
\[
\sqrt{N}A_{f}(\theta(e))_{\perp}^{\prime}f_{N}(\theta(e))=\sqrt{N}%
g_{f,T}(\theta(e)),
\]
so using the large sample behavior of $\sqrt{N}g_{f,T}(\theta(e))$ stated in
the proof of Theorem 4, we have that the large sample behavior of $\sqrt
{N}A_{f}(\theta(e))_{\perp}^{\prime}f_{N}(\theta(e))$ for a (large) sample of
size $N$ results as:
\[%
\begin{array}
[c]{c}%
\sqrt{N}A_{f}(\theta(e))_{\perp}^{\prime}f_{N}(\theta(e))=\left[  e^{2}%
\sigma^{2}\binom{\iota_{p}}{0}+B(N)^{\prime}\left(
\begin{array}
[c]{c}%
\varepsilon_{a}\\
\varepsilon_{b}\\
\varepsilon_{d}%
\end{array}
\right)  \right]  +o_{p}(1).
\end{array}
\]
The large sample behavior of $\sqrt{N}h_{N}(\theta_{0,N})A_{f}(\theta
(e))_{1}^{\prime}f_{N}(\theta(e))$ result from Theorem 1 (the representation
theorem) and accords with, since by Assumption 2a $\sqrt{N}h_{N}(\theta
_{0,N})\rightarrow0,$
\[%
\begin{array}
[c]{cc}%
\sqrt{N}h_{N}(\theta_{0,N})A_{f}(\theta(e))_{1}^{\prime}f_{N}(\theta(e))= &
A_{f}(\theta(e))_{1}^{\prime}A_{f}(\theta(e))\bar{\psi}+o_{p}(1),
\end{array}
\]
where $\bar{\psi}=\psi-h_{N}(\theta_{0,N})\sigma_{1,n}\iota_{T-1}\psi_{c},$ so
upon combining:
\[%
\begin{array}
[c]{cc}%
\left[  \sqrt{N}(h_{N}(\theta_{0,N})A_{f}(\theta(e))_{1}\text{ }\vdots\text{
}A_{f}(\theta(e))_{\perp})^{\prime}f_{N}(\theta(e))\right]  = & \left[
\begin{array}
[c]{c}%
A_{f}(\theta(e))_{1}^{\prime}A_{f}(\theta(e))\bar{\psi}\\
e^{2}\sigma^{2}\binom{\iota_{p}}{0}+B(N)^{\prime}\left(
\begin{array}
[c]{c}%
\varepsilon_{a}\\
\varepsilon_{b}\\
\varepsilon_{d}%
\end{array}
\right)
\end{array}
\right]  +o_{p}(1).
\end{array}
\]
We next focus on the components of $\left[  (h_{N}(\theta_{0,N})A_{f}%
(\theta(e))_{1}\text{ }\vdots\text{ }A_{f}(e)_{\perp})^{\prime}\hat{V}%
_{ff}(e)(h_{N}(\theta_{0,N})A_{f}(e)_{1}\text{ }\vdots\text{ }A_{f}(e)_{\perp
})\right]  .$ Since $g_{f,T}(\theta(e))$ does not depend on the initial
observations $y_{i1},$ the (normalized) covariance of $A_{f}(\theta
(e))_{1}^{\prime}f_{N}(\theta(e))$ and $A_{f}(\theta(e))_{\perp}^{\prime}%
f_{N}(\theta(e))$ equals zero:
\[%
\begin{array}
[c]{cc}%
h_{N}(\theta_{0,N})A_{f}(\theta(e))_{1}^{\prime}\hat{V}_{ff}(\theta
(e))A_{f}(\theta(e))_{\perp}= & o_{p}(1).
\end{array}
\]
Under Assumption 2a also:
\[%
\begin{array}
[c]{ll}%
h_{N}(\theta_{0,N})^{2}A_{f}(\theta(e))_{1}^{\prime}\hat{V}_{ff}%
(\theta(e))A_{f}(\theta(e))_{1} & =A_{f}(\theta(e))_{1}^{\prime}A_{f}%
(\theta(e))\Lambda A_{f}(\theta(e))^{\prime}A_{f}(\theta(e))_{1}+o_{p}(1),\\
A_{f}(\theta(e))_{\perp}^{\prime}\hat{V}_{ff}(e)A_{f}(\theta(e))_{\perp} &
=B(N)^{\prime}V_{abd}B(N)+o_{p}(1),
\end{array}
\]
where
\[%
\begin{array}
[c]{rl}%
\Lambda= & var\left(  \lim_{N\rightarrow\infty}\bar{\psi}\right) \\
= & diag(\bar{\sigma}^{2})+\left[  \lim_{N\rightarrow\infty}\left(
h_{N}(\theta_{0,N})^{2}\sigma_{1,n}^{2}\right)  \right]  \iota_{T-1}%
\iota_{T-1}^{\prime}var(c_{i})
\end{array}
\]
so%
\[%
\begin{array}
[c]{l}%
(h_{N}(\theta_{0,N})A_{f}(\theta(e))_{1}\text{ }\vdots\text{ }A_{f}%
(\theta(e))_{\perp})^{\prime}\hat{V}_{ff}(\theta(e))(h_{N}(\theta_{0,N}%
)A_{f}(\theta(e))_{1}\text{ }\vdots\text{ }A_{f}(\theta(e))_{\perp})=\\
\left(
\begin{array}
[c]{cc}%
A_{f}(\theta(e))_{1}^{\prime}A_{f}(\theta(e))\Lambda A_{f}(\theta(e))^{\prime
}A_{f}(\theta(e))_{1} & 0\\
0 & B(N)^{\prime}V_{abd}B(N)
\end{array}
\right)  +o_{p}(1).
\end{array}
\]
Because $h_{N}(\theta_{0,N})A_{f}(\theta(e))_{1}^{\prime}f_{N}(\theta(e)) $
and $A_{f}(\theta(e))_{\perp}^{\prime}f_{N}(\theta(e))$ are uncorrelated under
Assumption 2a,
\[%
\begin{array}
[c]{c}%
\left[  (h_{N}(\theta_{0,N})A_{f}(\theta(e))_{1}\text{ }\vdots\text{ }%
A_{f}(\theta(e))_{\perp})^{\prime}\hat{V}_{ff}(\theta(e))(h_{N}(\theta
_{0,N})A_{f}(\theta(e))_{1}\text{ }\vdots\text{ }A_{f}(\theta(e))_{\perp
})\right]
\end{array}
\]
convergences to a block diagonal matrix so we obtain the large sample behavior
of $\sqrt{N}((h_{N}(\theta_{0,N})A_{f}(\theta(e))_{1}$ $\vdots$ $A_{f}%
(\theta(e))_{\perp})^{\prime}\hat{V}_{ff}(\theta(e))\allowbreak(h_{N}%
(\theta_{0,N})A_{f}(\theta(e))_{1}$ $\vdots$ $A_{f}(\theta(e))_{\perp}%
))^{-1}(h_{N}(\theta_{0,N})A_{f}(\theta(e))_{1}$ $\vdots$ $A_{f}%
(\theta(e))_{\perp})^{\prime}f_{N}(\theta(e)):$
\[%
\begin{array}
[c]{l}%
\sqrt{N}((h_{N}(\theta_{0,N})A_{f}(\theta(e))_{1}\text{ }\vdots\text{ }%
A_{f}(\theta(e))_{\perp})^{\prime}\hat{V}_{ff}(\theta(e))(h_{N}(\theta
_{0,N})A_{f}(\theta(e))_{1}\text{ }\vdots\text{ }A_{f}(\theta(e))_{\perp
}))^{-1}\times\\
(h_{N}(\theta_{0,N})A_{f}(\theta(e))_{1}\text{ }\vdots\text{ }A_{f}%
(\theta(e))_{\perp})^{\prime}f_{N}(\theta(e))\\
=\left(
\begin{array}
[c]{c}%
\left[  A_{f}(\theta(e))_{1}^{\prime}A_{f}(\theta(e))\Lambda A_{f}%
(\theta(e))^{\prime}A_{f}(\theta(e))_{1}\right]  ^{-1}A_{f}(\theta
(e))_{1}^{\prime}A_{f}(\theta(e))\bar{\psi}\\
\left(  B(N)^{\prime}V_{abd}B(N)\right)  ^{-1}\left(  e^{2}\sigma^{2}%
\binom{\iota_{p}}{0}+B(N)^{\prime}\left(
\begin{array}
[c]{c}%
\varepsilon_{a}\\
\varepsilon_{b}\\
\varepsilon_{d}%
\end{array}
\right)  \right)
\end{array}
\right)  +o_{p}(1).
\end{array}
\]

\textbf{2. }To obtain the large sample behavior of $q_{N}(\theta(e))$ under
Assumptions 1 and 2a, we characterize the behavior of the different components
of
\[%
\begin{array}
[c]{c}%
(h_{N}(\theta_{0,N})A_{f}(\theta(e))_{1}\text{ }\vdots\text{ }A_{f}%
(\theta(e))_{\perp})^{\prime}q_{N}(\theta(e))
\end{array}
\]
for which we use the representation of $q_{N}(\theta(e))$ in Theorem 1 (and
Theorem 1$^{\ast}$).

Under DGPs according with Assumptions 1 and 2a, $\sqrt{N}h_{N}(\theta
_{0,N})A_{f}(\theta(e))_{1}^{\prime}q_{N}(\theta(e))$ is characterized by%
\[%
\begin{array}
[c]{l}%
\sqrt{N}h_{N}(\theta_{0,N})A_{f}(\theta(e))_{1}^{\prime}q_{N}(\theta(e))=\\
A_{f}(\theta(e))_{1}^{\prime}\left[  A_{q}(\theta(e))\bar{\psi}+h_{N}%
(\theta_{0,N})\sqrt{N}(\mu_{q}(\theta(e),\bar{\sigma}^{2})-\right. \\
\left.  A_{q}(\theta(e))\iota_{T-1}d_{2}\right]  +o_{p}(1),
\end{array}
\]
which converges to%
\[
A_{f}(\theta(e))_{1}^{\prime}A_{q}(\theta(e))\bar{\psi}%
\]
since under Assumption 2a:%
\[%
\begin{array}
[c]{c}%
\sqrt{N}h_{N}(\theta_{0,N})(\mu_{q}(\theta(e),\bar{\sigma}^{2})-A_{q}%
(\theta(e)))\iota_{T-1}d_{2}\rightarrow0,
\end{array}
\]
which results from Assumption 1b and $h_{N}(\theta_{0,N})\sqrt{N}%
\rightarrow0.$

Regarding $A_{f}(\theta(e))_{\perp}^{\prime}q_{N}(\theta(e)),$ we distinguish
between the AS and Sys moment conditions. For the Sys moment conditions:%
\[%
\begin{array}
[c]{l}%
\sqrt[4]{N}A_{f}(\theta(e))_{\perp}^{\prime}q_{N}(\theta(e))=\sqrt[4]%
{N}\left(
\begin{array}
[c]{c}%
G_{f,T}(\theta(e))^{\prime}q_{N}(\theta(e))\\
G_{2,T}^{\prime}q_{N}(\theta(e))
\end{array}
\right) \\
=-\left(
\begin{array}
[c]{c}%
e\sigma^{2}\iota_{p}\\
0
\end{array}
\right)  +\frac{1}{\sqrt[4]{N}}\left(
\begin{array}
[c]{c}%
\frac{1}{h_{N}(\theta_{0,N})}G_{f}(\theta(e))^{\prime}A_{q}(\theta
(e))\bar{\psi}+\varepsilon_{aq}\\
\varepsilon_{bq}%
\end{array}
\right)  +o_{p}(N^{-1/4}),
\end{array}
\]
for which we used the representation for $q_{N}(\theta(e))$ that results from
Theorem 1* in the Appendix which includes $B_{q}(\theta)\psi_{uu}$, since for
the Sys moment conditions $G_{2,T}^{\prime}A_{q}(\theta(e))=0,$ $G_{2,T}%
^{\prime}\mu(\theta(e),\bar{\sigma}^{2})=0,$ $G_{f,T}(\theta(e))^{\prime}%
A_{q}(\theta(e))\iota_{T-1}=0,$ $G_{f,T}(\theta(e))^{\prime}\mu(\theta
(e),\bar{\sigma}^{2})=-\frac{e}{\sqrt[4]{N}}\sigma^{2}\iota_{p}$ and
$\varepsilon_{aq}=G_{f}(\theta(e))^{\prime}B_{q}(\theta(e))\psi_{uu}$ and
$\varepsilon_{bq}=G_{2,T}^{\prime}B_{q}(\theta(e))\psi_{uu}$ are mean zero
normal random variables that capture the remaining random parts.

For the AS moment conditions:%
\[%
\begin{array}
[c]{l}%
\sqrt[4]{N}A_{f}(\theta(e))_{\perp}^{\prime}q_{N}(\theta(e))=\\
-\left(
\begin{array}
[c]{c}%
e(2\sigma^{2}-d_{2})\iota_{p}\\
0
\end{array}
\right)  +\frac{1}{\sqrt[4]{N}}\left(
\begin{array}
[c]{c}%
\frac{1}{h_{N}(\theta_{0,N})}G_{f}(\theta(e))^{\prime}A_{q}(\theta
(e))\bar{\psi}+\varepsilon_{aq}\\
\varepsilon_{bq}%
\end{array}
\right)  +o_{p}(N^{-1/4})
\end{array}
\]
since for the AS moment conditions $G_{2,T}^{\prime}A_{q}(\theta(e))=0,$
$G_{2,T}^{\prime}\mu(\theta(e),\bar{\sigma}^{2})=0,$ $G_{f,T}(\theta
(e))^{\prime}A_{q}(\theta(e))\iota_{T-1}=\frac{e}{\sqrt[4]{N}}\iota_{p},$
$G_{f}(\theta(e))^{\prime}\mu(\theta(e),\bar{\sigma}^{2})=-\frac{2e}%
{\sqrt[4]{N}}\sigma^{2}\iota_{p}$ and $\varepsilon_{aq}=G_{f}(\theta
(e))^{\prime}B_{q}(\theta(e))\psi_{cu}$ and $\varepsilon_{bq}=G_{2,T}^{\prime
}B_{q}(\theta(e))\psi_{cu}$ are mean zero normal random variables that capture
the remaining random parts.

Overall, the large sample behavior of $A_{f}(\theta(e))_{\perp}^{\prime}%
q_{N}(\theta(e))$ for both the AS and Sys moment conditions reads:%
\[%
\begin{array}
[c]{l}%
\sqrt[4]{N}A_{f}(\theta(e))_{\perp}^{\prime}q_{N}(\theta(e))=\left[  -\left(
\begin{array}
[c]{c}%
\bar{e}\iota_{p}\\
0
\end{array}
\right)  +\frac{1}{\sqrt[4]{N}}\left(
\begin{array}
[c]{c}%
\frac{1}{h_{N}(\theta_{0,N})}G_{f}(\theta(e))^{\prime}A_{q}(\theta
(e))\bar{\psi}+\varepsilon_{aq}\\
\varepsilon_{bq}%
\end{array}
\right)  \right]  +o_{p}(N^{-1/4}),
\end{array}
\]
where for
\[%
\begin{array}
[c]{rrl}%
\text{\textbf{Sys:}} & \bar{e}= & e\sigma^{2},\\
\text{\textbf{AS:}} & = & e\left[  2\sigma^{2}-d_{2}\right]  .
\end{array}
\]
Combining our results for the two components:%
\[%
\begin{array}
[c]{l}%
(\sqrt{N}h_{N}(\theta_{0,N})A_{f}(\theta(e))_{1}\text{ }\vdots\text{ }%
\sqrt[4]{N}A_{f}(\theta(e))_{\perp})^{\prime}q_{N}(\theta(e))\\
=(\sqrt{N}h_{N}(\theta_{0,N})A_{f}(\theta(e))_{1}\text{ }\vdots\text{
}\sqrt[4]{N}(G_{f,T}(\theta(e))\text{ }\vdots\text{ }G_{2,T})\ )^{\prime}%
q_{N}(\theta(e))\\
=\left(
\begin{array}
[c]{c}%
A_{f}(\theta(e))_{1}^{\prime}A_{q}\\
\binom{\frac{1}{h_{N}(\theta_{0,N})\sqrt[4]{N}}G_{f}(\theta(e))^{\prime}A_{q}%
}{0}%
\end{array}
\right)  \bar{\psi}+\left(
\begin{array}
[c]{c}%
0\\
\binom{-\bar{e}\iota_{p}+\frac{1}{\sqrt[4]{N}}\varepsilon_{aq}}{\frac
{1}{\sqrt[4]{N}}\varepsilon_{bq}}%
\end{array}
\right)  +o_{p}(N^{-1/4}),
\end{array}
\]
where it is again important to incorporate the higher order components. We can
also specify the above convergence as%
\[%
\begin{array}
[c]{l}%
\sqrt{N}(h_{N}(\theta_{0,N})A_{f}(\theta(e))_{1}\text{ }\vdots\text{ }%
A_{f}(\theta(e))_{\perp})^{\prime}q_{N}(\theta(e))\\
=\left(
\begin{array}
[c]{c}%
A_{f}(\theta(e))_{1}^{\prime}A_{q}\\
\binom{\frac{1}{h_{N}(\theta_{0,N})}G_{f}(\theta(e))^{\prime}A_{q}}{0}%
\end{array}
\right)  \bar{\psi}+\left(
\begin{array}
[c]{c}%
0\\
\binom{-\sqrt[4]{N}\bar{e}\iota_{p}+\varepsilon_{aq}}{\varepsilon_{bq}}%
\end{array}
\right)  +o_{p}(1),
\end{array}
\]

\textbf{3. }We next determine the behavior of $\hat{V}_{\theta f}(\theta(e)):$%
\[%
\begin{array}
[c]{l}%
(h_{N}(\theta_{0,N})A_{f}(\theta(e))_{1}\text{ }\vdots\text{ }A_{f}%
(\theta(e))_{\perp})^{\prime}\hat{V}_{\theta f}(\theta(e))(h_{N}(\theta
_{0,N})A_{f}(\theta(e))_{1}\text{ }\vdots\text{ }A_{f}(\theta(e))_{\perp})=\\
\left(
\begin{array}
[c]{c}%
A_{f}(\theta(e))_{1}^{\prime}A_{q}\\
\binom{\frac{1}{h_{N}(\theta_{0,N})}G_{f}(\theta(e))^{\prime}A_{q}}{0}%
\end{array}
\right)  \Lambda\left(
\begin{array}
[c]{c}%
A_{f}(\theta(e))_{1}^{\prime}A_{f}(\theta(e))\\
0\\
0
\end{array}
\right)  ^{\prime}+\\
\left(
\begin{array}
[c]{cc}%
0 & 0\\
0 & \binom{V_{aq,abd}B(N)}{V_{bq,abd}B(N)}%
\end{array}
\right)  +o_{p}(1),
\end{array}
\]
with $V_{aq,abd},$ $V_{aq,abd}$ the covariance between $\varepsilon_{aq}$ and
$(\varepsilon_{a}^{\prime}$ $\vdots$ $\varepsilon_{b}^{\prime}$ $\vdots$
$\varepsilon_{d}^{\prime})^{\prime}$ and $\varepsilon_{bq}$ and $(\varepsilon
_{a}^{\prime}$ $\vdots$ $\varepsilon_{b}^{\prime}$ $\vdots$ $\varepsilon
_{d}^{\prime})^{\prime}$ respectively, which results directly from the
specifications in Theorem 1 (and 1* in the Appendix) and those above.

Combining with the large sample behavior of $\sqrt{N}((h_{N}(\theta
_{0,N})A_{f}(\theta(e))_{1}$ $\vdots$ $A_{f}(\theta(e))_{\perp})^{\prime}%
\hat{V}_{ff}(\theta(e))$\newline$(h_{N}(\theta_{0,N})A_{f}(\theta(e))_{1}$
$\vdots$ $A_{f}(\theta(e))_{\perp}))^{-1}(h_{N}(\theta_{0,N})A_{f}%
(\theta(e))_{1}$ $\vdots$ $A_{f}(\theta(e))_{\perp})^{\prime}f_{N}%
(\theta(e)),$ we have:
\begin{align*}
&  \sqrt{N}(h_{N}(\theta_{0,N})A_{f}(\theta(e))_{1}\text{ }\vdots\text{ }%
A_{f}(\theta(e))_{\perp})^{\prime}\hat{V}_{\theta f}(\theta(e))\hat{V}%
_{ff}(\theta(e))^{-1}f_{N}(\theta(e))\\
&
\begin{array}
[c]{cl}%
= & \sqrt{N}(h_{N}(\theta_{0,N})A_{f}(\theta(e))_{1}\text{ }\vdots\text{
}A_{f}(\theta(e))_{\perp})^{\prime}\hat{V}_{\theta f}(\theta(e))(h_{N}%
(\theta_{0,N})A_{f}(\theta(e))_{1}\text{ }\vdots\text{ }A_{f}(\theta
(e))_{\perp})\\
& ((h_{N}(\theta_{0,N})A_{f}(\theta(e))_{1}\text{ }\vdots\text{ }A_{f}%
(\theta(e))_{\perp})^{\prime}\hat{V}_{ff}(\theta(e))(h_{N}(\theta_{0,N}%
)A_{f}(\theta(e))_{1}\text{ }\vdots\text{ }A_{f}(\theta(e))_{\perp}))^{-1}\\
& (h_{N}(\theta_{0,N})A_{f}(\theta(e))_{1}\text{ }\vdots\text{ }A_{f}%
(\theta(e))_{\perp})^{\prime}f_{N}(\theta(e))\\
= & \left(
\begin{array}
[c]{c}%
A_{f}(\theta(e))_{1}^{\prime}A_{q}\\
\binom{\frac{1}{h_{N}(\theta_{0,N})}G_{f}(\theta(e))^{\prime}A_{q}}{0}%
\end{array}
\right)  \bar{\psi}+\left(
\begin{array}
[c]{c}%
0\\
\binom{V_{aq,abd}B(N)}{V_{bq,abd}B(N)}%
\end{array}
\right)  \times\\
& \left(  B(N)^{\prime}V_{abd}B(N)\right)  ^{-1}\left(  e^{2}\sigma^{2}%
\binom{\iota_{p}}{0}+B(N)^{\prime}\left(
\begin{array}
[c]{c}%
\varepsilon_{a}\\
\varepsilon_{b}\\
\varepsilon_{d}%
\end{array}
\right)  \right)  +o_{p}(1).
\end{array}
\end{align*}

\textbf{4. }For the large sample behavior of $\hat{D}_{N}(\theta(e)),$ we next
combine the behaviors of $\sqrt{N}(h_{N}(\theta_{0,N})A_{f}(\theta(e))_{1}$
$\vdots$ $A_{f}(\theta(e))_{\perp})^{\prime}q_{N}(\theta(e))$ constructed
under \textbf{2 }and $\sqrt{N}(h_{N}(\theta_{0,N})A_{f}(\theta(e))_{1}$
$\vdots$ $A_{f}(\theta(e))_{\perp})^{\prime}\hat{V}_{\theta f}(\theta
(e))$\allowbreak$\hat{V}_{ff}(\theta(e))^{-1}\allowbreak f_{N}(\theta(e))$
which is constructed under \textbf{3. }Upon combining them, the large sample
behavior of $\sqrt[4]{N}(h_{N}(\theta_{0,N})A_{f}(\theta(e))_{1}$ $\vdots$
$A_{f}(\theta(e))_{\perp})^{\prime}\hat{D}_{N}(\theta(e))$ results as%
\[%
\begin{array}
[c]{l}%
\sqrt[4]{N}(h_{N}(\theta_{0,N})A_{f}(\theta(e))_{1}\text{ }\vdots\text{ }%
A_{f}(\theta(e))_{\perp})^{\prime}\hat{D}_{N}(\theta(e))\\
=\frac{1}{\sqrt[4]{N}}\left\{  \left[  \sqrt{N}(h_{N}(\theta_{0,N}%
)A_{f}(\theta(e))_{1}\text{ }\vdots\text{ }A_{f}(\theta(e))_{\perp})^{\prime
}q_{N}(\theta(e))-\right.  \right. \\
\qquad\left.  \sqrt{N}(h_{N}(\theta_{0,N})A_{f}(\theta(e))_{1}\text{ }%
\vdots\text{ }A_{f}(\theta(e))_{\perp})^{\prime}\hat{V}_{\theta f}%
(\theta(e))\hat{V}_{ff}(\theta(e))^{-1}f_{N}(\theta(e))\right] \\
=\frac{1}{\sqrt[4]{N}}\left\{  \left(
\begin{array}
[c]{c}%
A_{f}(\theta(e))_{1}^{\prime}A_{q}\\
\binom{\frac{1}{h_{N}(\theta_{0,N})}G_{f}(\theta(e))^{\prime}A_{q}}{0}%
\end{array}
\right)  \bar{\psi}+\left(
\begin{array}
[c]{c}%
0\\
\binom{-\sqrt[4]{N}\bar{e}\iota_{p}+\varepsilon_{aq}}{\varepsilon_{bq}}%
\end{array}
\right)  -\right. \\
\left(
\begin{array}
[c]{c}%
A_{f}(\theta(e))_{1}^{\prime}A_{q}\\
\binom{\frac{1}{h_{N}(\theta_{0,N})}G_{f}(\theta(e))^{\prime}A_{q}}{0}%
\end{array}
\right)  \bar{\psi}-\left(
\begin{array}
[c]{c}%
0\\
\binom{V_{aq,abd}B(N)}{V_{bq,abd}B(N)}%
\end{array}
\right)  \times\\
\left.  \left(  B(N)^{\prime}V_{abd}B(N)\right)  ^{-1}\left(  e^{2}\sigma
^{2}\binom{\iota_{p}}{0}+B(N)^{\prime}\left(
\begin{array}
[c]{c}%
\varepsilon_{a}\\
\varepsilon_{b}\\
\varepsilon_{d}%
\end{array}
\right)  \right)  \right\}  +o_{p}(1),\\
=\left(
\begin{array}
[c]{c}%
0\\
-\binom{\iota_{p}}{0}\bar{e}%
\end{array}
\right)  +\frac{1}{\sqrt[4]{N}}\left(
\begin{array}
[c]{c}%
0\\
\nu
\end{array}
\right)  +o_{p}(N^{-1/4})\\
=-\left(
\begin{array}
[c]{c}%
0\\
\binom{\iota_{p}}{0}%
\end{array}
\right)  \bar{e}+o_{p}(1)
\end{array}
\]
where we have rescaled since all the higher order terms have dropped out and
which shows that the additional components in Theorem 1$^{\ast}$ compared to
Theorem 1 do not affect the limit behavior of $\hat{D}_{N}(\theta(e))$ up to
order $N^{-1/4}.$ The specification of $\nu$ is:%
\[%
\begin{array}
[c]{l}%
\nu=-\left(  \binom{V_{aq,abd}B(N)}{V_{bq,abd}B(N)}\right)  \left(
B(N)^{\prime}V_{abd}B(N)\right)  ^{-1}\binom{\iota_{p}}{0}e^{2}\sigma^{2}+\\
\qquad\left[  \left(
\begin{array}
[c]{c}%
\varepsilon_{aq}\\
\varepsilon_{bq}%
\end{array}
\right)  -\binom{V_{aq,abd}B(N)}{V_{bq,abd}B(N)}\left(  B(N)^{\prime}%
V_{abd}B(N)\right)  ^{-1}B(N)^{\prime}\left(
\begin{array}
[c]{c}%
\varepsilon_{a}\\
\varepsilon_{b}\\
\varepsilon_{d}%
\end{array}
\right)  \right]  ,
\end{array}
\]
which is independent of the limit behavior of $\sqrt{N}g_{f,T}(\theta(e)).$

We obtain the limit behavior of $\sqrt{N}\hat{D}_{N}(\theta(e))^{\prime}%
\hat{V}_{ff}(\theta(e))^{-1}D_{N}(\theta(e))$ from:%
\begin{align*}
&  \sqrt{N}\hat{D}_{N}(\theta(e))^{\prime}\hat{V}_{ff}(\theta(e))^{-1}\hat
{D}_{N}(\theta(e))\\
&
\begin{array}
[c]{cl}%
= & \left[  \sqrt[4]{N}(h_{N}(\theta_{0,N})A_{f}(\theta(e))_{1}\text{ }%
\vdots\text{ }A_{f}(\theta(e))_{\perp})^{\prime}\hat{D}_{N}(\theta(e))\right]
^{\prime}\times\\
& ((h_{N}(\theta_{0,N})A_{f}(\theta(e))_{1}\text{ }\vdots\text{ }A_{f}%
(\theta(e))_{\perp})^{\prime}\hat{V}_{ff}(\theta(e))(h_{N}(\theta_{0,N}%
)A_{f}(\theta(e))_{1}\text{ }\vdots\text{ }A_{f}(\theta(e))_{\perp}%
))^{-1}\times\\
& \left[  \sqrt[4]{N}(h_{N}(\theta_{0,N})A_{f}(\theta(e))_{1}\text{ }%
\vdots\text{ }A_{f}(\theta(e))_{\perp})^{\prime}\hat{D}_{N}(\theta(e))\right]
\\
= & \left[  \binom{\iota_{p}}{0}\bar{e}+\frac{1}{\sqrt[4]{N}}\nu\right]
^{\prime}\left(  B(N)^{\prime}V_{abd}B(N)\right)  ^{-1}\left[  \binom
{\iota_{p}}{0}\bar{e}+\frac{1}{\sqrt[4]{N}}\nu\right]  +o_{p}(1)\\
= & \bar{e}^{2}\binom{\iota_{p}}{0}^{\prime}\left(  B(N)^{\prime}%
V_{abd}B(N)\right)  ^{-1}\binom{\iota_{p}}{0}+o_{p}(1)
\end{array}
\end{align*}
and%
\begin{align*}
&  N^{\frac{3}{4}}\hat{D}_{N}(\theta(e))^{\prime}\hat{V}_{ff}(\theta
(e))^{-1}f_{N}(\theta(e))\\
&
\begin{array}
[c]{cl}%
= & \left[  \sqrt[4]{N}(h_{N}(\theta_{0,N})A_{f}(\theta(e))_{1}\text{ }%
\vdots\text{ }A_{f}(\theta(e))_{\perp})^{\prime}\hat{D}_{N}(\theta(e))\right]
^{\prime}\times\\
& ((h_{N}(\theta_{0,N})A_{f}(\theta(e))_{1}\text{ }\vdots\text{ }A_{f}%
(\theta(e))_{\perp})^{\prime}\hat{V}_{ff}(\theta(e))(h_{N}(\theta_{0,N}%
)A_{f}(\theta(e))_{1}\text{ }\vdots\text{ }A_{f}(\theta(e))_{\perp}%
))^{-1}\times\\
& \sqrt{N}\left[  (h_{N}(\theta_{0,N})A_{f}(\theta(e))_{1}\text{ }\vdots\text{
}A_{f}(\theta(e))_{\perp})^{\prime}f_{N}(\theta(e))\right] \\
= & \left[  \left(  B(N)^{\prime}V_{abd}B(N)\right)  ^{-\frac{1}{2}}{}%
^{\prime}\left[  \binom{\iota_{p}}{0}\bar{e}+\frac{1}{\sqrt[4]{N}}\nu\right]
\right]  ^{\prime}\times\\
& \left(  B(N)^{\prime}V_{abd}B(N)\right)  ^{-\frac{1}{2}}\left(  e^{2}%
\sigma^{2}\binom{\iota_{p}}{0}+B(N)^{\prime}\left(
\begin{array}
[c]{c}%
\varepsilon_{a}\\
\varepsilon_{b}\\
\varepsilon_{d}%
\end{array}
\right)  \right)  +o_{p}(1)\\
= & \bar{e}\binom{\iota_{p}}{0}^{\prime}\left(  B(N)^{\prime}V_{abd}%
B(N)\right)  ^{-1}\left(  e^{2}\sigma^{2}\binom{\iota_{p}}{0}+B(N)^{\prime
}\left(
\begin{array}
[c]{c}%
\varepsilon_{a}\\
\varepsilon_{b}\\
\varepsilon_{d}%
\end{array}
\right)  \right)  +o_{p}(1).
\end{array}
\end{align*}

Upon combining the behavior of the above two components, we obtain the large
sample behavior of the KLM statistic to test H$_{p}:\theta(e)=1+\frac
{e}{\sqrt[4]{N}}$ under Assumptions 1 and 2a which can for samples of (large)
size $N$ be specified as:
\[%
\begin{array}
[c]{l}%
\text{KLM(}\theta(e))\\%
\begin{array}
[c]{cl}%
= & Nf_{N}(\theta(e))^{\prime}\hat{V}_{ff}(\theta(e))^{-1}\hat{D}_{N}%
(\theta(e))\left[  \hat{D}_{N}(\theta(e))^{\prime}\hat{V}_{ff}(\theta
(e))^{-1}\hat{D}_{N}(\theta(e))\right]  ^{-1}\\
& \hat{D}_{N}(\theta(e))^{\prime}\hat{V}_{ff}(\theta(e))^{-1}f_{N}%
(\theta(e))\\
= & \left[  N^{\frac{3}{4}}\hat{D}_{N}(\theta(e))^{\prime}\hat{V}_{ff}%
(\theta(e))^{-1}f_{N}(\theta(e))\right]  ^{\prime}\left[  \sqrt{N}\hat{D}%
_{N}(\theta(e))^{\prime}\hat{V}_{ff}(\theta(e))^{-1}\hat{D}_{N}(\theta
(e))\right]  ^{-1}\\
& \left[  N^{\frac{3}{4}}\hat{D}_{N}(\theta(e))^{\prime}\hat{V}_{ff}%
(\theta(e))^{-1}f_{N}(\theta(e))\right] \\
= & \left(  e^{2}\sigma^{2}\binom{\iota_{p}}{0}+B(N)^{\prime}\left(
\begin{array}
[c]{c}%
\varepsilon_{a}\\
\varepsilon_{b}\\
\varepsilon_{d}%
\end{array}
\right)  \right)  ^{\prime}\left(  B(N)^{\prime}V_{abd}B(N)\right)
^{-1}\binom{\iota_{p}}{0}\bar{e}\\
& \left[  \bar{e}^{2}\binom{\iota_{p}}{0}^{\prime}\left(  B(N)^{\prime}%
V_{abd}B(N)\right)  ^{-1}\binom{\iota_{p}}{0}\right]  ^{-1}\\
& \bar{e}\binom{\iota_{p}}{0}^{\prime}\left(  B(N)^{\prime}V_{abd}B(N)\right)
^{-1}\left(  e^{2}\sigma^{2}\binom{\iota_{p}}{0}+B(N)^{\prime}\left(
\begin{array}
[c]{c}%
\varepsilon_{a}\\
\varepsilon_{b}\\
\varepsilon_{d}%
\end{array}
\right)  \right) \\
= & \left[  \kappa+\eta\right]  ^{\prime}\left[  \kappa+\eta\right]
+o_{p}(1)\\
\sim & \chi^{2}(\delta(N),1)+o_{p}(1),
\end{array}
\end{array}
\]
where $\bar{e}$ cancels out since it is a scalar, $\kappa=\left(  \binom
{\iota_{p}}{0}^{\prime}\left(  B(N)^{\prime}V_{abd}B(N)\right)  ^{-1}%
\binom{\iota_{p}}{0}\right)  ^{\frac{1}{2}}e^{2}\sigma^{2},$ $\eta=\left(
\binom{\iota_{p}}{0}^{\prime}\left(  B(N)^{\prime}V_{abd}B(N)\right)
^{-1}\binom{\iota_{p}}{0}\right)  ^{-\frac{1}{2}}\binom{\iota_{p}}{0}^{\prime
}\left(  B(N)^{\prime}V_{abd}B(N)\right)  ^{-1}B(N)^{\prime}\left(
\begin{array}
[c]{c}%
\varepsilon_{a}\\
\varepsilon_{b}\\
\varepsilon_{d}%
\end{array}
\right)  \sim N(0,1)$ and
\[%
\begin{array}
[c]{rl}%
\delta(N)= & (e\sigma)^{4}\binom{\iota_{p}}{0}^{\prime}(B(N)^{\prime}%
V_{abd}B(N))^{-1}\binom{\iota_{p}}{0}%
\end{array}
\]
on the right hand side of the above specification depends on $N$, which is
important to obtain an accurate approximation because of the quartic root
convergence rates. \bigskip

\newpage

\noindent\textbf{References}

\begin{description}
\item Ahn, S.C. and P. Schmidt (1995). Efficient estimation of models for
dynamic panel data.\textit{ Journal of Econometrics} \textbf{68}, 5--27.

\item Alvarez, J. and M. Arellano (2004). Robust Likelihood Estimation of
Dynamic Panel Data Models. CEMFI, Working Paper.

\item Anderson, T.W. and C. Hsiao (1981). Estimation of Dynamic Models with
Error Components. \textit{Journal of the American Statistical Association}
\textbf{76}, 598--606.

\item Anderson, T.W. and C. Hsiao (1982). Formulation and estimation of
dynamic models using panel data. \textit{Journal of Econometrics} \textbf{18}, 47--82.

\item Anderson, T.W. and H. Rubin (1949). Estimation of the Parameters of a
Single Equation in a Complete Set of Stochastic Equations. \textit{The Annals
of Mathematical Statistics} \textbf{21}, 570--582.

\item Andrews, D.W.K., M.J. Moreira and J.H. Stock (2006). Optimal Two-Sided
Invariant Similar Tests for Instrumental Variables Regression.
\textit{Econometrica} \textbf{74}, 715--752.

\item Andrews, I. (2016). Conditional Linear Combination Tests for Weakly
Identified Models. \textit{Econometrica} \textbf{84}, 2155--2182.

\item Andrews, I. and A. Mikusheva (2016). Conditional inference with a
functional nuisance parameter. \textit{Econometrica} \textbf{84}, 1571--1612.

\item Arellano, M. and O. Bover (1995). Another look at the instrumental
variable estimation of error-components models. \textit{Journal of
Econometrics} \textbf{68}, 29--51.

\item Arellano, M. and S. Bond (1991). Some Tests of Specification for Panel
Data: Monte Carlo Evidence and an Application to Employment Equations.
\textit{Review of Economic Studies} \textbf{58}, 277--297.

\item Blundell, R. and S. Bond (1998). Initial conditions and moment
restrictions in dynamic panel data models. \textit{Journal of Econometrics}
\textbf{87}, 115--143.

\item Bond, S. and F. Windmeijer (2005). Reliable Inference for GMM
Estimators? Finite Sample Properties of Alternative Test Procedures in Linear
Panel Data Models. \textit{Econometric Reviews} \textbf{24}, 1--37.

\item Bond, S., C. Nauges and F. Windmeijer (2005). Unit Roots: Identification
and Testing in Micro Panels. CEMMAP Working Paper CWP07/05.

\item Bun, M.J.G. and F. Windmeijer (2010). The weak instrument problem of the
system GMM estimator in dynamic panel data models. \textit{Econometrics
Journal} \textbf{13}, 95--126.

\item Davidson, R. and J.G. MacKinnon (2002). Graphical Methods for
Investigating the Size and Power of Hypothesis Tests. \textit{The Manchester
School }\textbf{66}$,$1-26.

\item Dhaene, G. and K. Jochmans (2016). Likelihood inference in an
autoregression with fixed effects. \textit{Econometric Theory} \textbf{31}, 1178--1215.

\item Dovonon, P. and A.R. Hall (2018). The asymptotic properties of GMM and
indirect inference under second-order identification. \textit{Journal of
Econometrics} \textbf{205}, 76--111.

\item Dovonon, P. and E. Renault (2013). Testing for common conditionally
heteroskedastic factors. \textit{Econometrica} \textbf{81}, 2561--2586.

\item Dovonon, P. A.R. Hall and F. Kleibergen (2020). Inference in
Second-Order Identified Models. \textit{Journal of Econometrics} \textbf{218}, 346--372.

\item Dufour, J.-M. (1997). Some Impossibility Theorems in Econometrics with
Applications to Structural and Dynamic Models. \textit{Econometrica}
\textbf{65}, 1365--388.

\item Dufour, J.-M. and M. Taamouti (2005). Projection-based statistical
inference in linear structural models with possibly weak instruments.
\textit{Econometrica} \textbf{73}, 1351--1365.

\item Engle, R.F. and C.W.J. Granger (1987). Co-integration and error
correction: Representation, estimation and testing. \textit{Econometrica}
\textbf{55}, 251--276.

\item Guggenberger, P., F. Kleibergen, and S. Mavroeidis (2019). A more
powerful Anderson Rubin test in linear instrumental variables regression.
\textit{Quantitative Economics} \textbf{10}, 487--526.

\item Guggenberger, P. , F. Kleibergen, S. Mavroeidis, and L. Chen (2012). On
the asymptotic sizes of subset Anderson-Rubin and Lagrange multiplier tests in
linear instrumental variables regression. \textit{Econometrica} \textbf{80}, 2649--2666.

\item Hahn, J., J. Hausman and G. Kuersteiner (2007). Long Difference
Instrumental Variable Estimation for Dynamic Panel Models with fixed effects.
\textit{Journal of Econometrics} \textbf{140}, 574--617.

\item Han, C. and P.C.B. Phillips (2010). GMM estimation for Dynamic Panels
with Fixed Effects and Strong Instruments at Unity. \textit{Econometric
Theory} \textbf{26}, 119--151.

\item Hansen, L.P. (1982). Large Sample Properties of Generalized Method
Moments Estimators. \textit{Econometrica} \textbf{50}, 1029--1054.

\item Hansen, L.P., J. Heaton and A. Yaron (1996). Finite Sample Properties of
Some Alternative GMM Estimators. \textit{Journal of Business and Economic
Statistics} \textbf{14}, 262--280.

\item Hsiao, C., M.H. Pesaran and A.K. Tahmiscioglu (2002). Maximum Likelihood
Estimation of Fixed Effects Dynamic Panel Data Models covering Short Time
Periods. \textit{Journal of Econometrics} \textbf{109}, 107--150.

\item Johansen, S. (1991). Estimation and hypothesis testing of cointegration
vectors in Gaussian vector autoregressive models. \textit{Econometrica}
\textit{59}, 1551--1580.

\item Kleibergen, F. (2005). Testing parameters in GMM without assuming that
they are identified. \textit{Econometrica} \textbf{73}, 1103--1124.

\item Kleibergen, F. (2021). Efficient size correct subset inference in
homoskedastic linear instrumental variables regression. \textit{Journal of
Econometrics} \textbf{221}, 78--96.

\item Kleibergen, F. and Z. Zhan (2020). Robust Inference for
Consumption-Based Asset Pricing. \textit{Journal of Finance} \textbf{75}, 507--550.

\item Kleibergen, F., L. Kong and Z. Zhan (2020). Identification Robust
Testing of Risk Premia in Finite Samples. Working Paper, University of Amsterdam.

\item Kruiniger, H. (2002). On the estimation of panel regression models with
fixed effects. Manuscript, Queen Mary University.

\item Kruiniger, H. (2009). GMM Estimation and Inference in Dynamic Panel Data
Models with Persistent Data. \textit{Econometric Theory} \textbf{25}, 1348--1391.

\item Madsen, E. (2003). GMM Estimators and Unit Root Tests in the AR(1) Panel
Data Model. Working Paper, Centre for Applied Micro Econometrics, University
of Copenhagen.

\item Moreira, M.J., (2003). A Conditional Likelihood Ratio Test for
Structural Models. \textit{Econometrica} \textbf{71}, 1027--1048.

\item Newey, W.K. and F. Windmeijer (2009). Generalized method of moments with
many weak moment conditions. \textit{Econometrica} \textbf{77}, 687--719.

\item Nickell, S.J. (1981). Biases in dynamic models with fixed effects.
\textit{Econometrica} \textbf{49}, 1417--1426.

\item Phillips, P.C.B. (1989). Partially Identified Econometric Models.
\textit{Econometric Theory} \textbf{5}, 181--240.

\item Phillips, P.C.B. (2018). Dynamic Panel Anderson-Hsiao Estimation with
Roots near Unity. \textit{Econometric Theory} \textbf{34}, 253--276.

\item Staiger, D. and J.H. Stock (1997). Instrumental Variables Regression
with Weak Instruments. \textit{Econometrica} \textbf{65} 557--586.

\item Stock, J.H. and J.H. Wright (2000). GMM with Weak Identification.
\textit{Econometrica} \textbf{68}, 1055--1096.
\end{description}

\end{document}